%% file: manuscript.tex
\documentclass[preprint]{aastex}

\usepackage{url}
\usepackage{lscape}
\usepackage[version=3]{mhchem}

\newcommand{\Msun}{~\textrm{M}_\sun}

\slugcomment{Accepted by The Astrophysical Journal}

\shorttitle{Molecular Clouds in the Milky Way}
\shortauthors{Rice, Goodman, Bergin, Beaumont, Dame}

\begin{document}

\title{A Uniform Catalog of Molecular Clouds in the Milky Way}

\author{Thomas S. Rice\altaffilmark{1},
Alyssa A. Goodman\altaffilmark{2},
Edwin A. Bergin\altaffilmark{1},\\
Christopher Beaumont\altaffilmark{3},
and
T. M. Dame\altaffilmark{2},
}

\vspace{0.5cm}

\affil{1: Department of Astronomy, University of Michigan, 
          311 West Hall, 1085 South University Avenue, Ann Arbor, MI 48109, USA; tsrice@umich.edu}
\affil{2: Harvard-Smithsonian Center for Astrophysics, 
          60 Garden Street, Cambridge, MA 02138, USA}
\affil{3: Counsyl, 180 Kimball Way, South San Francisco, CA 94080 }

\begin{abstract}

The all-Galaxy CO survey of \citet*{dame01} is by far the most uniform, large-scale Galactic CO survey.
Using a dendrogram-based decomposition of this survey, 
we present a catalog of 1064 massive molecular clouds throughout the Galactic plane. 
This catalog contains $2.5 \times 10^8$ solar masses, or $25^{+10.7}_{-5.8} \%$ of the Milky Way's estimated \ce{H2} mass.
We track clouds in some spiral arms through multiple quadrants.
The power index of Larson's first law, the size-linewidth relation, is consistent with $0.5$ in all regions -- possibly due to an observational bias -- 
but clouds in the inner Galaxy systematically have significantly ($\sim 30\%$) higher linewidths at a given size, indicating that their linewidths are set in part by Galactic environment.
The mass functions of clouds in the inner Galaxy versus the outer Galaxy are both qualitatively and quantitatively distinct.
The inner Galaxy mass spectrum is best described by a truncated power-law with a power index of $\gamma=-1.6\pm0.1$ and an upper truncation mass $M_0 = (1.0 \pm 0.2) \times 10^7 \Msun$,
while the outer Galaxy mass spectrum is better described by a non-truncating power law 
with $\gamma=-2.2\pm0.1$ and an upper mass $M_0 = (1.5 \pm 0.5) \times 10^6 \Msun$,
indicating that the inner Galaxy is able to form and host substantially more massive GMCs than the outer Galaxy.
Additionally, we have simulated how the Milky Way would appear in CO from extragalactic perspectives, for comparison with CO maps of other galaxies.

\end{abstract}

\keywords{
keywords
}

\clearpage

\section{Introduction} 

\input{intro.tex}

\section{Data and Methods}
\label{sec:methods}

\input{data_methods.tex}

\clearpage
\section{Quadrant-by-quadrant Results}\footnote{Please refer to Figure~\ref{fig:allquad_topdown} for a reminder on the longitude ranges corresponding to the so-called ``Galactic Quadrants.''}

\label{sec:quadbyquad}

\input{quadrants.tex}

\section{All-Galaxy Analysis}
\label{sec:all_galaxy_analysis}

\input{analysis.tex}

\clearpage
\section{Caveats and Robustness}
\label{sec:caveats}

\input{caveats.tex}

\clearpage
\section{Conclusions}
\label{sec:conclusions}

\input{conclusions.tex}

\acknowledgments

This research made use of Astropy, a community-developed core Python
package for Astronomy \citep{astropy13}.

\bibliography{manuscript}
  
\clearpage

\end{document}

%% file: intro.tex

Giant molecular clouds (GMCs) are the fundamental building blocks of star formation in disk galaxies; studying their properties and structure clarifies our understanding of the initial conditions of star formation. 
In particular, a key question in star formation is whether cloud evolution and stellar formation are independent of local conditions, or whether the local and global Galactic environment plays a strong role in how the molecular gas within GMCs is processed into stars. Studying GMCs, especially throughout the Galaxy, establishes constraints on these initial conditions, enabling subsequent detailed studies.

Our primary understanding of GMCs derives from studies of our own Galaxy.
We have learned that
molecular star-forming gas (primarily \ce{H2}) is concentrated in large, discrete clouds with a low volume-filling factor, with masses $10^4-10^6 \Msun$ and above; 
gas is concentrated along spiral arms; 
clouds have virialized internal motions, producing an observed relation between cloud size and spectral line-width \citep{larson81}; 
and that the mass distribution function of GMCs has a negative slope, with power law index $\sim -1.5$ previously measured in the Galaxy (see recent review by \citealt{heyer15} as well as classic reviews by \citealt{blitz93} and \citealt{combes91}).
To date, the only well-sampled survey of molecular gas throughout the entire Galactic plane has been the CfA-Chile survey presented by
\citeauthor*{dame01}~(\citeyear{dame01}; hereafter DHT).
This all-Galaxy CO survey is by far the most uniform, large-scale CO survey ever performed in the Galaxy, wherein all regions were observed using nearly identical techniques, instruments, research teams, and with similar sensitivity and resolution.

Our position in the Galaxy makes systematic studies of GMCs difficult.
The challenges of sky coverage, distance determination, and blending of cloud emission have hampered efforts to make a uniform all-Galaxy catalog of giant molecular clouds; as such, the catalogs available are limited in sky coverage and scope.
\citet{dame86}, \citet{scoville87}, and \citet{solomon87} made cloud catalogs from first-quadrant CO surveys. (Dame used the Columbia CO survey, while Scoville \& Solomon used the U. Mass-Stony Brook CO survey).
The more recent, high-resolution Galactic Ring Survey allowed for first-quadrant clouds to be identified on smaller scales in the Galactic plane 
\citep{roman-duval09,roman-duval10,rathborne09}. 
A catalog of the fourth quadrant was created by 
\citet{garcia14}
from the CfA-Chile southern CO survey.
\citet{sodroski91} made a catalog of 35 GMCs in the outer Galaxy using a preliminary CO survey of the region with an angular resolution of only 0\fdg5 (Dame et al. 1987). 
A limited region of the outer Galaxy was mapped at much higher angular resolution by the FCRAO outer Galaxy survey with an associated cloud catalog
(\citealt{heyer01}, \citealt{brunt03}).
Thus, while there are many regional maps of clouds in the Galaxy, there exists no algorithmically produced catalog throughout all observed parts of the Galaxy prior to this work.

Owing to the difficulty of reliably describing Galactic CO emission on a cloud-by-cloud basis (i.e., by making catalogs and analyzing the clouds therein), much of what we know about the Galactic distribution of molecular gas comes from global descriptions of all the molecular gas in the Galaxy. 
An axisymmetric modeling technique was performed by
\citet{bronfman88} and updated by \citet{heyer15} to derive the \ce{H2} surface mass density as a function of Galactic radius.
The non-axisymmetric analyses of 
\citet{clemens88}
and
\citet{nakanishi06}
were based on the assumption that the inner Galaxy CO emission was distributed in a fairly smooth disk which produces a recognizable double-Gaussian distribution in latitude at any point in longitude and velocity.
Unfortunately, these ``smoothed'' or ``statistical'' techniques  do not take advantage of a key property of molecular gas: it is concentrated into large discrete GMCs amenable to individual identification. 

Progress in making an all-Galaxy catalog of GMCs has stalled due to the technical challenges of emission blending, uncertainties in rotation curves, and resolving ambiguous kinematic distances. 
Meanwhile, advancements in instrumentation and observing techniques -- as well as a face-on perspective on the clouds, which eliminates distance ambiguities and significantly reduces blending -- have allowed extragalactic CO studies to advance rapidly.
The past 15 years have seen a proliferation of detailed, resolved studies of giant molecular clouds in nearby external galaxies, such as the Large Magellanic Cloud (e.g., \citealt{fukui08}), Small Magellanic Cloud 
\citep{mizuno01}, 
M33 \citep{engargiola03,rosolowsky03}, 
M31 (e.g., \citealt{nieten06}), 
IC 10 \citep{leroy06},
and M51 (e.g., \citealt{colombo14}). 
\citet{fukui10} presented a general summary of much of this extragalactic work.
The great advantage of extragalactic surveys is that once the requisite sensitivity and resolution is available, observations of these galaxies can freely identify GMCs without encountering serious problems of velocity crowding, cloud blending, and the kinematic distance ambiguity.

We present here a new identification and analysis of the distribution of Milky Way molecular clouds.
Our goal is to provide a more complete catalog using advanced structure-finding techniques and improvements brought about by recent advances in our understanding of the Milky Way's spiral structure \citep{dame08,dame11_arm}, distribution of stellar populations \citep{churchwell09}, and kinematics \citep{reid09,reid14}.
In particular, the BeSSeL survey \citep{reid09,reid14} of trigonometric parallaxes toward masers embedded in star-forming regions has enabled a significantly improved measurement of the Galactic rotation curve (including the Sun's motion with respect to the Galaxy), allowing for a refinement of kinematically derived distances. 
Complementary to advances in our understanding of the Galaxy are 
new techniques for analyzing spectral line datacubes of molecular gas emission. 
The dendrogram technique introduced by \citet{rosolowsky08} and \citet{goodman09} uses the hierarchical nesting of molecular emission line data to derive structural properties in a spectral line datacube, allowing simultaneous study of properties of objects at many scales.
Dendrogram-based methods make it feasible to produce a new catalog of molecular clouds throughout the Galaxy. 
In this study, we use dendrograms to decompose the Milky Way's CO emission into ``cloud'' structures, and we assign distances to those clouds using ancillary information, such as new distances based on maser observations, and constraints based on line-width size relations and Galactic latitude. 
We analyze data from the CfA-Chile survey of DHT.
The above-mentioned dendrogram technique allows us to decompose Galactic CO emission into discrete clouds more effectively than previous attempts. 

This paper presents the catalog, the methods used to create it, and global analyses of the cloud properties throughout the Galaxy.
Section \ref{sec:methods} summarizes the data and techniques.
In Section \ref{sec:quadbyquad} we present the catalog and a summary of its properties in each quadrant.
Section \ref{sec:all_galaxy_analysis} contains global analyses of the total mass in the catalog, the spatial distribution of clouds, their size-linewidth relation, and their mass function; we discuss these results in relation to prior Galactic and extragalactic work. 
In Section \ref{sec:distribution}, we present simulated images of the Milky Way's CO emission as it would be viewed by an extragalactic observer.
In Section \ref{sec:caveats}, we present a discussion of the robustness of our dendrogram-based analysis and its relevant caveats.
We present our conclusions in Section \ref{sec:conclusions}.
The full catalog of molecular clouds
is available at \url{https://dataverse.harvard.edu/dataverse/dendrogal_catalog} with DOI:10.7910/DVN/EWY90X.

%% file: data_methods.tex

\begin{figure}
  \plotone{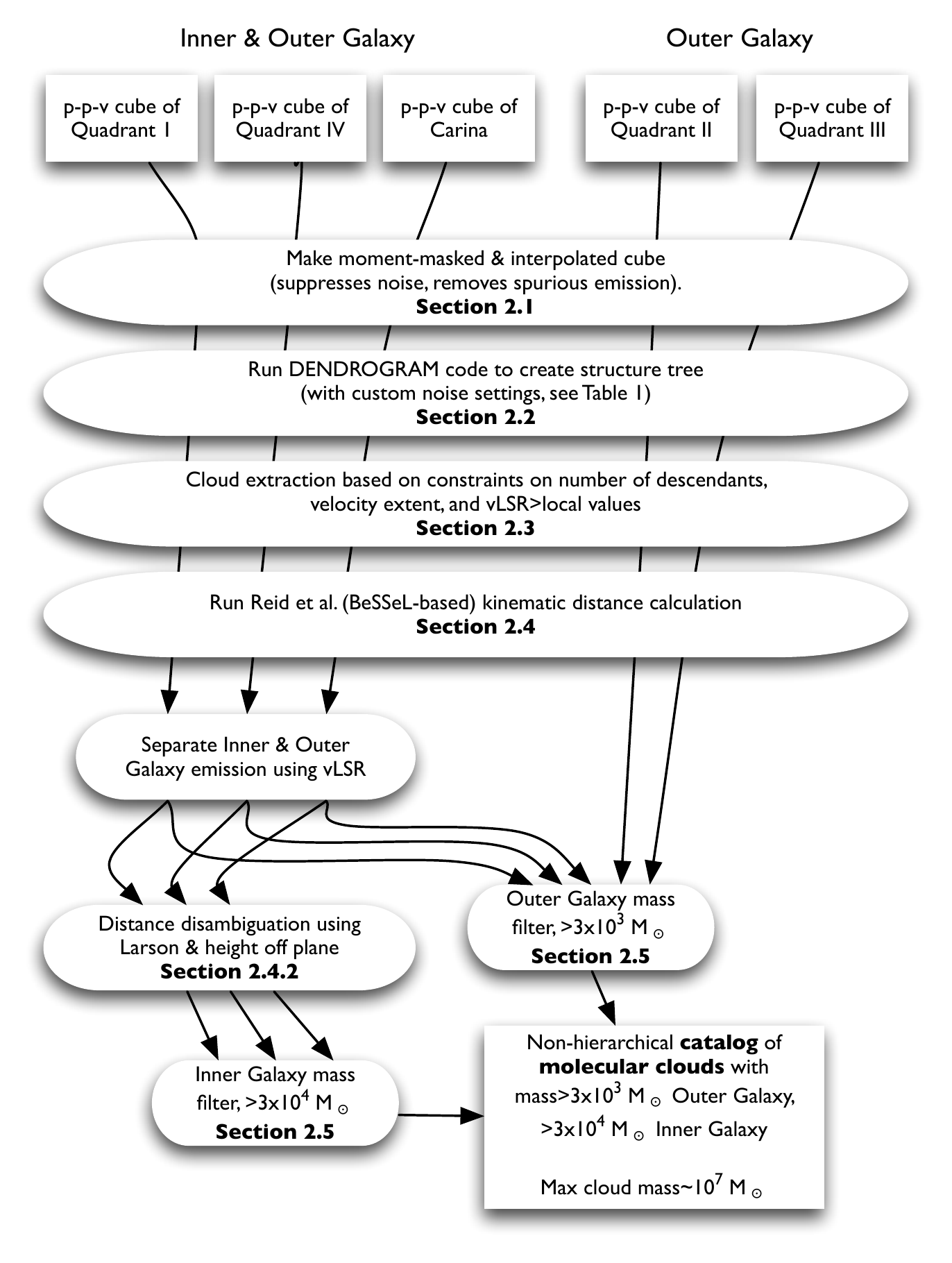}
  \caption{A flow chart explaining the methodology used in this paper. \label{fig:flowchart}}
\end{figure}

A flowchart summary of the methods in this paper is shown as Figure~\ref{fig:flowchart}. 
In summary, we perform the dendrogram technique on several of the DHT Galactic plane surveys in order to identify cloud-like emission structures throughout the Galaxy.
Distances to these objects are computed according to a maser-calibrated rotation curve. 
The kinematic distance ambiguity is resolved in a statistical sense using two classic 
techniques, one based on the observed size-linewidth relation of molecular clouds and 
the other on an assumed constant scale height for the clouds (Section \ref{sec:disambig}).

\subsection{Data and Data Manipulation}
\label{sec:data}

The data analyzed here consist of a number of Galactic plane surveys carried out with twin 1.2 m telescopes located in the Northern and Southern hemisphere, which were conducted between 1980 and 2001. 
Specifically,
the 1st quadrant (Survey \#8 from DHT), spanning Galactic longitudes $13\degr > l > 75\degr$; 
the 2nd quadrant (Survey \#17), spanning $72\degr > l > 206\degr$;
the 3rd quadrant (Survey \#31; first presented by \citealt{may93}), effectively spanning $194\degr > l > 280\degr$;
the Carina Survey (\#33; first presented by \citealt{grabelsky87}), effectively spanning $280\degr > l > 300\degr$, and
the 4th quadrant (Survey \#36; first presented by \citealt{bronfman89}), spanning $300\degr > l > 348\degr$.
The detailed properties of these surveys are displayed in Table \ref{tab:surveys}, as well as their Digital Object Identifiers (DOIs).
The spectral resolution is either 0.65 or 1.3 km s$^{-1}$, and the effective angular resolution for each survey is 0\fdg125, except for the high-latitude ($|b| > 1\degr$) region of the 4th quadrant survey, which has effective resolution of 0\fdg25.

\begin{deluxetable}{ccc|cccc}
\tablecaption{ Survey data properties 
\label{tab:surveys} }
\tablewidth{0pt}
\tablehead{
\colhead{ Name } & \colhead{ Survey \# } & \colhead{ DOI } & \colhead{ $l_{\rm{min}} (\degr)$ } & \colhead{ $l_{\rm{max}} (\degr)$ } & \colhead{ $\Delta v$ (km s$^{-1}$) } & \colhead{ noise/channel (K) }  \\
}
\startdata
Quadrant I   &  8 &  \texttt{10904/10027}    &   13 &  75 & 0.65 & 0.18 \\
Quadrant II  & 17 &  \texttt{10904/10019}    &   72 & 206 & 0.65 & 0.31 \\
Quadrant III & 31 &  \texttt{10904/10047}    &  194 & 280 & 1.30 & 0.12 \\
Carina       & 33 &  \texttt{10904/10049}    &  280 & 300 & 1.30 & 0.17 \\
Quadrant IV  & 36 &  \texttt{10904/10052}    &  300 & 348 & 1.30 & 0.12 \\
\enddata
\tablecomments{Survey parameters from DHT. The processed versions of these surveys that we created have a DOI:10.7910/DVN/ENBNTT and are hosted at \url{https://dataverse.harvard.edu/dataset.xhtml?persistentId=doi:10.7910/DVN/ENBNTT}.}
\end{deluxetable}

These surveys were obtained as complete data products from the Radio Telescope Data Center (\url{https://dataverse.harvard.edu/dataverse/rtdc}, with individual DOIs in Table~\ref{tab:surveys}).
The surveys there are available as $v-l-b$ FITS cubes with various levels of processing: (a) raw data; (b) raw, with missing values filled in via interpolation\footnote{\label{note:interp} The interpolation is applied as follows: when a spectrum is missing one or two pixels, they are filled in via linear interpolation; then, single missing pixels in each spatial plane are filled via linear interpolation, first in the $l$ direction, then along $b$.}; 
or (c) moment-masked, to suppress noise\footnote{This technique is called ``moment-masking'' due to the use of a smoothed version of the data in order to produce a ``mask'' used to remove noise. For a detailed explanation of moment masking, see \citet{dame11_moment}. These moment-masked cubes do not have missing values filled via interpolation.}.
In each region, we prepared the datacube for dendrogram analysis by starting with the moment-masked data, and creating a ``moment-masked-interpolated'' cube by interpolating according to the interpolation recipe described in Footnote~\ref{note:interp}.
The resulting datacubes are thus contiguous (i.e., do not have missing values or ``holes'') \textit{and} moment masked, to suppress noise in almost emission-free regions; both steps facilitate the following dendrogram analysis.
These ``moment-masked-interpolated'' data, which may be considered ``type (d)'' per the above scheme, are hosted at the following Dataverse link: \url{https://dataverse.harvard.edu/dataset.xhtml?persistentId=doi:10.7910/DVN/ENBNTT}.

\subsection{Dendrograms and Basic Catalog Properties}
\label{sec:dendrograms}

We use the dendrogram technique, introduced by \citet{rosolowsky08}, to segment emission data in these surveys\footnote{The website \url{http://dendrograms.org} hosts dendrogram codes, and also offers an  ``Illustrated Description of the Core Algorithm,'' explaining how dendrogramming works.}.
A dendrogram is a topological representation of the significant local maxima in $N$-dimensional intensity data and the way these local maxima are connected along contours (or isosurfaces in higher-dimensional data) of constant intensity. 
Dendrograms are abstractions of how isosurfaces nest within each other. 
Algorithmically, a dendrogram is constructed as follows:
starting at the highest intensity value in a datacube, all regions within the data that exceed the given intensity value threshold are identified, and then the nominal intensity threshold is iteratively decreased.
Every time a new, unconnected region appears above the intensity threshold, it is added to the dendrogram as an independent ``leaf''.
When the intensity threshold drops and two previously-unconnected regions merge, the merged region is added to the dendrogram as a parent (``branch'') structure containing the two previously independent structures (``leaves'').
The intensity threshold continues to decrease, with structures (both ``leaves'' and ``branches'') merging into each other, creating new ``branches'' until the intensity threshold reaches zero; the inventory of structures and their connectedness thus produced is the complete dendrogram.
The dendrogram maps each pixel in the data to the lowest-intensity structure containing it; this indexing shows how different regions are nested within each other.

Dendrograms have become a prominent technique for the analysis of both real and simulated molecular line emission data (e.g., \citealt{goodman09}, \citealt{shetty10}, \citealt{beaumont13}, \citealt{burkhart13}, and \citealt{storm14}).
Compared to the previously-used structure finding technique CLUMPFIND \citep{williams94},
dendrograms have been shown to be more robust against both noise and user-set cloud-finding parameters \citep{pineda09,goodman09}.
In particular, \citet{pineda09} found that the CLUMPFIND-derived mass spectrum power index was highly dependent on the user-selected step-size.
An advantage to dendrograms is that the intrinsic large-scale emission structures in the data are still preserved even when noise degrades small-scale structure, and dendrograms can be ``pruned'' to exclude low-significance features and thereby suppress the influence of noise (e.g., as demonstrated explicitly in \citealt{burkhart13}).

\begin{figure}
  \plotone{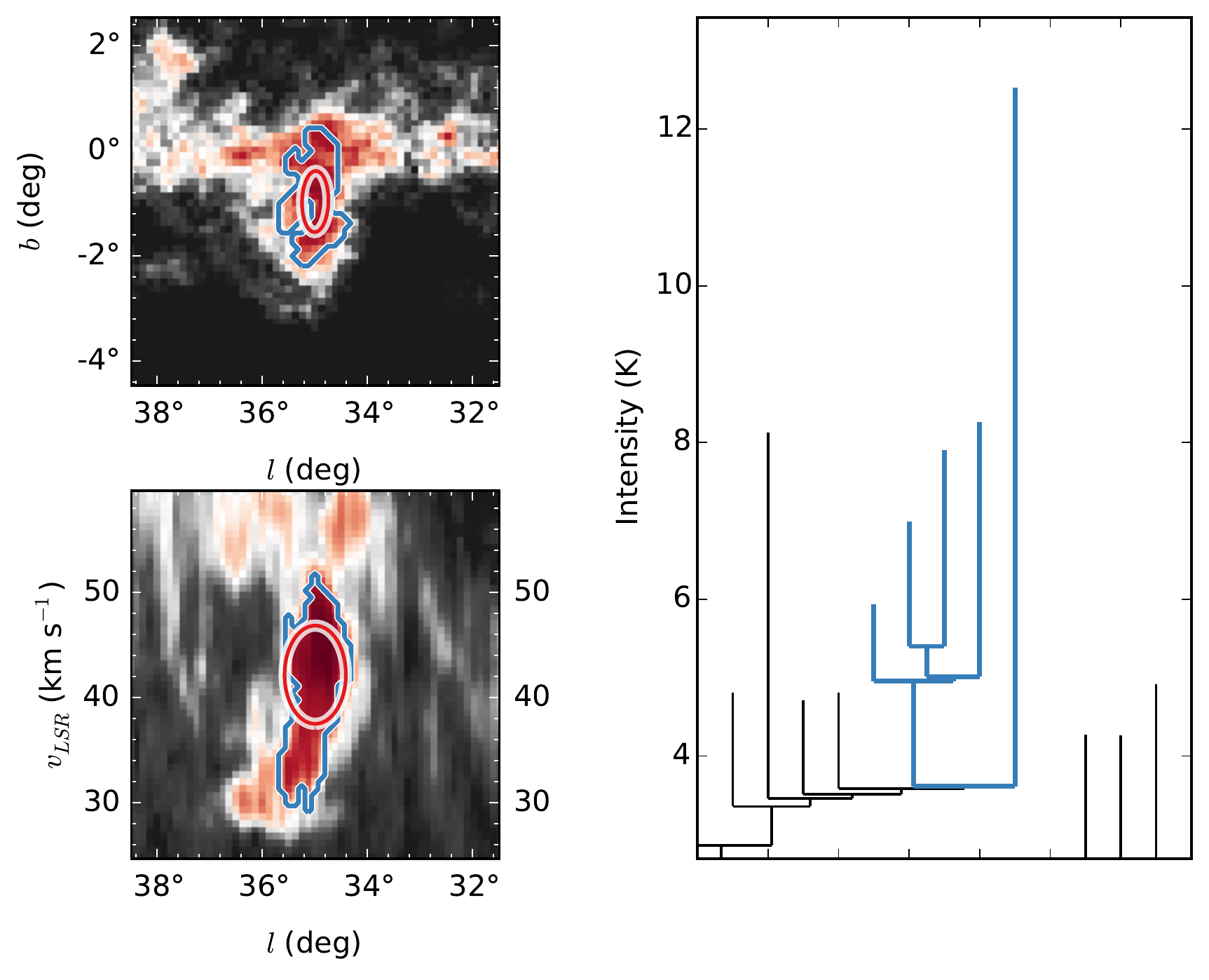}
  \caption{
  Thumbnail showing a single cloud from the first quadrant. The pixels corresponding to the highlighted dendrogram structures are outlined in the blue contour (in projection); a representative ellipse is drawn in red, with semimajor axis length equal to the second moment along each relevant dimension.
  All catalog properties for clouds are computed using the pixels inside the ``blue'' contour region.
  We find this cloud to lie at the near distance of $2.62 \pm 0.4$ kpc, implying a mass of $(4.8 \pm 2.0) \times 10^{5} \Msun$.
  \label{fig:quad1_thumb}}
\end{figure}

\begin{figure}
  \plotone{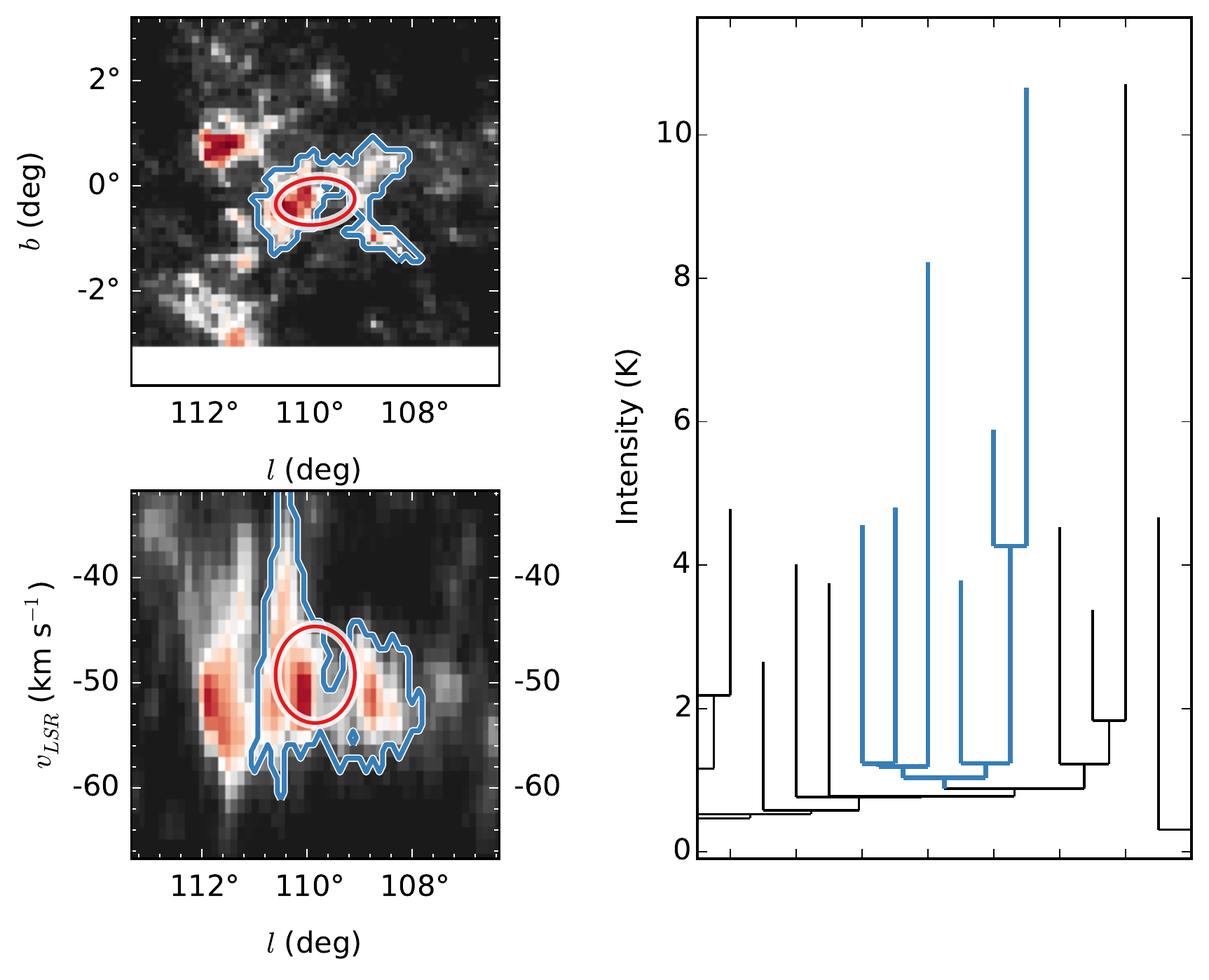}
  \caption{
  Thumbnail showing a single cloud from the second quadrant. 
  Ellipses and contours are shown as in Figure~\ref{fig:quad1_thumb}.
  We estimate this cloud to lie at $3.93^{+0.59}_{-0.58}$ kpc, with a mass of $(1.0 \pm 0.4) \times 10^6 \Msun$.
  \label{fig:quad2_thumb}}
\end{figure}

\begin{figure}
  \plotone{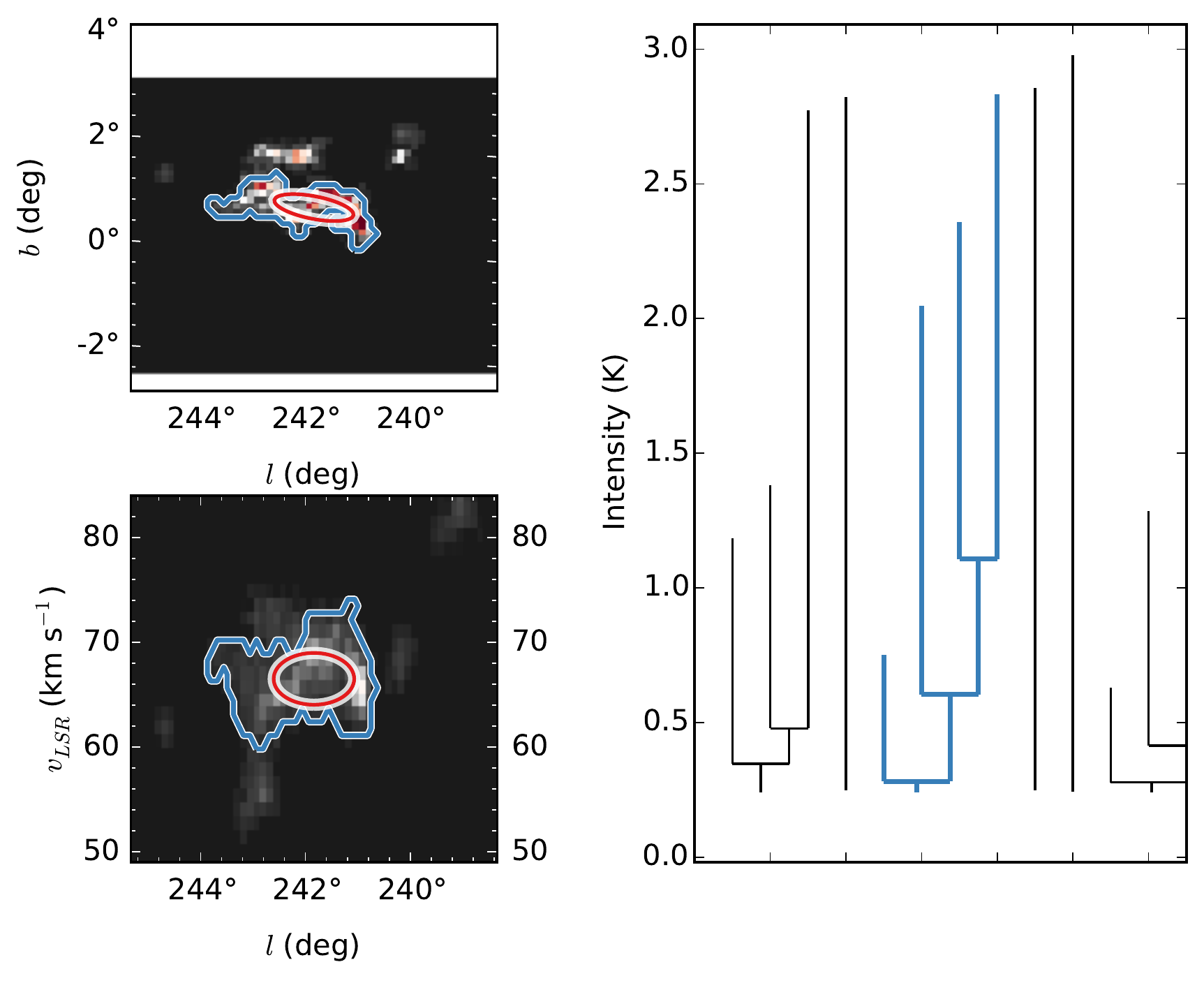}
  \caption{
  Thumbnail showing a single cloud from the third quadrant. 
  Ellipses and contours are shown as in Figure~\ref{fig:quad1_thumb}.
  We estimate this cloud to lie at $5.55^{+0.74}_{-0.69}$ kpc, with a mass of $(2.5 \pm 1.0) \times 10^5 \Msun$.
  \label{fig:quad3_thumb}}
\end{figure}

\begin{figure}
  \plotone{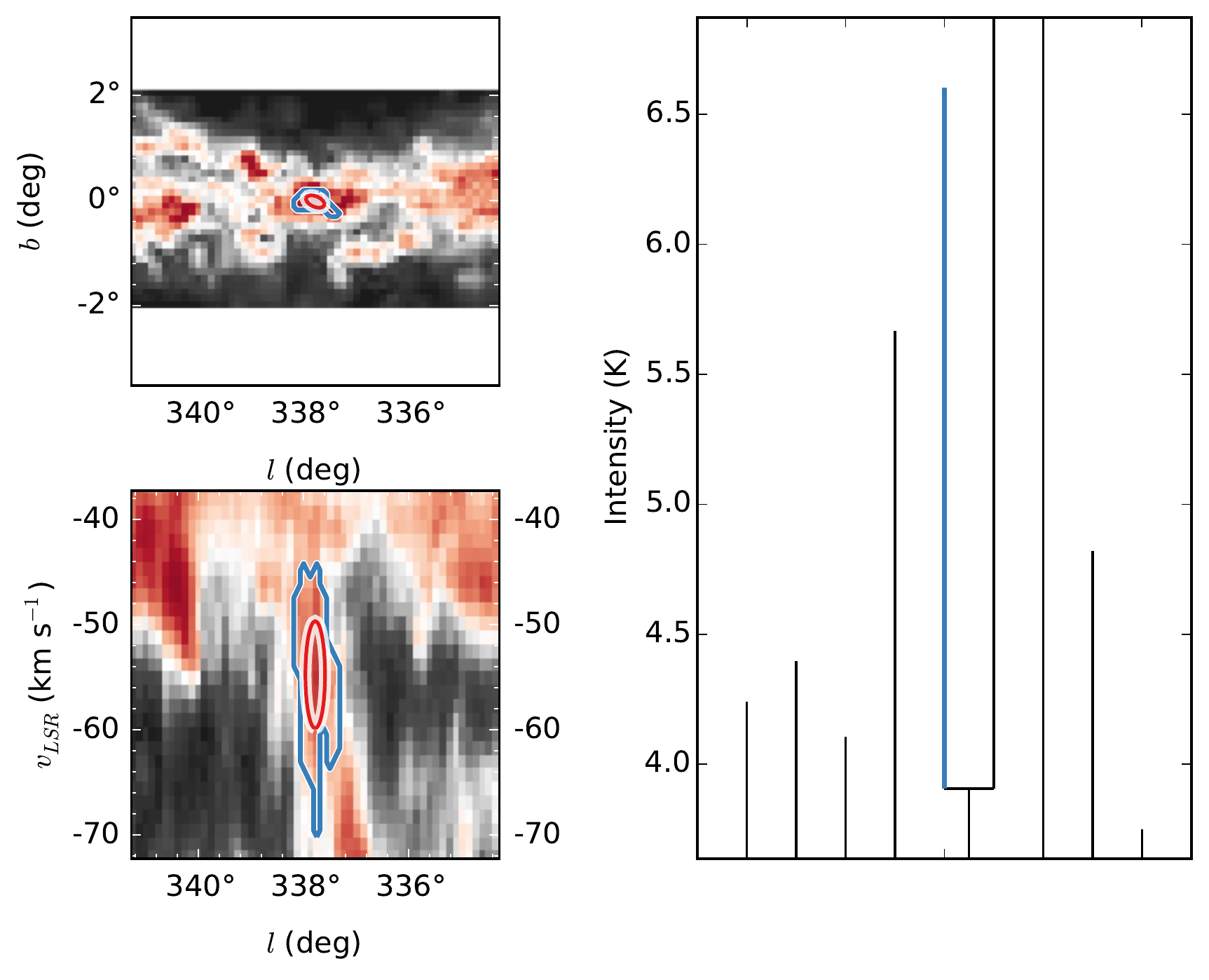}
  \caption{
  Thumbnail showing a single cloud from the fourth quadrant. 
  Ellipses and contours are shown as in Figure~\ref{fig:quad1_thumb}.
  We estimate this cloud to lie at the far distance of $11.8^{+0.38}_{-0.36}$ kpc, with a mass of $(2.6\pm0.8) \times 10^6 \Msun$.
  \label{fig:quad4_thumb}}
\end{figure}

Example dendrograms for individual clouds in the Galactic plane can be seen in Figures \ref{fig:quad1_thumb}, \ref{fig:quad2_thumb}, \ref{fig:quad3_thumb}, and \ref{fig:quad4_thumb}.
Each selected cloud is different; the diversity of clouds and their corresponding dendrograms shown in these figures is meant to highlight the versatility of the dendrogram technique for identifying clouds in isolated environments (Figs. \ref{fig:quad2_thumb} and \ref{fig:quad3_thumb}), even when they have varying amounts of substructure;
and for identifying clouds in crowded environments, as in Figs. \ref{fig:quad1_thumb} and \ref{fig:quad4_thumb}, for clouds at  near or far distances.

While the dendrogram algorithm described above is conceived of as operating on a continuously sampled, noiseless data set, real data has noise and discrete sampling.
The following parameters of dendrogram computation are thus introduced to reduce the impact of noise on the dendrogram; otherwise, every local maximum due to noise would be identified as a dendrogram leaf.
\begin{itemize}
  \item $N_{min}$: A minimum number of ``pixels'' (volume elements in position-position-velocity space) needed for a given region to be counted as a ``structure'';
  \item $(\Delta T)_{min}$: A minimum intensity contrast required to count two regions as ``separate'', i.e., a local maximum will only be counted as a separate structure if it is at least $(\Delta T)_{min}$ brighter than the intensity contour that separates it from a potentially-neighboring structure;
  \item $T_{min}$: A minimum absolute intensity threshold required to identify any region as part of the dendrogram.
\end{itemize}
For a contiguous, moment-masked cube with the resolution and noise properties described in Section \ref{sec:data}, we find that a choice of $N_{min} \sim 20$, $T_{min} \sim \sigma_{noise}$, $(\Delta T)_{min} \sim \sigma_{noise}$ effectively removes all spurious structures while preserving structure information relevant to a cloud decomposition.
A choice of $T_{min}= \sigma_{noise}$ is reasonable given that the moment-masking procedure is effective at removing spurious emission.
This choice of selection criteria has the trade-off of limiting our ability to detect small clouds at large distances.

Once a dendrogram algorithm has produced an index of structures in the data, this index can be used to catalog properties of each structure, such as integrated intensity, centroid position, spatial position angle, spatial extent, and spectral linewidth. 
The following conventions come from \citet{rosolowsky08} and \citet{rosolowsky06}, and are computed using all pixels corresponding to a given structure.
These properties are computed by taking the zeroth, first, and second moments of the intensity along the relevant axes. 
The orientation of the structure on the sky is determined by first computing its three moments of inertia (i.e. its second moments along the axes of the data), which gives the vector of its greatest elongation, and then projecting this vector onto the sky plane to determine the major axis direction. 
The minor axis is then perpendicular to the major axis.
In this convention, pixels have size $\delta \theta_l \delta \theta_b \delta v$ (corresponding to the sampling size along the $l$, $b$, and $v$ axes), and are indexed by $i$, where the intensity in a given pixel is written $T_i$:

\begin{itemize}
  \item Total flux $F$, computed as a zeroth moment:
\begin{equation}
   F = \sum_i T_i \delta \theta_l \delta \theta_b \delta v;
\end{equation}
  \item Centroid position $\langle x \rangle$, computed as a first moment along each axis $x$:
\begin{equation}
  \langle x \rangle = \frac{\sum_i T_i x_i}{\sum_i T_i};
\end{equation}
  \item Spatial size along major and minor axes $\sigma_x$ computed along two axes: along the major axis, and perpendicular to it, to obtain $\sigma_{\rm{major}}$ and $\sigma_{\rm{minor}}$
\begin{equation}
  \sigma_x = \frac{\sum_i T_i (x_i - \langle x \rangle)^2}{\sum_i T_i}
\end{equation}
  \item Overall spatial size $\sigma_r = \sqrt{\sigma_{\rm{major}} \sigma_{\rm{minor}}}$, a geometric mean of the major and minor sizes;
  \item Line width $\sigma_v$: 
\begin{equation}
  \sigma_v= \frac{\sum_i T_i (v_i - \langle v \rangle)^2 }{ \sum_i T_i.}
\end{equation}
If there are ordered motions across the body (such as expansion or rotation), they would be included in this quantity.
\end{itemize}

The above parameters, particularly flux $F$, spatial size $\sigma_r$ and line width $\sigma_v$, are computed under the ``bijection'' paradigm (see \citealt{rosolowsky08}). 
In the bijection paradigm, the total flux of a structure is simply the sum of fluxes of all pixels within that structure.
This means that no constant background flux is removed (``clipping''), and no extrapolation is done to extend the measured size down to the ``zero-intensity contour''. 
For clouds in highly crowded regions, the choice to use the ``bijection'' paradigm may have some effect on the ultimately measured properties, as discussed in \citet{rosolowsky08}.

\begin{figure}
  \plotone{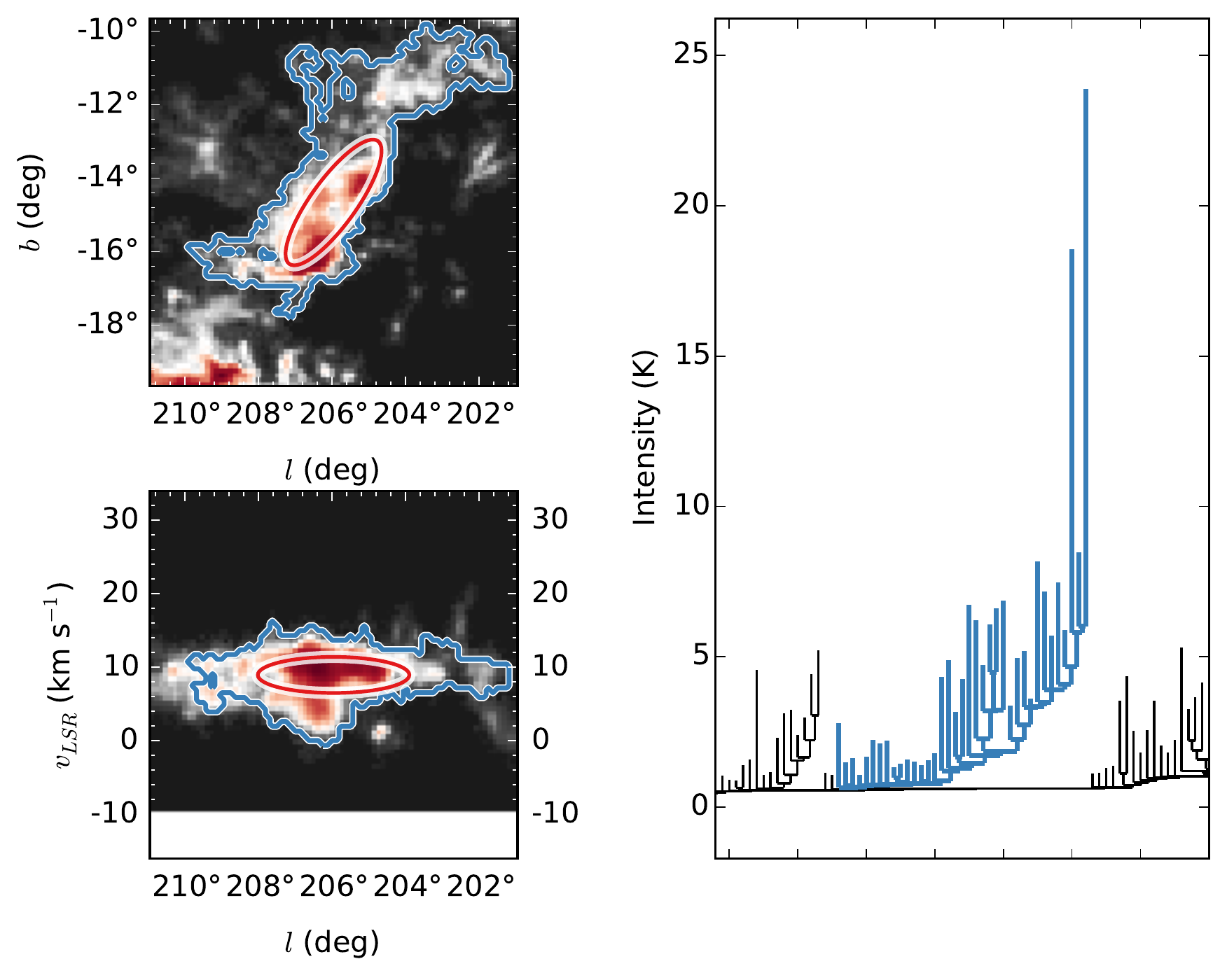}
  \caption{An example dendrogram extraction of Orion B: a nearby, well-studied giant molecular cloud.
  Top left: $(l, b)$ thumbnail of the cloud and its neighboring region as seen on the sky.
  Bottom left: $(l, v)$ thumbnail of the same region.
  Right: Dendrogram cutout, with Orion B's structures highlighted in blue.
  The pixels corresponding to the highlighted dendrogram structures are outlined in the blue contour (in projection); a representative ellipse is drawn in red, with semimajor axis length equal to the second moment along each relevant dimension (as calculated in Section \ref{sec:dendrograms}).
  Data come from DHT Survey \#27 (the Orion complex). \label{fig:orion_thumbnail}
  }
\end{figure}

An illustrative example dendrogram extraction of the Orion B molecular cloud (previously presented by \citealt{rosolowsky08}, from DHT Survey \#27, and analyzed by \citealt{wilson05}) appears in Figure~\ref{fig:orion_thumbnail}.
The figure's dendrogram panel (right) suggests that GMCs might be identified in dendrograms as self-contained, complex structures that branch out from the overall superstructure.
This suggestion is fleshed out into an algorithmic identifying criterion in the next subsection.

\subsection{Cloud Selection Criteria}
\label{sec:preliminary_clouds}

Especially in the inner Galaxy, where the emission is crowded and complex, it is difficult to isolate regions of significant emission corresponding to ``clouds''.
We have developed a technique that relies on dendrograms to identify what regions of emission correspond to GMCs, while avoiding the need to assume a background emission model.
We compute the following ``tree statistic'':

\begin{itemize}
  \item \url{n_descendants} : the number of structures that ``descend from'' a given branch, i.e., the number of substructures within a structure.
\end{itemize}

To identify the specific selection criteria to extract clouds from the CO emission dendrograms,
we made use of the first quadrant GMCs identified in \citet{dame86} as a way to anchor our cloud extraction criteria against known GMCs.
\citet{dame86} identified GMCs as bright, connected regions of emission that remained after a continuum background emission model of the inner Galaxy was subtracted.
The data used by 
\citet{dame86} was an early version of the first quadrant survey that spanned $l=12\degr-60\degr$ with $0 \fdg 125$ angular sampling within $|b| \leq 0\fdg 5$ and $0 \fdg 25$ sampling in $|b| > 0.5$, with rms sensitivity $T_R = 0.45 \textrm{ K}$ at a spectral resolution of $\Delta v = 1.3 \textrm{ km s}^{-1}$.
This catalog identified the largest known, bright, massive, well-defined molecular structures in the first quadrant, observed at similar resolution (albeit at lower sensitivity) to the data used in this study.

\begin{figure}
  \plotone{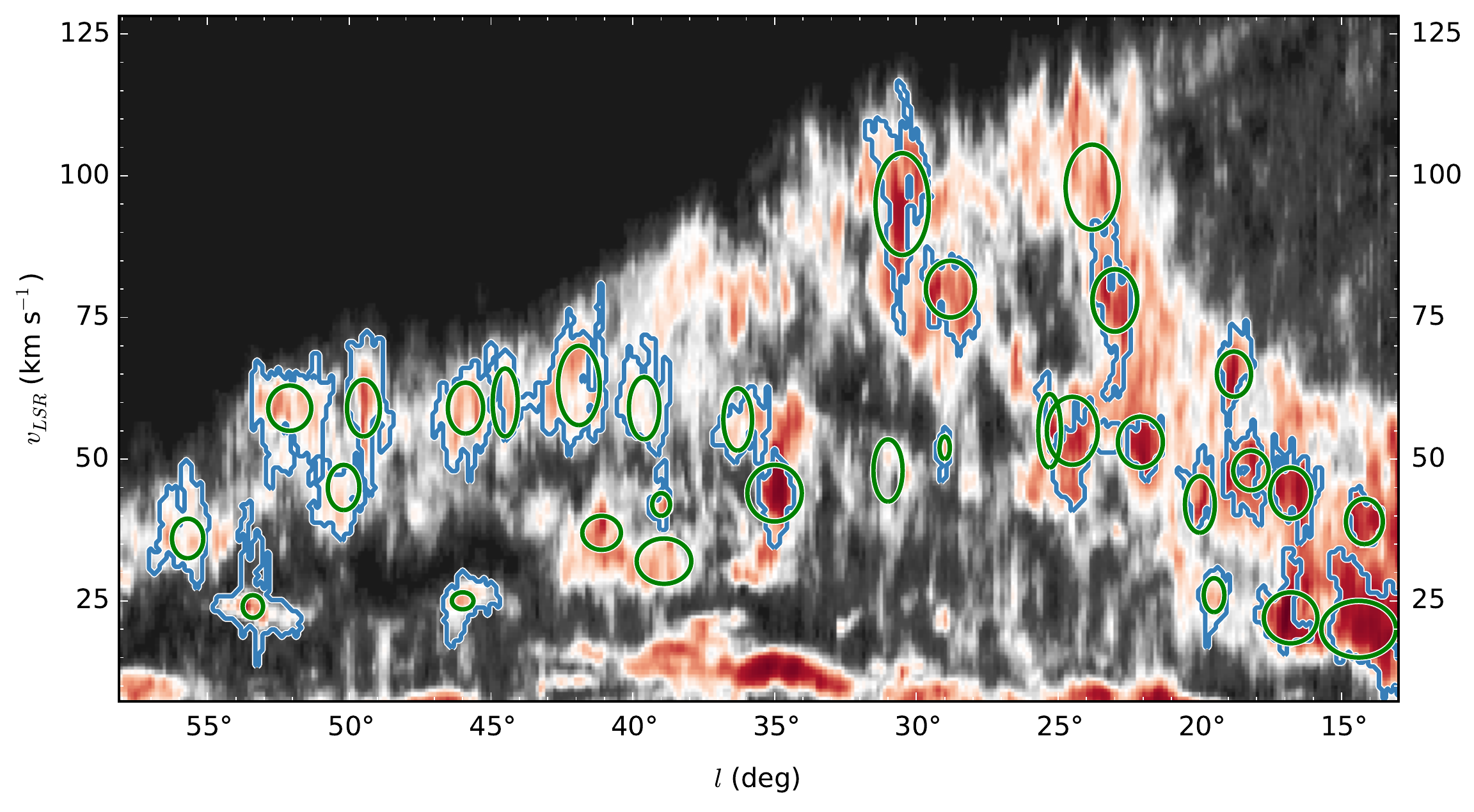}
  \epsscale{0.4}
  \plotone{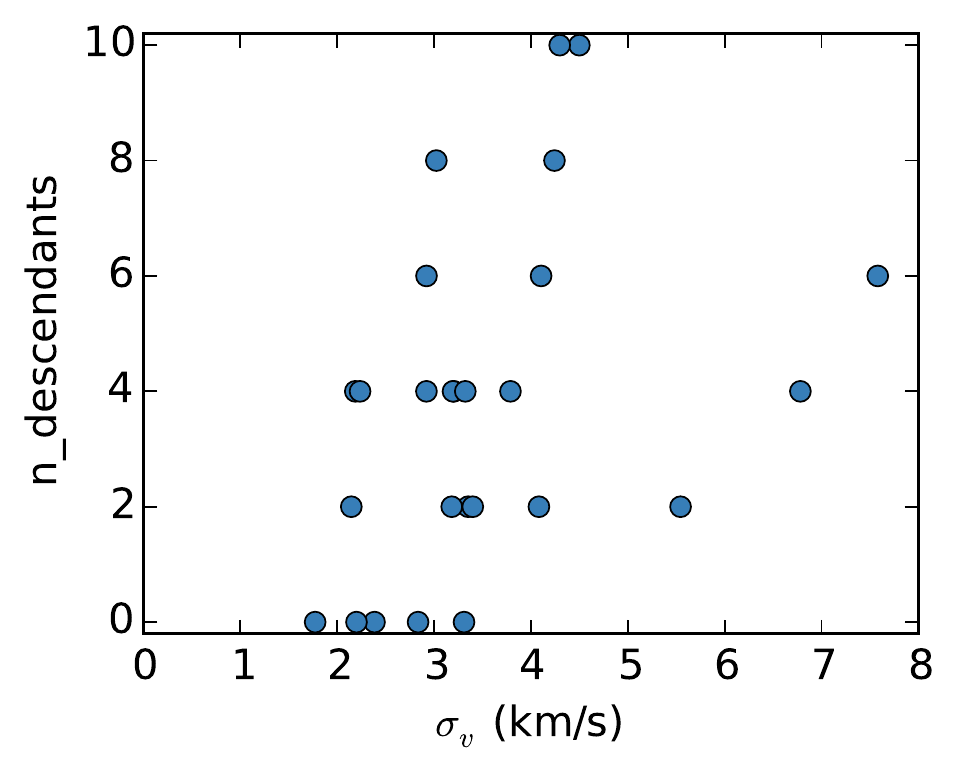}
  \epsscale{1} 
  \caption{Illustration of the selection of dendrogram emission structures corresponding to the \citet{dame86} clouds in the first quadrant, and their relevant properties. 
  \textit{Top}: zeroth moment map along the latitude axis of the first quadrant survey. Green ellipses show the location and extent of the \citet{dame86} clouds, while the blue contours show the extent of the emission region that is identified with each \citet{dame86} cloud. An accounting of non-matches between the blue contours and green ellipses is given in Footnote 4.
  \textit{Bottom}: Scatter plot of the 28 clouds, showing n-descendants versus $\sigma_v$. The cloud definitions presented in Section \ref{sec:preliminary_clouds} are derived in part from this analysis.
  \label{fig:dame_contours} }
\end{figure}

We directly identified dendrogram emission structures in the 1st quadrant corresponding to the \citet{dame86} clouds (Figure~\ref{fig:dame_contours}). 
That catalog has 32 clouds, of which 28 matched well to individual dendrogram structures; the remaining 4 \citet{dame86} clouds did not correspond to single structures\footnote{The \citet{dame86} clouds identified as (39,32) and (41,37) were not obvious decompositions of a large complex structure at $l \approx 40\degr$, $v \approx +35$ km s$^{-1}$. The emission corresponding to clouds (31,48) and (24,98) was, in each case, made of multiple resolved clouds that did not clearly merge into a larger structure.}.
By identifying the dendrogram structures corresponding to these known massive GMCs, we identified useful upper limits for cloud criteria.

As seen in Figure~\ref{fig:dame_contours}, an upper cutoff of \url{n_descendants}=10 and $\sigma_v = 10$ is sufficient to identify these massive structures.
We have investigated the result of setting \url{n_descendants} higher and lower. 
When it is set higher, the \citet{dame86} clouds begin to incorrectly merge together into large superstructures that are unlikely to be physically related; when \url{n_descendants} is lower, then the larger \citet{dame86} complexes are needlessly subdivided into smaller components.
Figure~\ref{fig:dame_contours} can be compared to Figure~\ref{fig:quad1_map} to see that this criterion is effective at identifying clouds throughout the first quadrant, and Figures \ref{fig:quad1_map}, \ref{fig:quad2_map}, \ref{fig:quad3_map}, \ref{fig:carina_map}, and \ref{fig:quad4_map} show maps in $l, v$ space of all clouds thus identified.

The above criterion is particularly critical in the inner Galaxy (i.e., the negative-velocity region of Quadrant IV and Carina, and the positive-velocity region of Quadrant I).
This definition ensures that large complexes of emission from physically un-related clouds are not merged into ``clouds'' in our cloud catalog.
In the outer Galaxy, emission tends to be uncrowded and unconnected, so the use of tree statistics is less crucial.

Finally, once a preliminary list of cloud structures was produced using the above criteria in each quadrant, we ``flattened''  the list by removing structures whose parent structures were also in the qualifying list, preventing double-counting of emission.

\subsection{Distances and Physical Catalog Properties}
\label{sec:distances}

Before physical properties like radius and mass can be assigned for the above-generated preliminary cloud list, distances must be assigned. 
Here, we rely upon the kinematic distance tool provided with \citet{reid09}, and adopt the Galactic and Solar parameters derived in \citet{reid14}, summarized in Table \ref{tab:rotationcurve}.
We use the Universal rotation curve of \citet{persic96}, which is discussed by \citet{reid14} as having the best accuracy in the entire range $0-16$ kpc from the Galactic Center.

To compute kinematic distances, we used a modified version of the Fortran program available as on-line material associated with \citet{reid09}, updated to use the \citet{persic96} rotation curve. 
This code will be made publicly available with an upcoming paper, \citet[submitted]{reid16_submitted}.
This software tool takes general Galactic parameters (e.g., as provided in Table \ref{tab:rotationcurve}), as well as information (sky position and $v_{LSR}$) about the sources for which to compute distances, and outputs the kinematic distance and an estimated error in the positive and negative directions.
The software internally transforms from the standard\footnote{The standard correction from the heliocentric frame to the LSR is to add 20 km s$^{-1}$ in the direction of $\alpha(1900)=18^{\textrm{h}}, \delta(1900)=+30^{\textrm{d}}$ (see e.g., \citealt{reid09})}  $v_{LSR}$ into a ``revised'' $v^r_{LSR}$ according to the Galactic parameters in Table \ref{tab:rotationcurve} in order to compute a more accurate kinematic distance; see \citet{reid09} for a detailed explanation of this prescription.

\subsubsection{Physical Properties}

Once distances were assigned, we computed the following physical parameters, in most cases following \citet{rosolowsky08}.
For a choice of distance $d$ to the object being studied, its radius is $R = \eta \sigma_r d$, with $\eta=1.9$ (following \citealt{rosolowsky08} and \citealt{solomon87}) representing the conversion between the CO-bright rms size of a cloud and its true radius. We calculate luminosity as 
\begin{equation}
  L = F d^2.
\end{equation}
\noindent A luminosity-based mass is computed as following:
\begin{equation}
   \frac{M_{lum}}{M_\sun} 
   = 
   \frac
   { X_{\textrm{CO}} }
   { 2\times 10^{20} [\textrm{cm}^{-2}/(\textrm{K km s}^{-1})] }  
   \times 
   4.4 
   \frac
   { L_{\textrm{CO}} }
   { 
     \textrm{K km s}^{-1} 
     \textrm{pc}^2 
   }
    = 4.4 X_2 L_{ \textrm{CO} }
    \label{eq:mass}
 \end{equation}
 \noindent where $X_\textrm{CO}$ is the CO-to-H$_2$ conversion factor \citep{bolatto13}. 
Here, $X_2$ is a normalization coefficient that reflects the choice of $X_{\rm{CO}}$ such that $X_2=1$ when $X_{\rm{CO}} =2 \times 10^{20} \rm{cm}^{-2} (\rm{K~km~s}^{-1})^{-1}$, which is the value we have adopted.
The currently accepted range of plausible $X_\textrm{CO}$ values is summarized in \citet{bolatto13}; a spread of $\pm 30\%$ is the best known range, and this spread has been incorporated into how we calculated the errors on cloud masses.

We also calculate the Cartesian ($x, y, z$) coordinates in both a solar-centric and galactocentric coordinate system. 
Note that in both cases, we use the usual right-handed coordinate systems.
The solar-centric coordinates are computed as follows, with the $x$ axis along the line connecting the Sun with the Galactic Center:

\begin{equation}
\label{eqn:local_coord}
\left( \begin{array}{c} x_\sun \\ y_\sun \\ z_\sun  \end{array} \right) =
\left( \begin{array}{c} d_{_\sun}\ \cos l\ \cos b \\ d_{_\sun}\ \sin l\ \cos b \\ d_{_\sun}\ \sin b \end{array} \right)~.
\end{equation}

\noindent To calculate galactocentric coordinates, we used the following equations from \citet{ellsworth-bowers13}, which take into account the $z_0 \approx 25$ pc height of the Sun above the Galactic midplane (\citealt{goodman14}, and references therein).

\begin{equation}
\label{eqn:xyz_gal}
\left( \begin{array}{c} x_\mathrm{gal} \\ y_\mathrm{gal} \\ z_\mathrm{gal}  \end{array} \right) =
\left( \begin{array}{c}
R_0\ \cos \theta - d_{_\sun}\ (\cos l\ \cos b\ \cos \theta + \sin b\ \sin \theta) \\
-d_{_\sun}\ \sin l\ \cos b \\
R_0\ \sin \theta - d_{_\sun}\ (\cos l\ \cos b\ \sin \theta - \sin b\ \cos \theta)
\end{array} \right) \ 
\end{equation}

\noindent where $\theta = \sin^{-1}(z_0/R_0)$.

\subsubsection{Distance Disambiguation}
\label{sec:disambig}

Along lines of sight towards the inner Galaxy, a well-known geometric ambiguity exists wherein there are two possible distance solutions for most given $v_{LSR}$ values.
We resolved this kinematic distance ambiguity (KDA), where it exists, by combining two pieces of information: the size-linewidth relationship and the latitude of the cloud in question.
In cases where neither distance provided a good fit, clouds were marked ``ambiguous'' for distance.

Clouds at high latitudes are more likely to be at the ``near'' distance (e.g., \citealt{fish03}). 
The half-width half-max (HWHM) scale height of the molecular gas layer in the Galaxy perpendicular to the Galactic plane is about 60 pc (\citealt{ellsworth-bowers13}, \citealt{bronfman88}); we have assumed a normal distribution with this HWHM, and have calculated the likelihood that $z_{gal}$ could have been drawn from this distribution for each of the two possible distances.
This gives a $p_{\textrm{latitude}}(\textrm{near})$ and a $p_{\textrm{latitude}}(\textrm{far})$.

The size-linewidth relationship (Larson's first law; \citealt{larson81}) can help resolve the distance ambiguity, as was done by \citet{solomon87}.
The two distance solutions gave two possible sizes $R$; these were each compared to a size-linewidth relationship of the form $\sigma_v = A R^\beta$.
In practice, we calculated the coefficients $A$ and $\beta$ from a size-linewidth fit of 93 unambiguously-distanced clouds in Quadrant II and III (Section \ref{sec:quadbyquad}) as a calibration.
This yielded a fit of $A=0.38\pm0.05$ and $\beta=0.49\pm0.04$, which we used as a preliminary size-linewidth distance discriminator; this $\sigma_v - R$ relation was then updated with an inner Galaxy size-linewidth fit.
We observed that outer Galaxy clouds of a given $\sigma_v$ would typically deviate by a factor 3 in $R$.
We assumed the scatter around this relationship was normally distributed in $\log_{10}(R)$, with a log-scatter of 0.5, and estimated the likelihood that each size could be drawn from the size-linewidth relationship, given a cloud's $\sigma_v$.
This gave a $p_{\rm{Larson}}(\textrm{near})$ and a $p_{\textrm{Larson}}(\textrm{far})$.

We then multiplied the likelihoods from the latitude and the size-linewidth relation together, for both the near and far distances (e.g., $p(\textrm{near}) = p_{\textrm{latitude}}(\textrm{near}) \times p_{\textrm{Larson}}(\textrm{near})$), and chose the distance with a higher combined likelihood. 
If neither distance had a $p \geq 0.05$, and the ratio between $p_{near}$ and $p_{far}$ (or vice versa) was smaller than 100, then the cloud's distance was marked as ``ambiguous''.

The above procedure was carried out once to identify a better fit for the inner Galaxy size-linewidth relation.
While we began with the outer Galaxy fit of $A=0.38$, $\beta=0.5$, we obtained a new fit of $A=0.5$, $\beta=0.5$ after the first iteration.
This new fit was stable to further iterations, so we ultimately chose it as the size-linewidth relationship for the purposes of disambiguating distances.

\begin{figure}
  \epsscale{0.9}
  \plotone{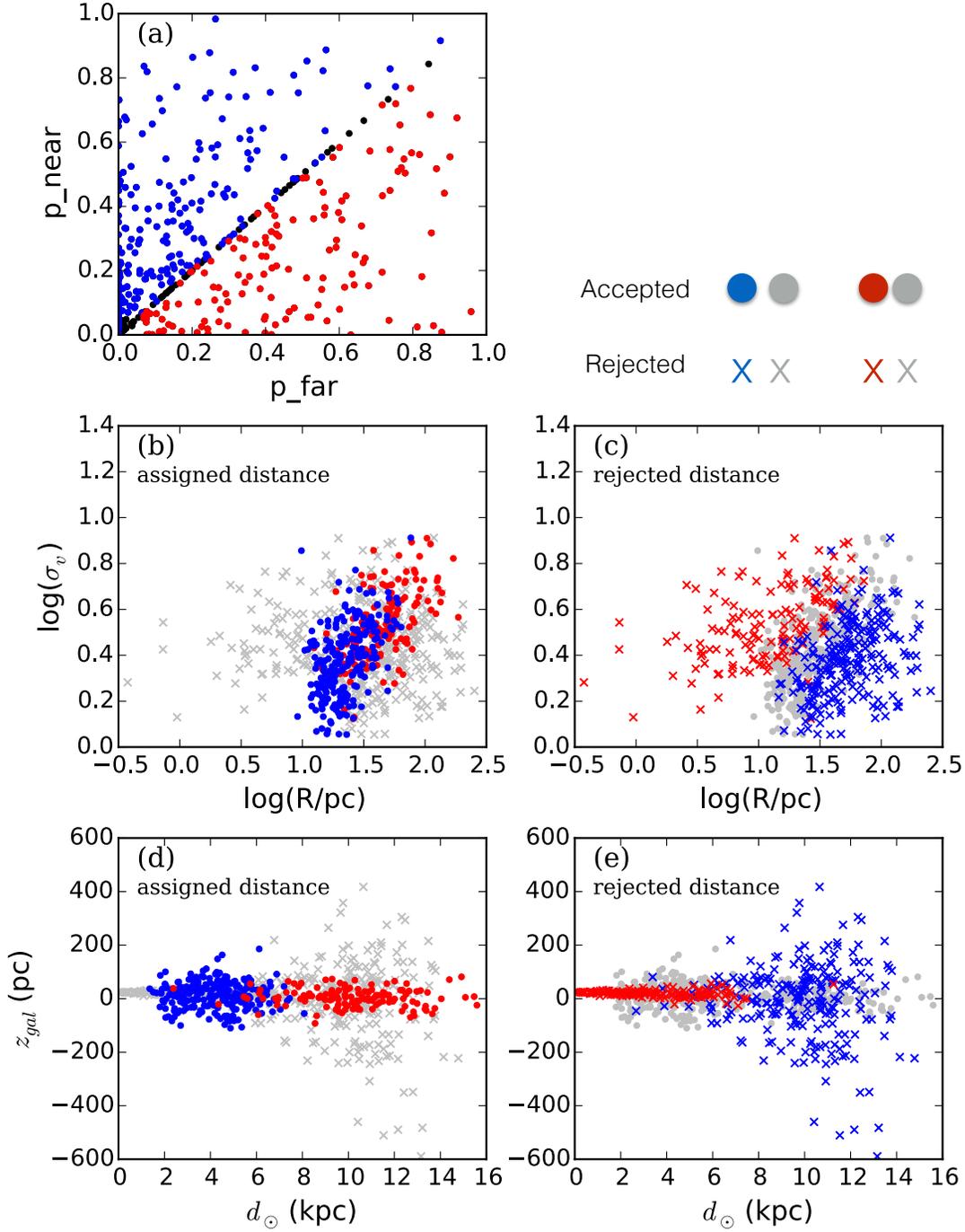}
  \caption{Illustration of our procedure to resolve the kinematic distance ambiguity in the inner Galaxy. 
  \textit{(a)}: Relative probabilities of the near versus far distances. Clouds assigned the near distance are represented with blue points, with red representing far distance clouds.
  \textit{(b, c)}: Linewidth-size plot, where each cloud has the size corresponding to both its ``near'' and ``far'' distance displayed. The size corresponding to the distance not chosen is marked with an ``x''. 
  \textit{(d, e)}: Plot showing cloud height $z$ above or below the Galactic midplane versus heliocentric distance. As in the middle row, the near $z$ and far $z$ is shown for each cloud, and the one corresponding to the distance not chosen is marked with an ``x''.
  \label{fig:probabilities}}
\end{figure}

An illustration showing the distance resolution for 376 inner Galaxy clouds appears in Figure~\ref{fig:probabilities}.
Of these clouds, 236 resolve to the near distance, while 140 resolve to the far distance.
Figure~\ref{fig:probabilities}bc shows where each inner Galaxy cloud fits into the size-linewidth relationship for both its near and far size; in many cases, the predicted size for one of the distances is orders of magnitude displaced from the expected relationship (i.e., a cloud radius of 0.3 pc or 250 pc), and it is clear that it is incorrect.
Likewise, Figure~\ref{fig:probabilities}de shows the projected height above the Galactic midplane, $z_{gal}$, for each distance.
For many nearby clouds, the ``far'' distance places them 100 pc or more from the midplane, which is unrealistically far.

\begin{figure}
  \plottwo{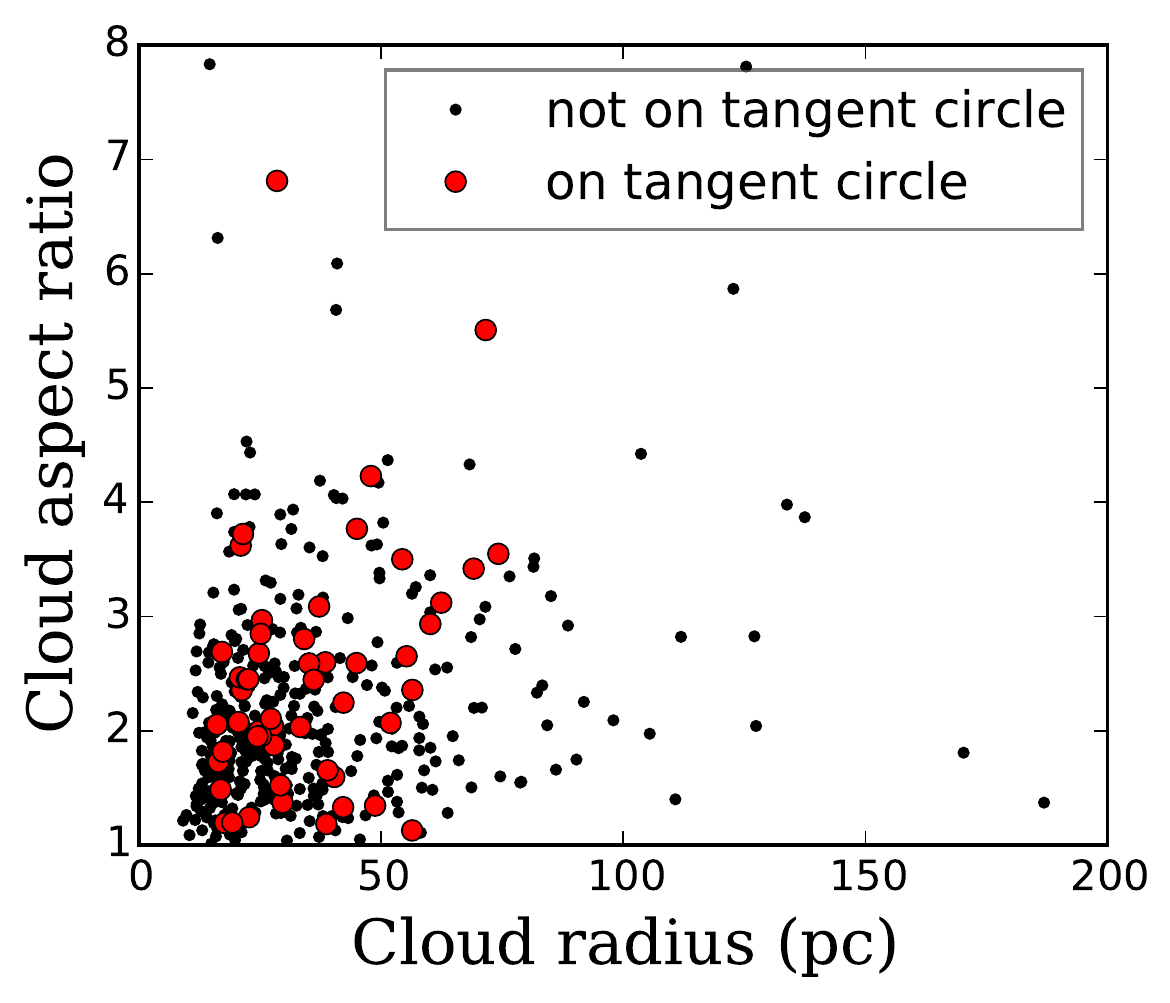}{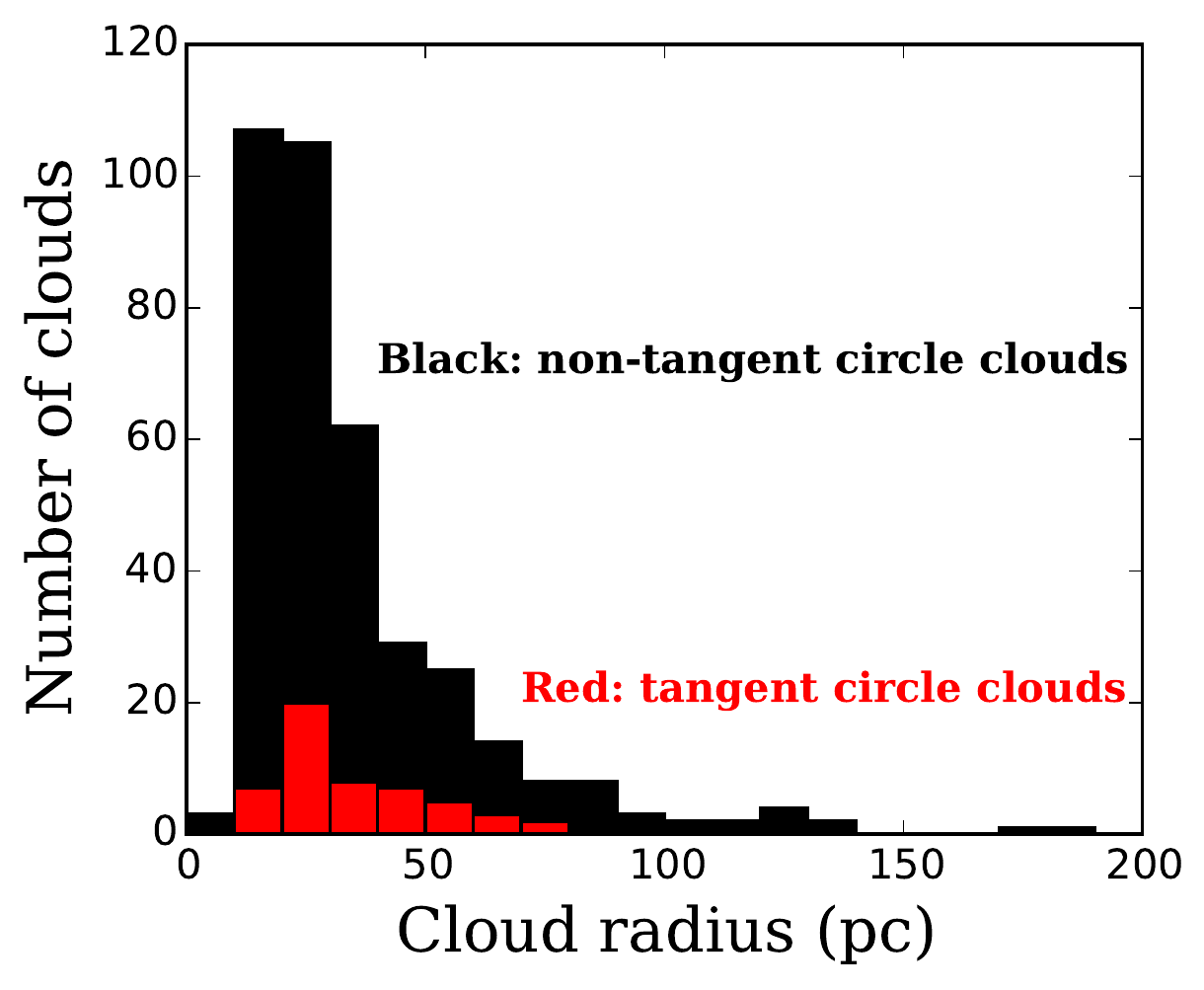}
  \caption{Investigation of possible bias in near/far disambiguation, by comparing all inner Galaxy clouds to those on the tangent circle (i.e., those that have no distance ambiguity). Left panel: Aspect ratio ($\sigma_{maj}/\sigma_{min}$) versus cloud radius, for inner Galaxy clouds. Right panel: distribution of cloud radii for inner Galaxy clouds. A handful of large ($R>100$ pc) clouds may indicate either incorrect distance assignment or blending of distant unresolved clouds. \label{fig:tangent}}
\end{figure}

To investigate whether there might be some statistical bias in the distance disambiguation, we compared the sizes and aspect ratios of inner Galaxy clouds as a whole to the subset of those which were moving at terminal velocities, i.e., those that landed on the tangent circle. 
These 52 tangent circle clouds have no distance ambiguity, as the terminal velocity at each longitude has only one associated distance.
We show in Figure \ref{fig:tangent} that the sample of inner Galaxy non-tangent clouds contains a handful of clouds with sizes more than 1.5$\times$ the largest tangent circle cloud.
This is a possible indication that among inner Galaxy clouds, there may be a slight statistical bias in the disambiguation of distances that might put some clouds at the ``far'' distance needlessly; 
alternatively, the lack of spatial resolution towards distant molecular complexes might cause multiple GMCs at the ``far'' distance to blend and merge into single ``clouds'' in this catalog.
From the data available we cannot distinguish between these two possibilities.
Otherwise, the distributions of sizes and aspect ratios for tangent and non-tangent inner Galaxy clouds are quite similar and do not show severe bias.

\begin{deluxetable}{ccc}
\tablecaption{ Galactic and Solar rotation curve parameters 
\label{tab:rotationcurve} }
\tablewidth{0pt}
\tablehead{
\colhead{ Parameter } & \colhead{ Value } & \colhead{ Definition }   \\
}
\startdata
$R_0$           & 8.34 kpc       & Distance of Sun from GC \\
$\Theta_0$      & 240.0 km s$^{-1}$     & Rotation speed of Galaxy at $R_0$ \\
$d\Theta/dR$    & $-0.2$ km s$^{-1}$ kpc$^{-1}$  & Derivative of $\Theta$ with $R$ \\
$U_\odot$       & 10.7 km s$^{-1}$      & Solar Motion toward GC \\
$V_\odot$       & 15.6 km s$^{-1}$      & Solar Motion in direction of Galactic rotation \\
$W_\odot$       &  8.9 km s$^{-1}$      & Solar Motion toward North Gal. Pole \\
$\overline{U_s}$&  2.9 km s$^{-1}$      & Average source peculiar motion toward GC \\
$\overline{V_s}$& $-1.6$ km s$^{-1}$      & Average source peculiar motion in direction of Galactic rotation \\
$\overline{W_s}$&  0.0 km s$^{-1}$      & Average source peculiar motion Ws toward North Gal. Pole \\
$\sigma_v$      &  7.0 km s$^{-1}$      & LSR velocity uncertainty (1-$\sigma$) \\
$a_1$           & 241 km s$^{-1}$       & Rotation rate for Universal rotation curve at fiducial radius \\
$a_2$           & 0.90           & Dimensionless parameter for Universal rotation curve \\
$a_3$           & 1.46           & Dimensionless parameter for Universal rotation curve \\
\enddata
\tablecomments{Galactic Rotation Parameters from \citet{reid14}.}
\end{deluxetable}

\subsection{Final Cloud Qualification}
\label{sec:qualification}

Once all of the preliminary clouds had been assigned distances and physical properties according to the above prescription, we selected likely realistic, massive clouds using the following criteria:

\begin{itemize}
 \item Most structures within $|v_{LSR}| \lesssim 10$ were excluded as ``local'' emission, unless (as with the Perseus Arm in the first quadrant) they were clearly identified with a spiral arm; the specific $v_{LSR}$ cut was quadrant-dependent, and listed in Section \ref{sec:quadbyquad}. 
 Local clouds are removed because of their highly inaccurate kinematic distances; we discuss in Section \ref{sec:distribution} how removing local emission affects our results.
 \item In the outer Galaxy, structures with an estimated mass below $3 \times 10^3 \Msun$ have been excluded from the cloud catalog. In the inner Galaxy, the minimum mass was raised to $3 \times 10^4 \Msun$, as crowding makes it difficult to identify any but the most massive clouds.
 \item Structures within $\sim 13\degr$ of the Galactic Center ($l=0\degr$) have been excluded due to unreliable kinematic distances.
 \item Structures which sat on the ``edges'' of any of the surveys were excluded as a data fidelity issue.
\end{itemize}

%% file: quadrants.tex

\subsection{First Quadrant}
\label{sec:quad1}

The first quadrant of the Galactic plane is the historically most-studied region of the Galaxy in CO, primarily because of its rich molecular cloud content and accessibility to northern-hemisphere observatories.
The first catalogs of molecular clouds in the first quadrant were compiled by 
\citet{dame86}, \citet{solomon87}, and \citet{scoville87}.
Surveys have continued to focus on the inner Galaxy in this quadrant, especially at higher resolution and using \ce{^{13}CO}; 
the recent FCRAO Galactic Ring Survey (GRS) identified hundreds of first-quadrant clouds in the Galactic plane
at $46 \arcsec$ resolution \citep{rathborne09,roman-duval10}.
In general, the clouds appearing in our catalog are larger and studied at lower spatial resolution than the aforementioned GRS clouds.
In contrast to the situation in the first quadrant inner Galaxy, no previous catalog has been made of clouds beyond the Solar circle in the first quadrant.

We used the data from the DHT08 first-quadrant survey here, which consists of 37,610 spectra between $13\degr <l<75\degr$ and roughly $-5 \degr < b < 5\degr$, sampled every $0\fdg 125$ on an $l$, $b$ grid. 
The spectral resolution is $\Delta v = 0.65 \textrm{ km s}^{-1}$.
As discussed in Section \ref{sec:methods}, we used the moment-masked survey provided by the RTDC, and then applied an additional interpolation step to ensure the data were contiguous.

\begin{landscape}

\begin{figure}
  \plotone{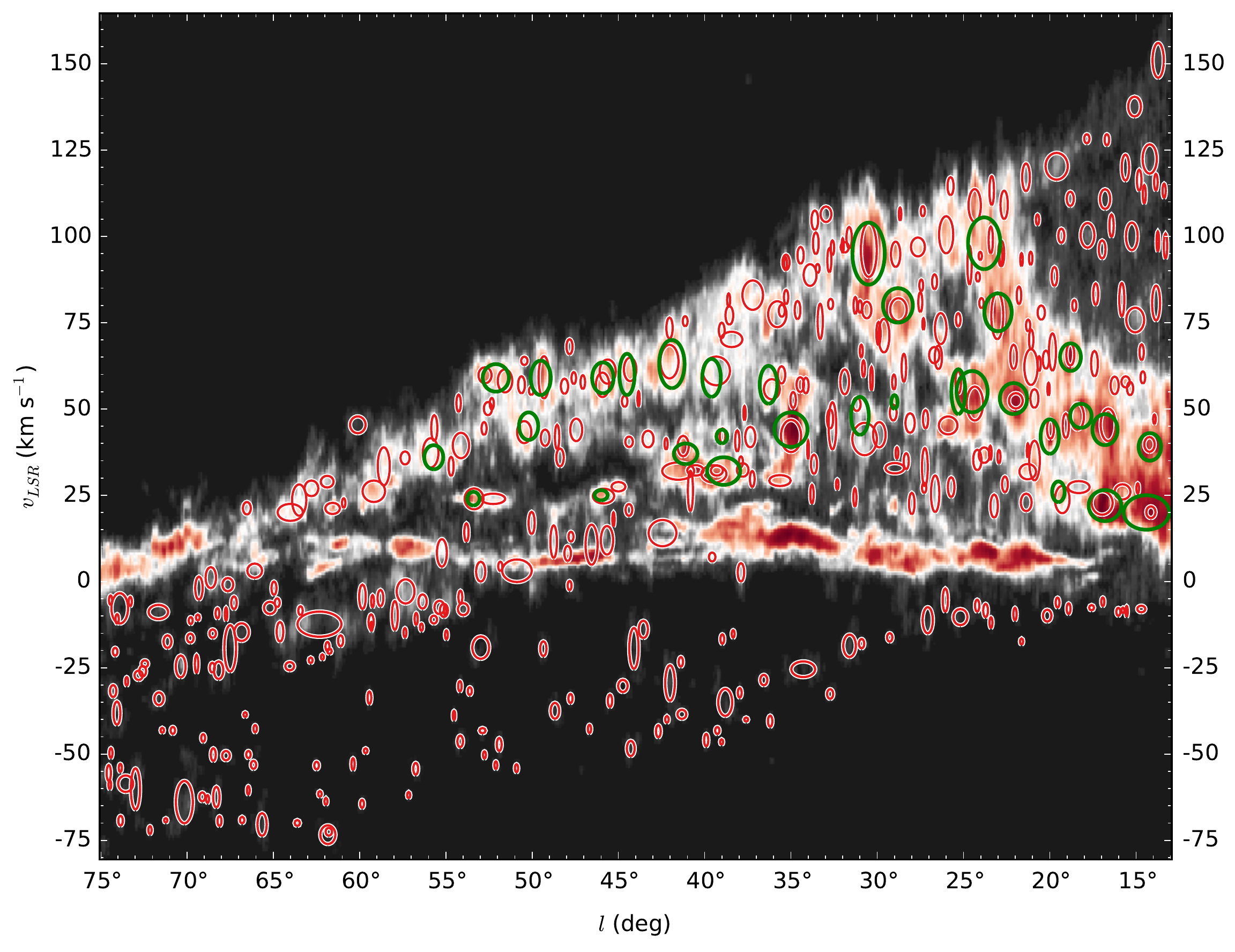}  
  \caption{Map of clouds identified in the first quadrant survey (DHT \#08),
  which shows the \citet{dame86} clouds superposed upon the clouds identified in this work for the purposes of comparison. 
  Green ellipses: \citet{dame86} clouds.
  Red ellipses: Clouds from this work.
  Background shows moment-0 integrated (along latitude) map of the DHT survey.  
  \label{fig:quad1_map}}
\end{figure}

\end{landscape}

389 clouds were identified from the first quadrant survey, with a total mass of $9.69 \times 10^7 \Msun$.
A position-velocity map of these is presented as Figure~\ref{fig:quad1_map}.
In this quadrant, the inner Galaxy emission appears at positive velocities, while that outside the solar circle appears at negative velocities. 
Because of the greater crowding in the inner Galaxy, the positive-velocity portion of this survey uses the stricter inner Galaxy criteria described in Section \ref{sec:qualification}. 
The Perseus arm is well-defined in $l,b,v$ space in this quadrant, and cuts through the region of velocity space otherwise labeled ``local''; we extracted clouds in this region when their high linewidths indicated non-local distances, and when they were not severely blended with local emission.

\subsubsection{Comparison with \citet{dame86} clouds}
\label{sec:quad1_comparison}

Here we compare between the clouds found in the first Galactic quadrant by \citet{dame86} and the clouds found by our approach.
As discussed in Section \ref{sec:preliminary_clouds}, the \citet{dame86} clouds are used to help tune the cloud definition criteria, but they are not explicitly put in the catalog by-hand. 
Figure~\ref{fig:quad1_map} demonstrates that effectively all of the emission from the \citet{dame86} clouds (the green ellipses) is recovered, often as single corresponding clouds in our catalog, or a few slightly smaller (but still massive) individual clouds.

The upper-right region, at longitudes below $l=25\degr$ and velocities above $v_{LSR} \approx 75 \textrm{ km s}^{-1}$,
has no clouds in the \citet{dame86} catalog because of the rarity of massive GMCs in this region due to the Galactic bar, which has swept out much of the molecular gas at $R_{gal} < 3$ kpc.
The remaining molecular clouds are lower mass, and were not detected with the lower-sensitivity \citet{dame86} survey.
Both the \citet{dame86} catalog and this one choose to omit most local emission, i.e. the dense horizontal bar near $v=0$.

\subsection{Second Quadrant}
\label{sec:quad2}

The second quadrant of the Galactic plane has received substantially less attention than the first quadrant, despite its ease of observability, perhaps because it contains much less molecular mass.
The first catalog of molecular clouds in the second quadrant was \citet{casoli84}, which identified clouds in a region $l=108\degr - 112 \degr$ and around $l=111\degr$ and $126 \degr$, covering 12 square degrees total.
A larger-scale outer Galaxy catalog, comprising clouds in both the second and third quadrants, is presented by \citet{sodroski91}, who identified 35 cloud complexes from an earlier lower-resolution CO survey of the entire Galactic plane \citep{dame87}. 
The FCRAO outer Galaxy survey observed longitudes $l=108-141 \degr$ at angular resolution $50\arcsec$, creating a catalog of clouds \citep{heyer01,brunt03}. 
Many of these clouds were very small, often smaller than 10 pc in radius and below $10^3 \Msun$ in mass.

We used the data from the DHT17 second-quadrant survey here, which consists of 146,944 spectra between $72\degr <l<205\degr$ and roughly $-2 \degr < b < 2\degr$, sampled every $0\fdg 125$. 
The spectral resolution is $\Delta v = 0.65 \textrm{ km s}^{-1}$.
As discussed in Section \ref{sec:methods} and just like the first quadrant (Section \ref{sec:quad1}), we used the moment-masked survey provided by the RTDC, and then applied an additional interpolation step to ensure the data were contiguous.

\begin{landscape}

\begin{figure}
  \plotone{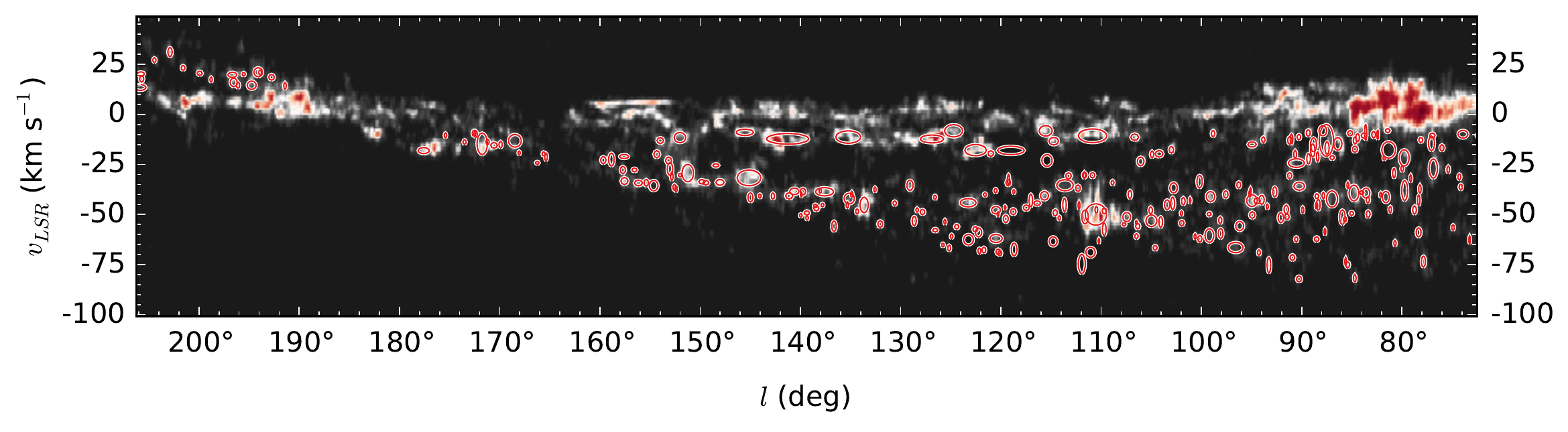}  
  \caption{Map of clouds identified in the second quadrant survey (DHT \#17).
  \label{fig:quad2_map}}
\end{figure}

\end{landscape}

276 clouds exceeding $3 \times 10^3 \Msun$ were identified in this quadrant, with a total mass of $1.21 \times 10^7 \Msun$.
A position-velocity map of these clouds is presented as Figure~\ref{fig:quad2_map}.
In this survey, all non-local emission in the second quadrant appears at negative velocities. 
The survey also extends a small amount into the third quadrant, where the non-local emission emerges at positive velocities.
Clouds are identified in these regions per the outer Galaxy criteria described in Section \ref{sec:qualification}.

\subsection{Third Quadrant}
\label{sec:quad3}

The third Galactic quadrant is largely inaccessible to northern observatories and contains relatively little molecular mass.
A catalog of clouds based on the CfA-Chile survey of this quadrant was presented by \citet*{may97}.
The \citet{sodroski91} catalog discussed above in Section \ref{sec:quad2} also contains clouds in this region.

We used the data from the DHT31 third-quadrant survey \citep{may93} here, which consists of 6,750 spectra between $194\degr <l<296\degr$ in a strip $2-4\degr$ wide in latitude (variable), sampled every $0\fdg 125$. 
The spectral resolution is $\Delta v = 1.3 \textrm{ km s}^{-1}$.
As discussed in Section \ref{sec:methods} and just like the first quadrant (Section \ref{sec:quad1}), we used the moment-masked survey provided by the RTDC, and then applied an additional interpolation step to ensure the data were contiguous.
Because this survey overlaps with the higher-resolution second quadrant survey at low longitudes, and the Carina survey (which has better velocity coverage) at high longitudes, we restrict our cloud selection to the longitude range of $l=205-280$ here.

\begin{landscape}

\begin{figure}
  \plotone{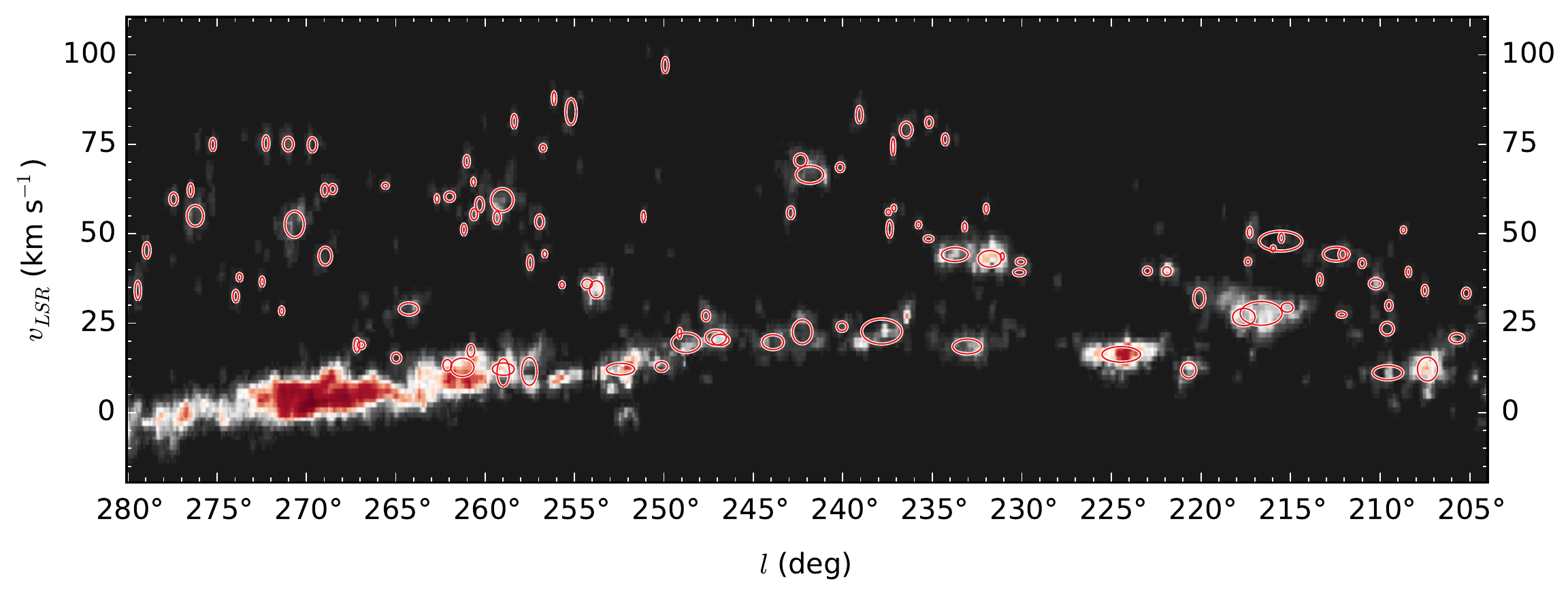}
  \caption{Map of clouds identified in the third quadrant survey (DHT \#31).
  \label{fig:quad3_map}}
\end{figure}

\end{landscape}

110 clouds exceeding $3 \times 10^3 \Msun$ are identified in this survey, totaling $4.24 \times 10^6 \Msun$.
An $l,v$ map of these clouds is shown as Figure~\ref{fig:quad3_map}.
In this survey, all non-local emission appears at positive velocities. 
Clouds are identified in these regions per the outer Galaxy criteria described in Section \ref{sec:qualification}.
Because the longitude coverage in this survey is substantially less than that of the second quadrant survey, it is expected that fewer clouds with less total mass would be found here.

\subsection{Fourth Quadrant}
\label{sec:quad4}

The fourth quadrant of the Galactic plane is as important scientifically as the first quadrant, as it encompasses the other half of the \ce{H2}-rich inner Galaxy.
It has received somewhat less attention than the first quadrant due to its unobservability from northern telescopes.
The GMCs in the fourth quadrant outside the solar circle were mapped and cataloged by \citet{grabelsky87,grabelsky88} and identified as belonging to the Carina spiral arm.
No catalog of massive molecular complexes was made within the solar circle in the fourth quadrant until recently, unlike the corresponding region in the first quadrant. 
\citet{garcia14} present the first GMC catalog in the inner Galaxy fourth quadrant using the DHT \#36 CfA-Chile survey \citep{bronfman89}, applying a background-subtracting method very similar to \citet{dame86} in order to identify 92 GMCs, of which the distance ambiguity was resolved for 87 GMCs.

In this region, we used the data from two surveys:
\begin{itemize}
  \item The DHT33 Carina survey \citep{grabelsky87}, which consists of 6,162 spectra between $270\degr <l<300\degr$ and roughly $-1 \degr < b < 1\degr$, sampled every $0\fdg 125$ on an $l$, $b$ grid. 
  \item The DHT36 fourth quadrant survey \citep{bronfman89}, which consists of 7,195 spectra between $300\degr <l<348\degr$ and $-2 \degr < b < 2\degr$, sampled every $0\fdg 125$ in $|b| \leq 1\degr$ and every $0\fdg 25$ at $1\degr<|b|<2\degr$.
\end{itemize}
The spectral resolution of these two surveys is $\Delta v = 1.3 \textrm{ km s}^{-1}$.
For the Carina survey, we used the moment-masked survey provided by the RTDC, and then applied an additional interpolation step to ensure the data were contiguous (as discussed in Section \ref{sec:methods}).
For the fourth quadrant survey, we instead started with the interpolated data because of its greater latitude coverage, which is desirable for a dendrogram analysis. 
From this, we created an ``interpolated-moment-masked'' cube from the interpolated data by applying moment masking. 
In this moment-masking step, we used a noise level appropriate for the lower-quality interpolated regions at higher latitude in order to avoid including noise inappropriately anywhere in the datacube.\footnote{In other words, the noise chosen for the purposes of moment masking is the noise in the interpolated region, which is twice the noise in the fully-sampled region.}
In this survey, which is interpolated at high latitudes, the noise is spatially correlated. 
Thus, we chose to use $T_{min}=3\sigma_{noise}$ instead of the usual $T_{min}=\sigma_{noise}$ when constructing its dendrogram in order to prevent spurious emission being counted as clouds.

\begin{figure}
  \plotone{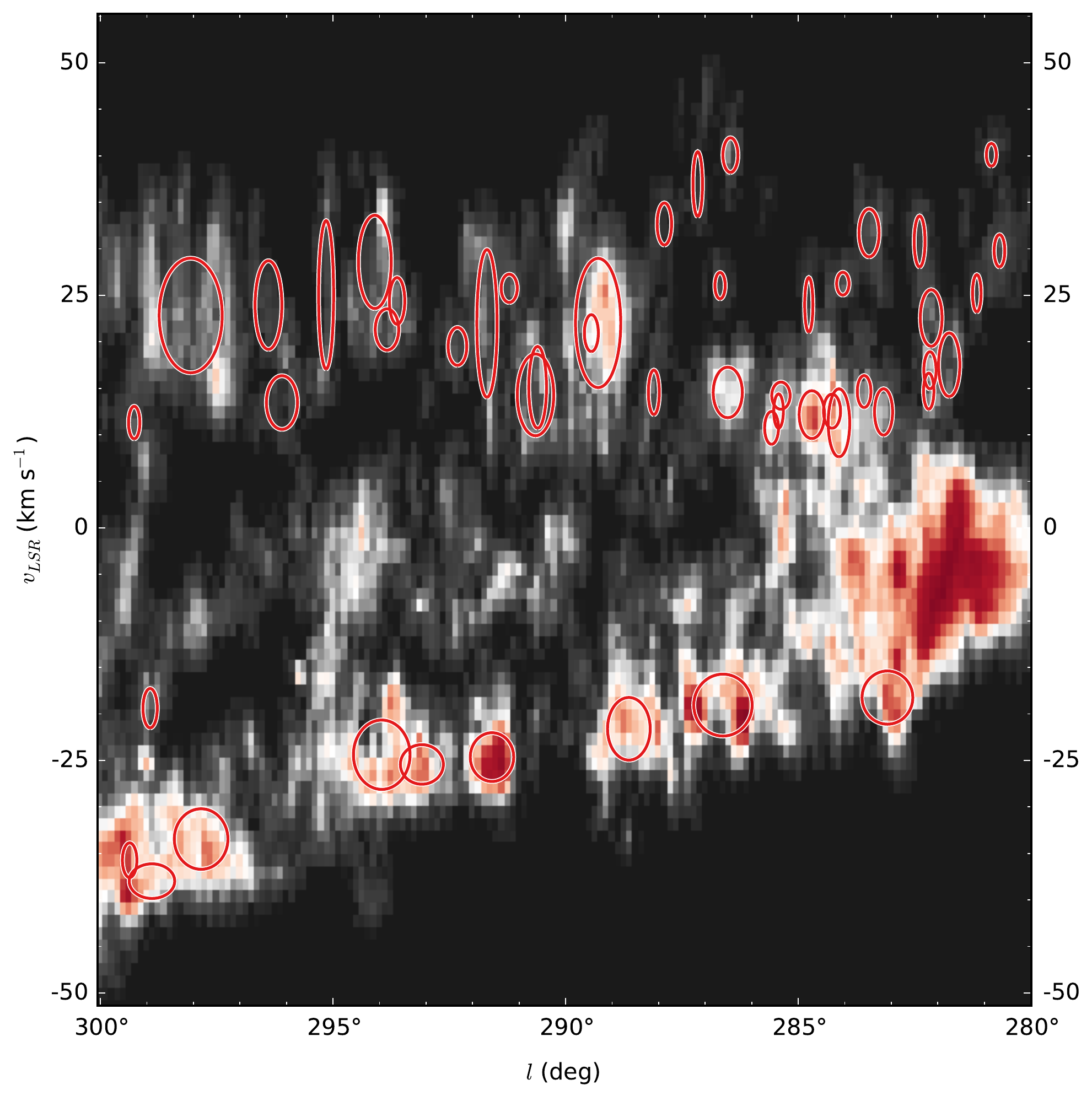}  
  \caption{Map of clouds identified in the Carina survey (DHT \#33).
  \label{fig:carina_map}}
\end{figure}

\begin{landscape}

\begin{figure}
  \plotone{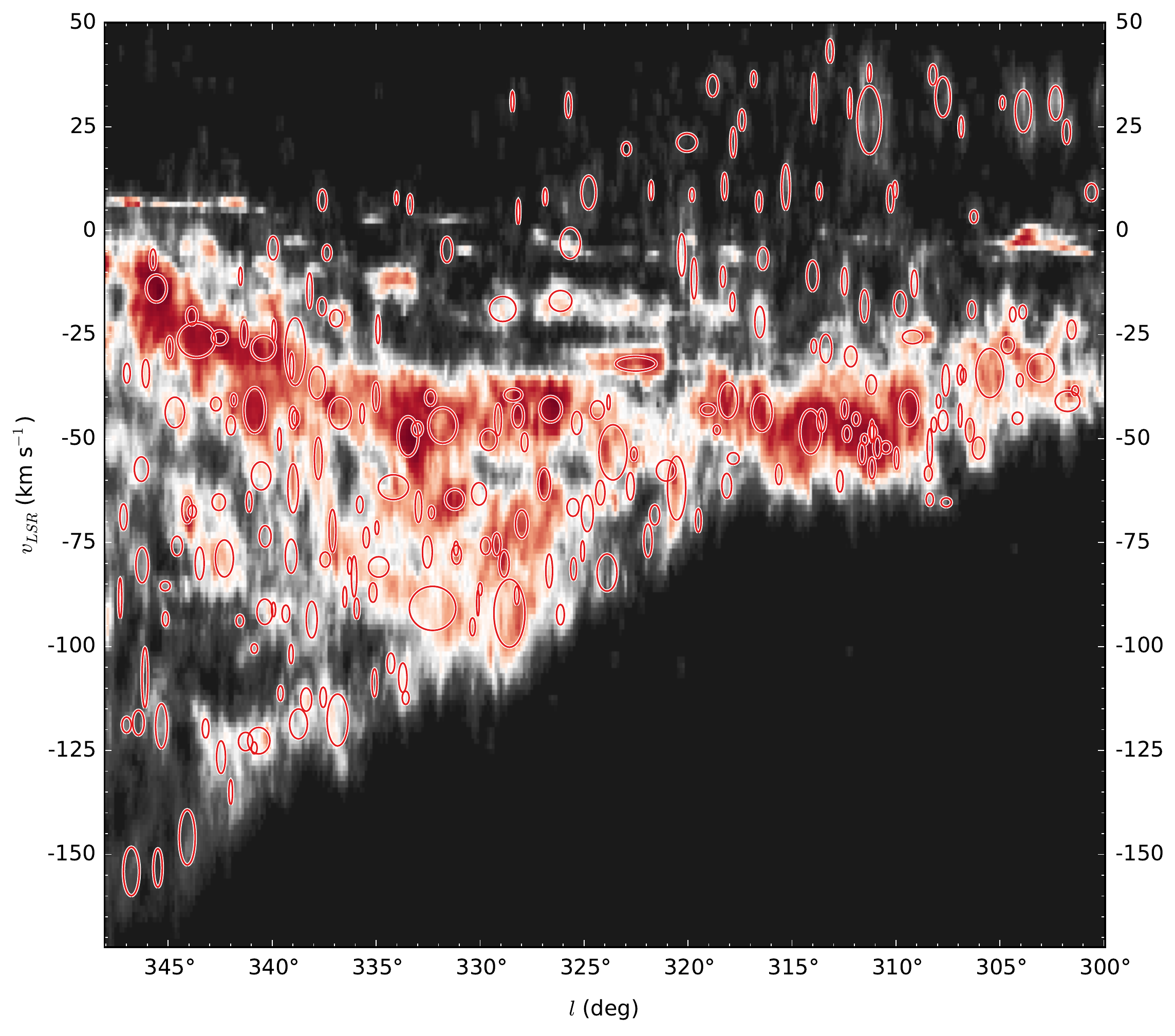}  
  \caption{Map of clouds identified in the Fourth quadrant survey (DHT \#36).
  \label{fig:quad4_map}}
\end{figure}

\end{landscape}

289 clouds were identified in these two surveys, totaling $1.31 \times 10^8 \Msun$.
They are shown in $l, v$ space in Figures \ref{fig:carina_map} and \ref{fig:quad4_map}.
In this quadrant, the inner Galaxy emission appears at negative velocities, while that outside the solar circle appears at positive velocities. 
Because of the greater crowding in the inner Galaxy, the negative-velocity portion of these surveys uses the stricter inner Galaxy criteria described in Section \ref{sec:qualification}.

%% file: analysis.tex

\input{test_table_thing.tex}

Our catalog of 1064 clouds is given as Table \ref{tab:catalog}.
In this section we present figures and analysis related to the global Galactic distribution of molecular clouds and their properties.
These are addressed in the following subsections: 
Section \ref{sec:totalmass} discusses the total mass in the catalog;
Section \ref{sec:distribution} addresses the distribution of clouds;
Section \ref{sec:larson} addresses the size-linewidth relation (Larson's First Law);
Section \ref{sec:cmf} addresses the cumulative cloud mass function.

\subsection{Total mass of clouds}
\label{sec:totalmass}

The best estimate for the molecular mass in the Galaxy is $(1.0 \pm 0.3) \times 10^9 \Msun$ 
\citep{heyer15},
derived primarily from the inner Galaxy work of \citet{bronfman88} and \citet{nakanishi06}, and the outer Galaxy work of \citet{pohl08} and \citet{digel90}.
As seen in Table \ref{tab:qq_results}, the total mass of the 1039 clouds with resolved distances is $2.45 \times 10^8 \Msun$.
A lower mass limit for the 25 clouds with ambiguous distances can be estimated by assuming the near distance for all of them;
the total `near-distance' mass for these clouds is $1.34 \times 10^6 \Msun$.
An upper limit (from the far distance masses) is $4.31\times10^6 \Msun$.
In neither case does this significantly impact the figure of $2.45 \times 10^8 \Msun$ for the resolved-distance clouds.
Thus we can claim that we have at least $25^{+10.7}_{-5.8} \%$ of the Galaxy's molecular mass in this catalog.

The total mass budget can be broken down into ``outer'' versus ``inner'' Galaxy.
\citet{heyer15} identify that the mass within the solar circle is
$\sim 6.9 \times 10^8 \Msun$.
Summing only the 428 inner Galaxy clouds yields $1.93 \times 10^8 \Msun$ in this catalog;
this is 28\% of the estimated inner Galaxy molecular mass.
We exclude a large section of the Galactic Ring due to our omission of emission within $12\degr$ of the Galactic Center; properly taking this missing mass into account increases the effective completeness of our catalog.
\citet{williams97} estimate that the first quadrant surveys they analyze contain $\frac{2}{5}$ of the inner Galaxy's molecular mass; when we include the fourth quadrant survey, which has similar coverage reflected across the Galactic Center, we bring these surveys to about $\frac45$ sensitivity. 
Using this correction, our 428 inner Galaxy clouds contain 35\% of the molecular gas mass in the surveyed regions.
The mass completeness of this Milky Way catalog in the inner Galaxy (35\%) can be compared to the M51 GMC catalog of \citet{colombo14}, to provide extragalactic context. 
In their resolved survey of CO in the inner 9 kpc of the spiral galaxy M51, \citet{colombo14} identified 1507 objects, but only associated 54\% of the CO emission with clouds.

The estimates of molecular mass in the inner Galaxy can be compared to previous Galactic GMC catalogs.
The first quadrant cloud catalogs of 
\citet{dame86}, \citet{solomon87}, and \citet{scoville87}, when summed over mass, each contain less than $20\%$ of the mass in the first quadrant inner Galactic region.
\citet{williams97} combined the above catalogs and integrated over the mass spectrum, recovering $40\%$ of the expected molecular mass in that particular region of the first quadrant.
This integration was not a sum over cloud masses, but rather an integration over an interpolated mass spectrum with parameters set by fitting the clouds to a truncated power-law mass spectrum, and is in this sense one step removed from a direct accounting of masses in clouds.
\citet{garcia14} measure $1.14 \pm 0.05 \times 10^8 \Msun$ in the observed region of the fourth quadrant;
in this study, we find $1.08\times10^8 \Msun$ in the same region, which is marginally consistent with an identical total mass.
By comparing these figures, we find that we recover a higher fraction of the mass in the first quadrant inner Galaxy than any previous catalog of clouds (30\% versus 20\%), and a comparable fraction (40\%) in the fourth quadrant inner Galaxy compared to the only catalog of GMCs there \citep{garcia14}.

Outside the solar circle, the estimated mass is $\sim 2.7 \times 10^8 \Msun$ \citep{heyer15}.
The 611 outer Galaxy clouds in our catalog all have unambiguous distances. 
Their total mass is $5.37 \times 10^7 \Msun$, which is 19.9\% of the estimated outer Galaxy molecular mass.
Perhaps as much as half of the outer Galaxy is missing from this catalog, as the distant half of the Galaxy is not observed with high sensitivity (and the Galactic Center region obscures much of the ``back'' part of the Galaxy);
taking this into account, the effective ``completeness'' of this catalog is even higher.
Also, it is possible that $X_{CO}$ is substantially higher in the outer Galaxy, which would add substantial mass to our catalog.
It should be further noted that the estimated outer Galaxy molecular mass presented by \citet{heyer15} is itself uncertain by $\pm 50\%$.

In this study, we have chosen to exclude most of the Galactic CO emission at low velocities and high latitudes, which typically corresponds to local clouds within 1 kpc of the Sun.
Notably, none of the major out-of-plane surveys which capture much bright and widespread local emission, such as Orion (DHT \#27), Taurus and Perseus (DHT \#21), the high-latitude second quadrant survey (DHT \#18), the Aquila Rift (DHT \#4), or Ophiuchus (DHT \#37), were included in making this catalog.
What are the consequences of excluding this local emission for our catalog and mass estimates?
\citet{dame87} analyzed this local emission using a lower-resolution version of the DHT CO survey.
They associated this emission with 25 clouds or cloud associations, and using distances from the literature, assigned distances and computed masses for these clouds.
They found, using an $X_{CO}$ factor of $2.7 \times 10^{20} \textrm{ cm}^{-2} \textrm{ K}^{-1}$ (or $X_2 = 1.35$, using the convention defined in Section \ref{sec:distances}), that the total mass of these clouds was $4.0 \times 10^6 \Msun$, with about half of this mass contained in five discrete objects: the Cygnus Rift, Cyg OB7, Cepheus, Orion B, and Mon OB1.
When the \citet{dame87} CO mass is scaled to $X_2=1$, the total local mass becomes $3.0 \times 10^6 \Msun$.
This mass represents only 0.3\% of the entire Galaxy's molecular mass, or 1.1\% of the outer Galaxy molecular mass.
If these clouds were included in this catalog's outer Galaxy mass estimate described above, they would increase its mass by 5\%, and bring the estimated fraction of outer Galaxy recovered mass from $19.9\%$ to $21.0\%$.
This mass is relatively inconsequential to the total mass estimates; thus, we are safe in excluding them from our catalog and the Galaxy-wide analyses described in this study.
Further, the distribution of clouds within 1 kpc does not affect our analysis of spiral structure given in Section \ref{sec:distribution}.

\begin{deluxetable}{lcc}
\tablecaption{ Quadrant-by-quadrant cloud results
\label{tab:qq_results} }
\tablewidth{0pt}
\tablehead{
\colhead{ Quadrant } & \colhead{ \# of clouds } & \colhead{ Mass ($M_\sun$) }   \\
}
\startdata
I               &  389     & 9.69$\times10^7$ \\
II              &  276     & 1.21$\times10^7$ \\
III             &  110     & 4.24$\times10^6$ \\
IV (no Carina)  &  239     & 1.22$\times10^8$ \\
IV (Carina-only)&   50     & 9.25$\times10^6$ \\
IV (combined)   &  289     & 1.31$\times10^8$ \\
\hline
inner           &  453     & 1.93$\times10^8$ \\
outer           &  611     & 5.16$\times10^7$ \\
\hline
All combined    & 1064     & 2.45$\times10^8$ \\

\enddata
\tablecomments{The quadrant surveys are described in Table \ref{tab:surveys}, and span varying amounts of longitude.}

\end{deluxetable}

\subsection{Distribution of clouds}
\label{sec:distribution}

The most obvious disadvantage that Galactic studies have relative to extragalactic studies is the difficulty in determining the position of molecular clouds in the $x, y$ plane of the galaxy in question.
In this study, we derive the Galactic $x, y, z$ positions of clouds via kinematic distances (discussed in Section \ref{sec:distances}, along with an explanation of how the twofold kinematic distance ambiguity is dealt with).

\begin{figure}
  \plotone{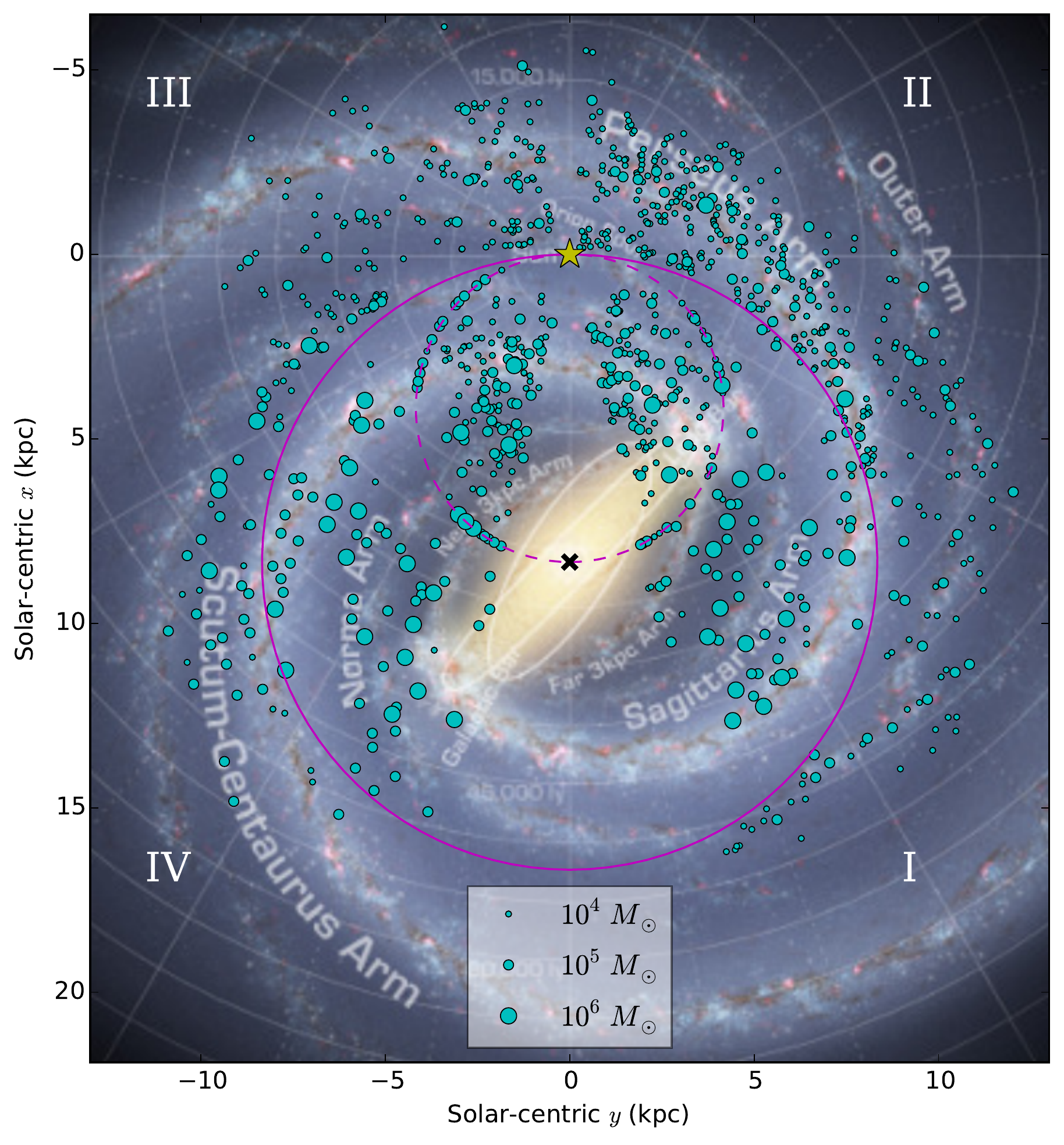}
  \caption{Top-down Galactic map of all clouds, with circle sizes indicating cloud masses. Background illustration \citep{churchwell09} produced by Robert Hurt of the Spitzer Science Center, reflecting the current understanding of Galactic structure.
  The solar circle is drawn in solid purple, and the tangent circle in the inner Galaxy is drawn in dashed purple.
  The Sun's location is marked with a yellow star, and the Galactic Center is marked with a black ``x''.
  \label{fig:allquad_topdown}}
\end{figure}

The $x, y$ position of clouds in the plane is shown in Figure~\ref{fig:allquad_topdown}, with clouds of increasing masses represented by circles of increasing size. 
The regions adjacent to the Galactic Center (within $12\degr$) are excluded from the present study, giving the appearance of a large ``wedge'' missing between the first and fourth quadrants.
A number of of clouds appear on the tangent circle (drawn with a dashed purple line in Figure~\ref{fig:allquad_topdown}) in the first and fourth quadrants. 
These clouds are systematically moving faster than the rotation curve would normally expect for circular motions. 
If the motions of clouds on the near side of the tangent circle in the fourth quadrant were corrected for this systematic offset, they would likely land on the Scutum-Centaurus arm.

Some knowledge of Galactic structure, especially spiral structure, can be gained from such a ``top-down'' $x, y$ representation of the positions of clouds in the Galaxy.
This analysis must be tempered by the following caveat, discussed in \citet{heyer15}:
the assignment of distances to clouds depends on these clouds following perfectly circular motions in the Galactic rotation curve measured by \citet{reid14}.
Especially in parts of the Galaxy affected by non-axisymmetric potentials (e.g. near the Galactic bar), or where large-scale streaming or shock motions occur, this assumption breaks down, introducing systematic errors into the derived distances.
Further, errors in resolving the kinematic distance ambiguity will place some clouds at significantly incorrect distances, also causing their masses to be mis-estimated, often by an order of magnitude or more where the near and far distances differ by more than a factor of 3. 
With the above caveats in mind, the face-on catalog view shown in Figure~\ref{fig:allquad_topdown} can be seen to show significant spiral structure\footnote{The background image in Figure \ref{fig:allquad_topdown}, taken from \citet{churchwell09}, has been re-scaled in this work such that the Sun-Galactic Center distance is exactly 8.34 kpc. Some slight differences in spiral structure between the data presented and the illustration should therefore be expected, especially far from the Sun. Note also that the \citet{churchwell09} illustration is itself an artist's approximation, built up from multiple datasets which used slightly different Sun-GC scaling.}. 

The \textit{Outer Arm} is clearly defined in both $l, v$ and $x, y$ space throughout the first quadrant, with some weak emission visible in the second quadrant.
In $l, v$ space, it is seen at negative velocities starting at $\sim 15\degr$ (near the innermost longitudes sampled in this study) and continuing as a continuous stripe at the most extreme negative velocities until $l \sim 95\degr$.
Some emission at extreme negative velocities between $l=110\degr$ and $160\degr$ may also be associated with the Outer Arm, but none of it is substantial enough to be identified as massive molecular clouds per the criteria outlined in Section \ref{sec:methods}.
In $x, y$ space, these clouds form a clearly identifiable lane of clouds emerging from behind the far side of the Galaxy at $R_{gal} \sim 7$ kpc, winding around towards the Sun and increasing to $R_{gal} \sim 11$ kpc.
None of the clouds in this visible section of the Outer Arm contains or exceeds $10^6 \Msun$ in mass.

The \textit{Perseus Arm} can be traced in the first, second, and third quadrants, spanning over $180 \degr$ on the visible sky.
This arm appears as high-linewidth clouds in the crowded positive-velocity inner Galaxy below $l=45\degr$, before emerging at negative velocities at $l=55\degr$, where it steadily grows in negative velocity into the second quadrant, before reaching a minimum $v_{LSR}\sim -70 \textrm{ km s}^{-1}$ at $l=120\degr$ and then increasing back towards $v_{LSR} = 0 \textrm{ km s}^{-1}$ near $l=180\degr$.
The Perseus Arm re-emerges at positive velocities at $l=190\degr$, increasing in velocity into the third quadrant where it can be followed until around $l=260\degr$.
The Perseus Arm is traced by many GMCs above $10^6 \Msun$, primarily within the Solar Circle.

The \textit{Sagittarius Arm} is less clearly delineated in the present study than the arms discussed above. 
In $l, v$ space, its primary clouds appear at positive velocities in the first quadrant. 
In $x, y$ space, a string of massive clouds in the first quadrant can be seen at the far distance.
This arm's fourth quadrant, outer Galaxy extension, known as the \textit{Carina Arm}, is very clearly defined in this catalog, especially outside the solar circle, where it can be clearly traced from $l=280\degr-330\degr$.
In $x, y$ space, this arm lies outside the Scutum-Centaurus Arm, forming an arc just outside the solar circle at nearly constant galactocentric radius. 

The \textit{Scutum-Centaurus Arm} can be followed from the first quadrant through much of the fourth.
Its first quadrant clouds appear at ``near'' distances, although a large spread in distances means the arm is difficult to cleanly delineate in $x, y$ space.
Into the fourth quadrant, it becomes somewhat easier to follow this arm, and a large number of clouds appear to be associated with it at both near and far distances.
The \textit{Norma Arm} is difficult to separate out among the clouds on the far side of the fourth quadrant.

From the molecular cloud catalog presented in this paper, we have produced synthetic CO ``images'' of what the Galaxy would look like in CO if observed remotely, and face-on.
This is particularly interesting for comparison with CO surveys of spiral galaxies at varying distances and angular resolutions.
Such an image suffers from the following biases and incompletenesses. 
Because the sensitivity to low-mass clouds decreases with distance, emission from low-mass clouds will be under-counted in the distant parts of the Galaxy; velocity crowding in the Inner Galaxy means that even massive clouds can go unaccounted for. 
The caveats from Section \ref{sec:distribution} regarding kinematic distances apply here as well.

To produce this image, we simply invert the luminosity-to-mass relationship (Equation \ref{eq:mass}):
\begin{equation}
  L_{CO} = M / (4.4 X_2)
\end{equation}
and assume all clouds follow a spherical\footnote{Clouds are not truly spherical, but because we cannot observe their shapes from a ``face-on'' perspective, we choose to approximate their appearance as spherical, with a radius given by $\sigma_r$ as computed in Section \ref{sec:methods}, for the purpose of this emission simulation.} Gaussian size profile, with $\sigma_r = R/\eta$, as was inversely done in Section \ref{sec:methods}.
The clouds' emission is calculated and placed on a face-on grid.
This simulated emission grid can then be convolved by a 2D Gaussian of arbitrary size to simulate the effect of a coarse angular resolution.

\begin{figure}
	\plottwo{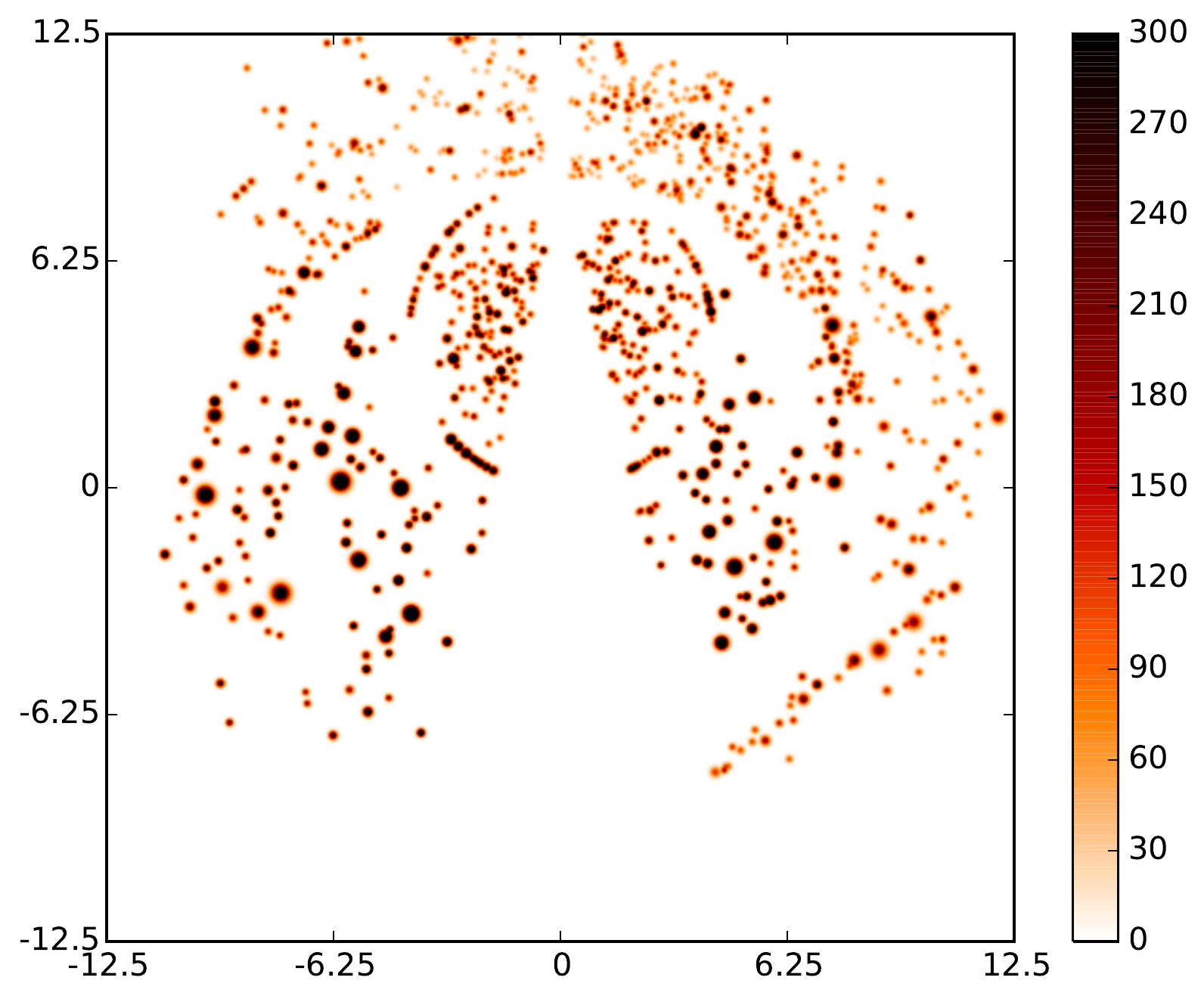}{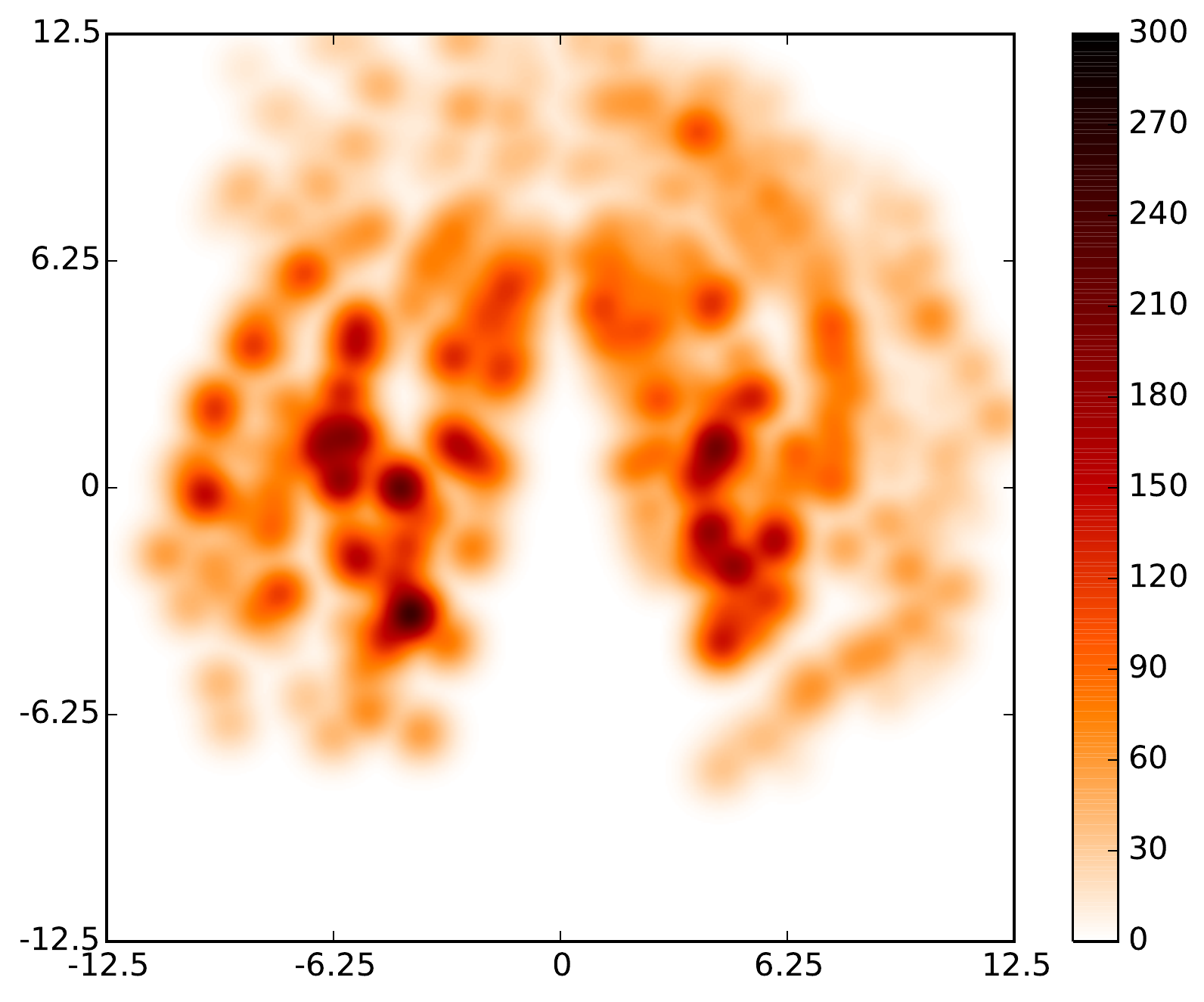}
	\caption{	Face-on ``simulated'' CO image of the Milky Way based on the clouds in this catalog, as seen by two hypothetical extragalactic observers:
  \textit{Left}: with 40 pc resolution, matching the PAWS survey of M51 \citep{schinnerer13}. 
  \textit{Right}: with 400 pc resolution, simulating a galaxy 10 times further than M51. 
  Units are in $\textrm{K km s}^{-1}$, and each image is displayed with a square root intensity scale.
	\label{fig:simulated_image}
	}
\end{figure}

We show two such images of the Milky Way in Figure~\ref{fig:simulated_image}:
one with 40 pc resolution, matching the PAWS survey of M51 \citep{schinnerer13};
and one with 400 pc resolution, simulating the effect of observing a galaxy 10 times further than M51 with the same instrument.
These images are integrated over velocity, and noiseless; we choose a circular beam (rather than an elliptical one) for simplicity.
Because many GMCs are resolved at the 40 pc resolution, Figure~\ref{fig:simulated_image}a appears largely identical to Figure~\ref{fig:allquad_topdown}.
It is interesting to note that even at 400 pc resolution (Figure~\ref{fig:simulated_image}b), 
spiral structure is clearly evident, with the Outer, Perseus, and Carina arms showing prominently in CO emission.

These maps can be compared to the CO maps of M51, M33, and the LMC presented in \citet{hughes13}, who compiled the survey data of \citet{schinnerer13}, \citet{rosolowsky07b}, and \citet{wong11}.
At large scales, the Milky Way emission does not have the immaculate grand-design spiral features seen in M51, especially inside the solar circle.
Nonetheless, the appearance of the Perseus, Outer, and Carina arms, and perhaps the first quadrant Sagittarius arm and fourth quadrant Scutum-Centaurus arm, are significant indicators of spiral structure in emission.
The comparison with these extragalactic surveys is hindered slightly by the fact that only the emission contained in catalogued clouds is included in our Milky Way emission simulation, whereas much diffuse inter-arm, non-cloud ``chaff'' is seen in M51;
because our ``face-on'' Galaxy image is generated from a cloud catalog, it necessarily excludes such diffuse emission.

\subsection{Line-width size relation}
\label{sec:larson}

The line-width size relation (otherwise known as Larson's 1st Law; \citealt{larson81}) is an often-measured property of molecular clouds. 
It states that the line-widths of clouds follow a power-law relationship with their physical radius:
\begin{equation}
	\sigma_v = A \times R^\beta.
\end{equation}
\citet{larson81} derived a power index $\beta$ near $\frac13$ and interpreted it as evidence for incompressible Kolmogorov turbulence. 
Studies shortly thereafter, including \citet{solomon87}, ruled out this interpretation with power indices nearer $\frac12$, and suggested that the relationship arises from clouds' virial equilibrium.
In this study, we fit the above functional form to the entire catalog and to specific sub-regions of the Galaxy.

Here we have performed the fit using orthogonal distance
regression (ODR)\footnote{A Python-based implementation of ODR is included in SciPy as \texttt{scipy.odr}}, 
which fits a curve to values with uncertainties in the x-coordinate (here, the x-coordinate is cloud radius).
The ODR code returns both a slope and an estimated error on the parameters of the curve.
The uncertainties in cloud radii are derived from the uncertainties in kinematic distances using the rotation curve model discussed in Section \ref{sec:distances}.
For the fits shown below, we restrict our sample to clouds with $|v_{LSR}|>20 \textrm{ km s}^{-1}$ located at least $20\degr$ from either the Galactic Center or anti-center, in order to minimize kinematic distance errors.

\begin{figure}
  \plottwo{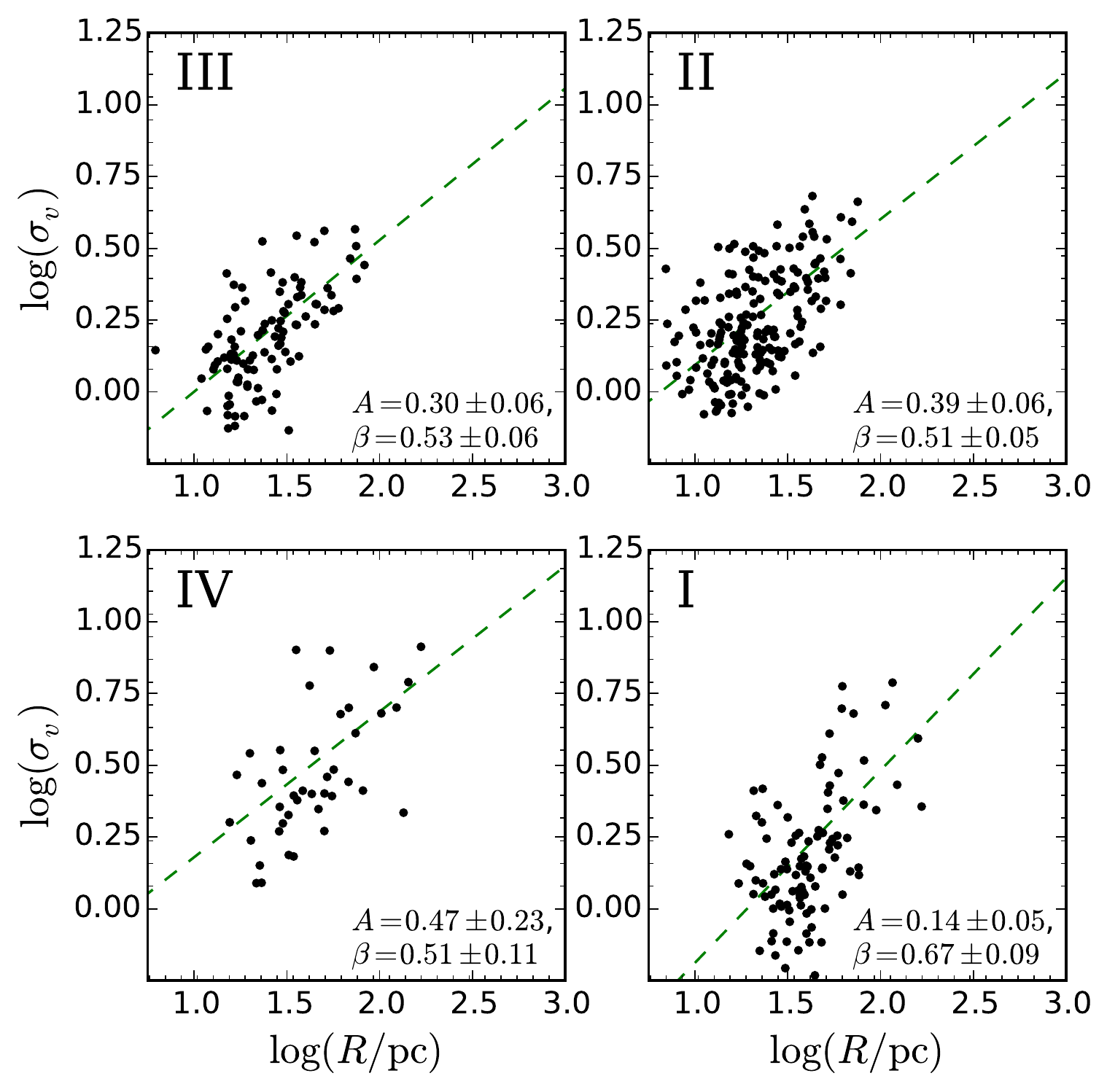}{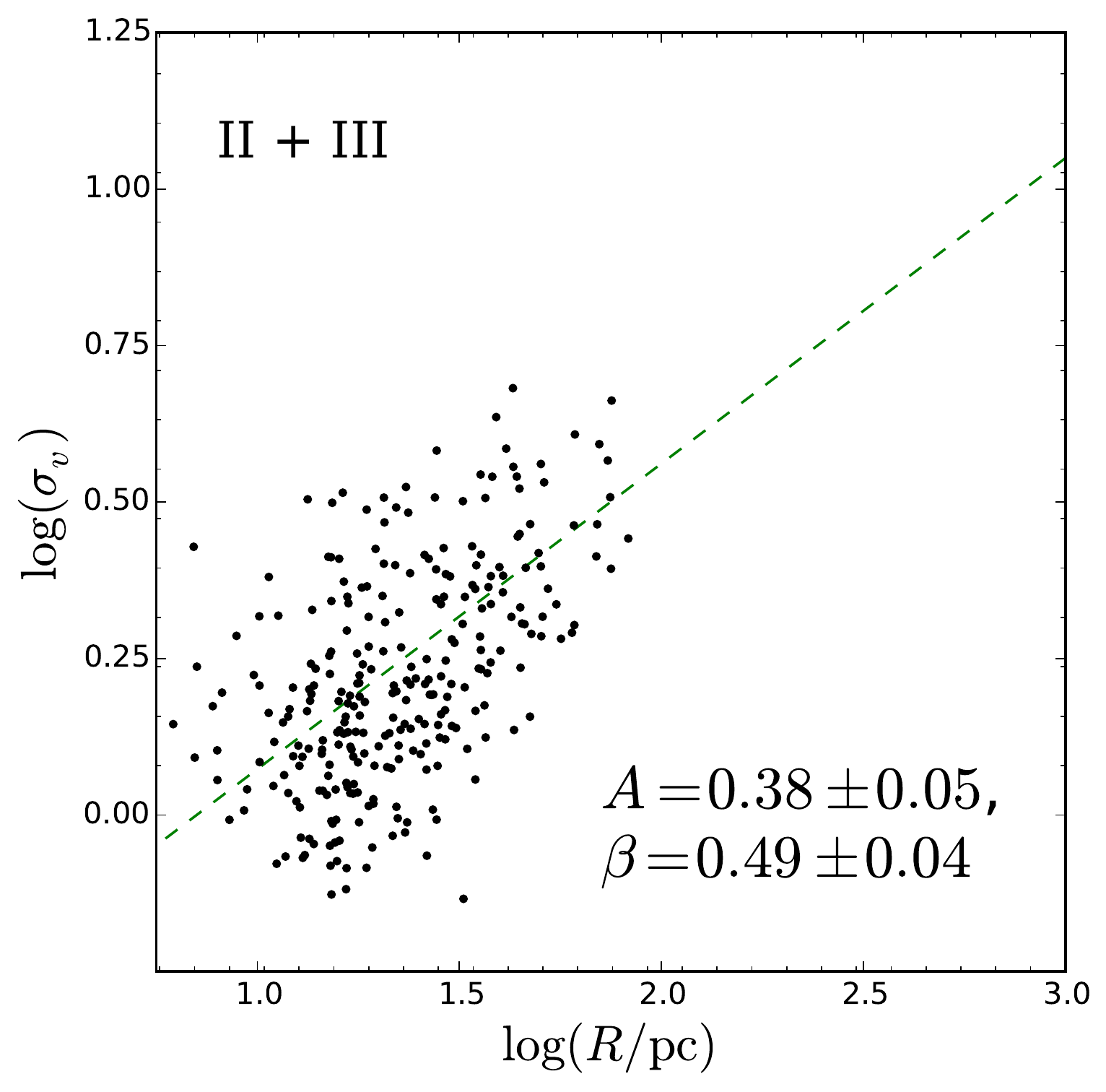}
  \caption{Size-linewidth relation $\sigma_v=A \times R^\beta$ for clouds beyond the solar circle.
  \textit{Left}: Quadrant-by-quadrant fit. \textit{Right}: Combined fit for the second and third quadrants, which contain the most reliable outer Galaxy clouds.
  \label{fig:outer_larson}}
\end{figure}

We show fits of outer Galaxy clouds in each quadrant in Figure~\ref{fig:outer_larson}a, and a combined outer Galaxy fit (from the second and third quadrants only, for robustness purposes) in Figure~\ref{fig:outer_larson}b.
The combined outer Galaxy fit is 
\begin{equation}
  \sigma_v = (0.38\pm0.05) R^{0.49\pm0.04},
\end{equation}
\noindent and in each quadrant, the power index $\beta$ is consistent with this combined $\beta \approx 0.5$ fit.
Clouds in the outer Galaxy do not suffer from the kinematic distance ambiguity, and can be considered independent measurements of the size-linewidth relation (within the uncertainties of the rotation curve chosen for kinematic distances).

\begin{figure}
  \plottwo{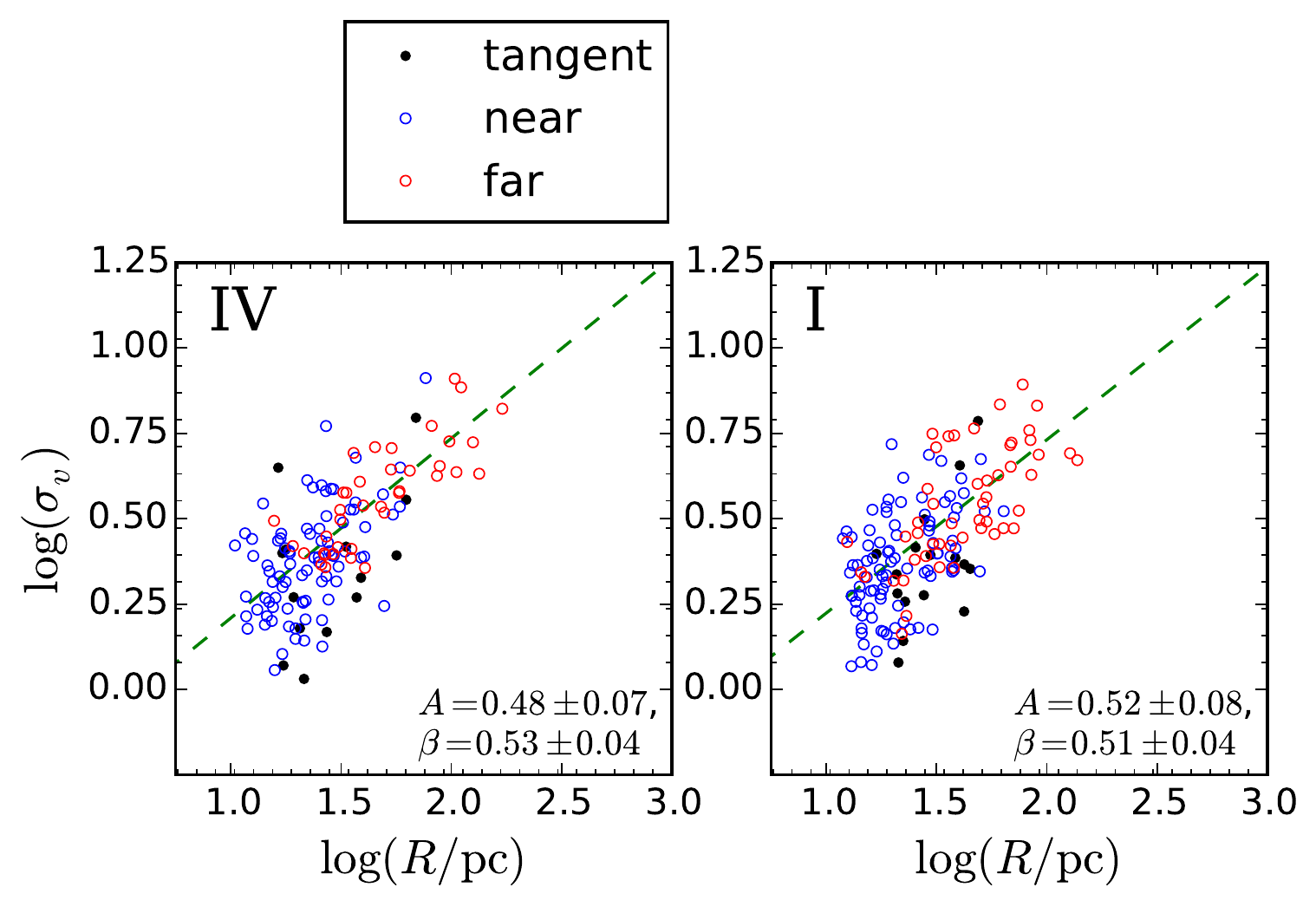}{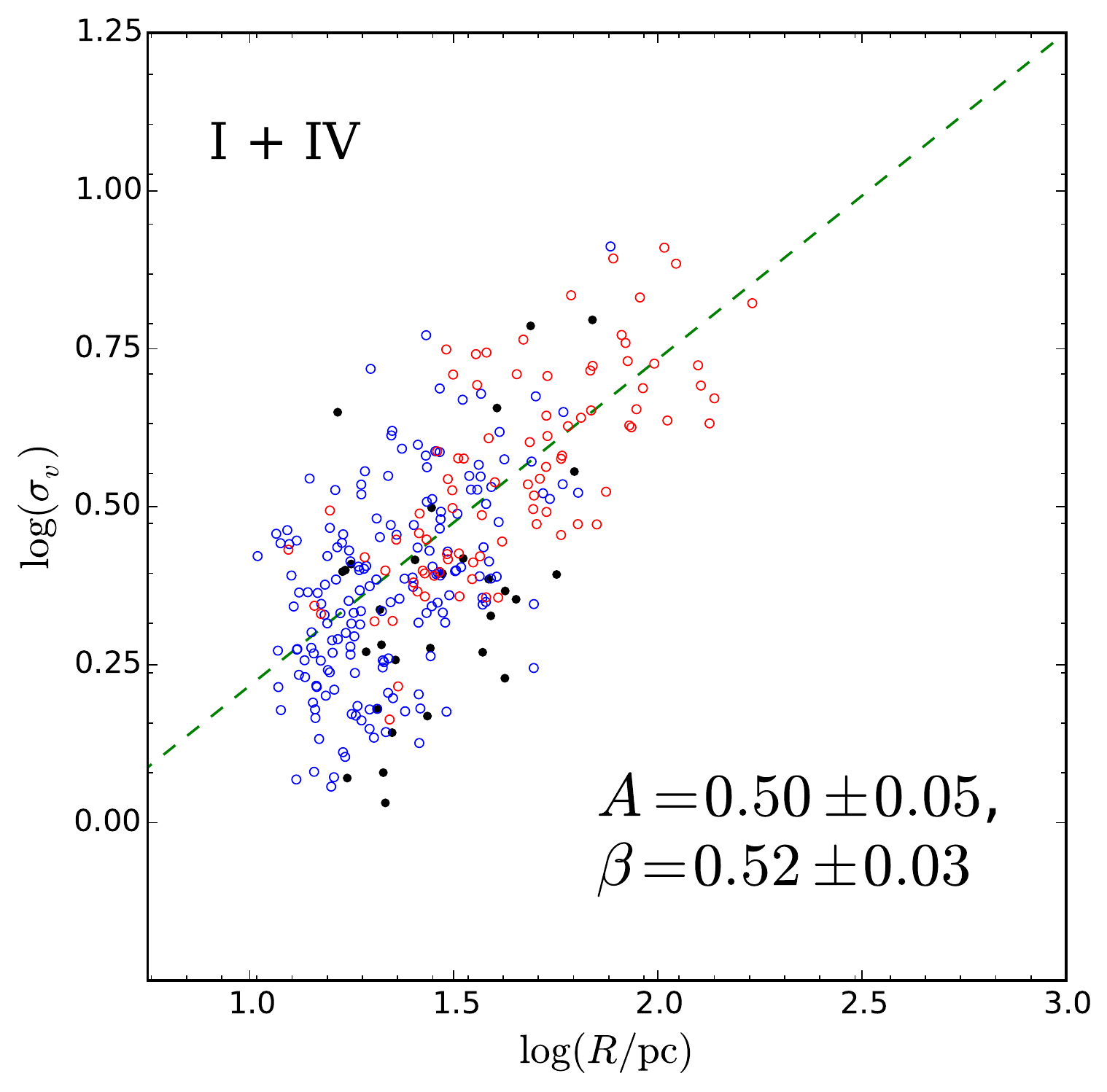}
  \caption{Size-linewidth relation $\sigma_v=A \times R^\beta$ for clouds inside the solar circle. 
  Blue and red points denote clouds resolved to the near and far distance, respectively; in these panels, the black points show clouds assigned to the tangent circle, where no ambiguity exists.
  \textit{Left}: Quadrant-by-quadrant fit. \textit{Right}: Combined fit for the first and fourth quadrants.
  \label{fig:inner_larson}}
\end{figure}

In the inner Galaxy, we fit the size-linewidth relation with 
\begin{equation}
  \sigma_v = (0.50 \pm 0.05) R^{0.52 \pm 0.03}.
\end{equation}
\noindent Clouds in the inner Galaxy are subject to the kinematic distance ambiguity; because we resolve the KDA partially using the size-linewidth relation, the resulting size-linewidth fits are not free from bias. 
Regardless, as noted in Section \ref{sec:distances}, a value of $A\approx 0.5$ and $\beta \approx 0.5$ emerges even when using the outer Galaxy fit (with a lower $A$) as a prior guess, giving this fit more confidence.
The fits to the inner first quadrant, inner fourth quadrant, and combined inner Galaxy are shown in Figure~\ref{fig:inner_larson}.

\begin{figure}
  \plotone{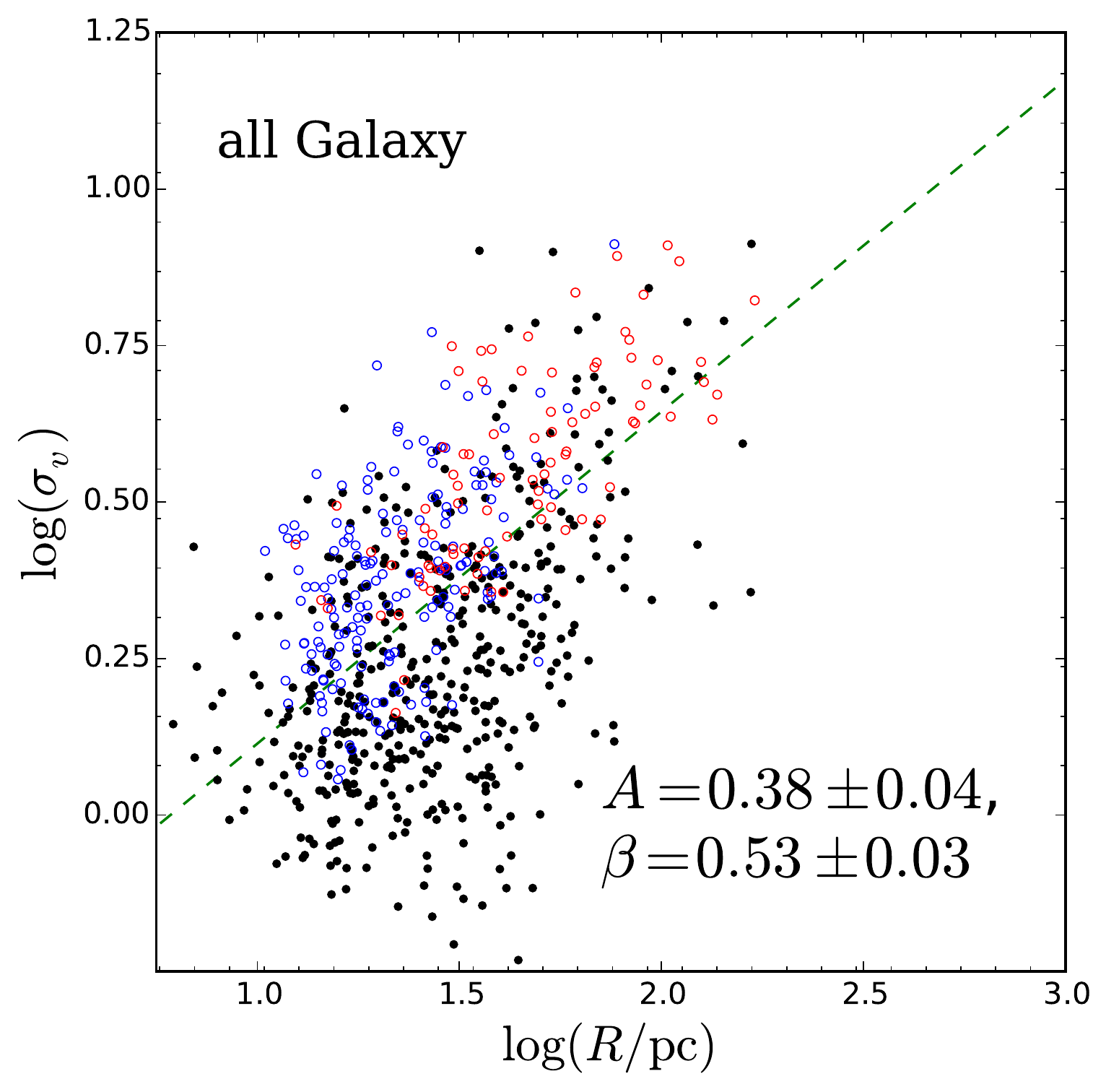}
  \caption{Size-linewidth relation $\sigma_v=A \times R^\beta$ for all clouds in all quadrants. 
  The black points, primarily showing the outer Galaxy clouds from Figure~\ref{fig:outer_larson}, have systematically lower linewidths, and fewer of them have large sizes. This fit should be taken as a blend of two measurably different populations, rather than as a universal size-linewidth relation.
  \label{fig:all_larson}}
\end{figure}

While the power index $\beta \approx 0.50$ is consistent between the inner and outer Galactic fits, the scaling coefficient $A$ is significantly higher in the inner Galaxy ($A=0.50\pm0.05$) than the outer Galaxy ($A=0.38\pm0.05$). 
A combined size-linewidth plot showing inner and outer Galactic clouds is shown in Figure~\ref{fig:all_larson}.
If linewidths are set purely by the internal virialization and/or turbulence of clouds, this distinction would not be expected; thus, this may be interpreted as evidence that the environment of the inner Galaxy itself encourages higher linewidths, through pressure confinement (cf. \citealt{meidt13}) or some other means.

The size-linewidth power index we measure, of $\beta \approx 0.50$ in most regions, is highly consistent with previous inner Galaxy measurements.
In the first quadrant inner Galaxy, \citet{dame86} and \citet{solomon87} both derive $\beta=0.50\pm0.05$, while \citet{scoville87} find the very similar $\beta=0.55\pm0.05$. 
The fourth quadrant study of \citet{garcia14} also finds a fit of $\beta=0.50\pm0.07$.

The size-linewidth relation in the outer Galaxy was previously measured by \citet{sodroski91} (using a combined catalog of 35 GMCs in the second and third quadrants) and \citet{may97} (using 177 clouds in the third quadrant).
\citet{sodroski91} measured a power index of $\beta=0.47\pm0.08$, while \citet{may97} found $\beta=0.45\pm0.04$.
\citet{heyer01} also report that the clouds in their catalog larger than 9 pc in radius are well-fit by a power index of $\beta \sim 0.5$.

A note of caution must be added regarding the power index $\beta = 0.5$ derived from the size-linewidth fits here.
A number of authors, including \citet{scalo87} and \citet{goodman98}, have noted that because single-molecule tracers are only able to probe a relatively small dynamic range in gas density, radio telescopes are observationally biased towards detecting a constant column density; when clouds at many different distances are observed, this effect automatically produces a size-linewidth power index of $\beta=0.5$ if clouds are in virial equilibrium.
In other words, $\beta=0.5$ is due primarily to a selection effect of the observational setup.
Because the $\beta$ values derived in this study are all highly consistent with 0.5, this caveat should be kept earnestly in mind.
Further, as discussed in the following paragraph, populations of extragalactic clouds, which are effectively observed all at the same distance and are therefore not subject to the above bias, do not recover significant correlations between size and linewidth, adding further evidence that this effect is due to observational biases.
The Larson's relationships were further explored by \citet{heyer09}, where they discussed how the measured quantities $\sigma_v/R^{1/2}$ depend explicitly on $\Sigma$, which may have been previously overlooked due to the narrow range of measured surface densities in previous, lower-resolution work.

In the extragalactic context of M51, \citet{colombo14} found that in all galactic regions, the size and linewidth of clouds were at best weakly correlated, even when only the highest signal-to-noise clouds were included.
\citet{colombo14} do note that inter-arm regions lack clouds with high $\sigma_v$ compared to spiral arms, and that the clouds in the central regions of M51 had systematically higher $R$ and $\sigma_v$.
Likewise, in a comparative study of clouds between the LMC, M33, and M51, \citet{hughes13} found that, on the whole, M51's clouds were larger and had higher linewidths than those in M33 and the LMC.
These situations are analogous to what is seen between the inner and outer Galaxy clouds in this study, as shown in Figure~\ref{fig:all_larson}, where the outer Galaxy clouds (black points) have systematically lower $\sigma_v$, and somewhat lower sizes, than the inner Galaxy clouds.
\citet{hughes13} suggest that the differences they observe between cloud populations can be explained by ISM pressure mediating cloud density and velocity dispersion; such effects are likely at play in the Milky Way clouds studied here.

\clearpage
\subsection{Cloud Mass Functions}
\label{sec:cmf}

\input{mass_function_fits_table.tex}

The cloud mass spectrum, which describes the distribution by number of clouds of different masses, is possibly related to the mass functions of clusters and stars \citep[and references therein]{kennicutt12}.
Regional variations of the mass spectrum can indicate differences in the mechanisms that influence cloud formation, evolution, and destruction \citep{rosolowsky05,colombo14}.
A cloud mass function is often expressed (in differential form) as a power law $\frac{dN}{dM} \propto M^\gamma$,
which, when integrated over mass, gives a cumulative mass function 
\begin{equation}
N(M' > M) = \left [ \left ( \frac{M}{M_0} \right )^{\gamma+1}  \right ],
\label{eq:pl_spectrum}
\end{equation}
\noindent where $\gamma$ is an index describing how mass is distributed amongst clouds.
$\gamma > -2$ indicates that the majority of mass is contained in massive clouds; $\gamma \approx -2$ means that mass is roughly equally distributed in all mass bins; and $\gamma < -2$ indicates that low-mass clouds contain the majority of mass.

Many authors (e.g. \citealt{williams97}, \citealt{colombo14}, and references therein) report that cloud mass functions steepen at the high mass end, motivating the following ``truncated power law'' mass spectrum definition:
\begin{equation}
	N(M' > M) = N_0 \left [ \left ( \frac{M}{M_0} \right )^{\gamma+1} - 1 \right ].
\label{eq:tpl_spectrum}
\end{equation}
\noindent This functional form, called a ``truncated power-law'' due to its upper limit $M_0$ above which the mass function drops quickly to zero, was introduced by 
\citet{williams97}. 
\citet{colombo14} use this functional form in their study of GMCs in M51 in order to study how the properties of GMCs vary by galactic region, and \citet{rosolowsky05} fit mass functions of both truncated and non-truncated power laws to regions of the Milky Way and other Local Group galaxies, discussing where either functional form is more appropriate.

To measure mass spectra in this study, we make use of the maximum-likelihood method of \citet{rosolowsky05} and \citet{rosolowsky07a}, implemented by Erik Rosolowsky in IDL as \texttt{mspecfit.pro}\footnote{This IDL code is available on-line at \url{https://github.com/low-sky/idl-low-sky}.}.
This fitting method was also used by \citet{colombo14} for mass spectra in M51.
\citet{williams97} discuss how the truncated power-law mass spectrum is only appropriate when $N_0 \gg 1$; for this reason, we fit both the truncated (Equation \ref{eq:tpl_spectrum}) and non-truncated (Equation \ref{eq:pl_spectrum}) power law mass spectra.
We carry out this measurement region-by-region, using same regions discussed in Section \ref{sec:larson}.
As in Section \ref{sec:larson}, we only use clouds with $|v_{LSR}|>20 \textrm{ km s}^{-1}$ located at least $20\degr$ from $l=0\degr$ or $l=180\degr$ in order to minimize kinematic distance errors, which have a substantial effect on estimated masses.
The smallest outer Galaxy clouds in this catalog are $3\times 10^3 \Msun$, so only clouds $3 \times 10^4 \Msun$ and above are included in the outer Galaxy fits;
likewise, because the inner Galaxy clouds are limited to $3 \times 10^4 \Msun$ and above, we include only clouds above $10^5 \Msun$ in the inner Galaxy fits.

\begin{figure}
  \plottwo{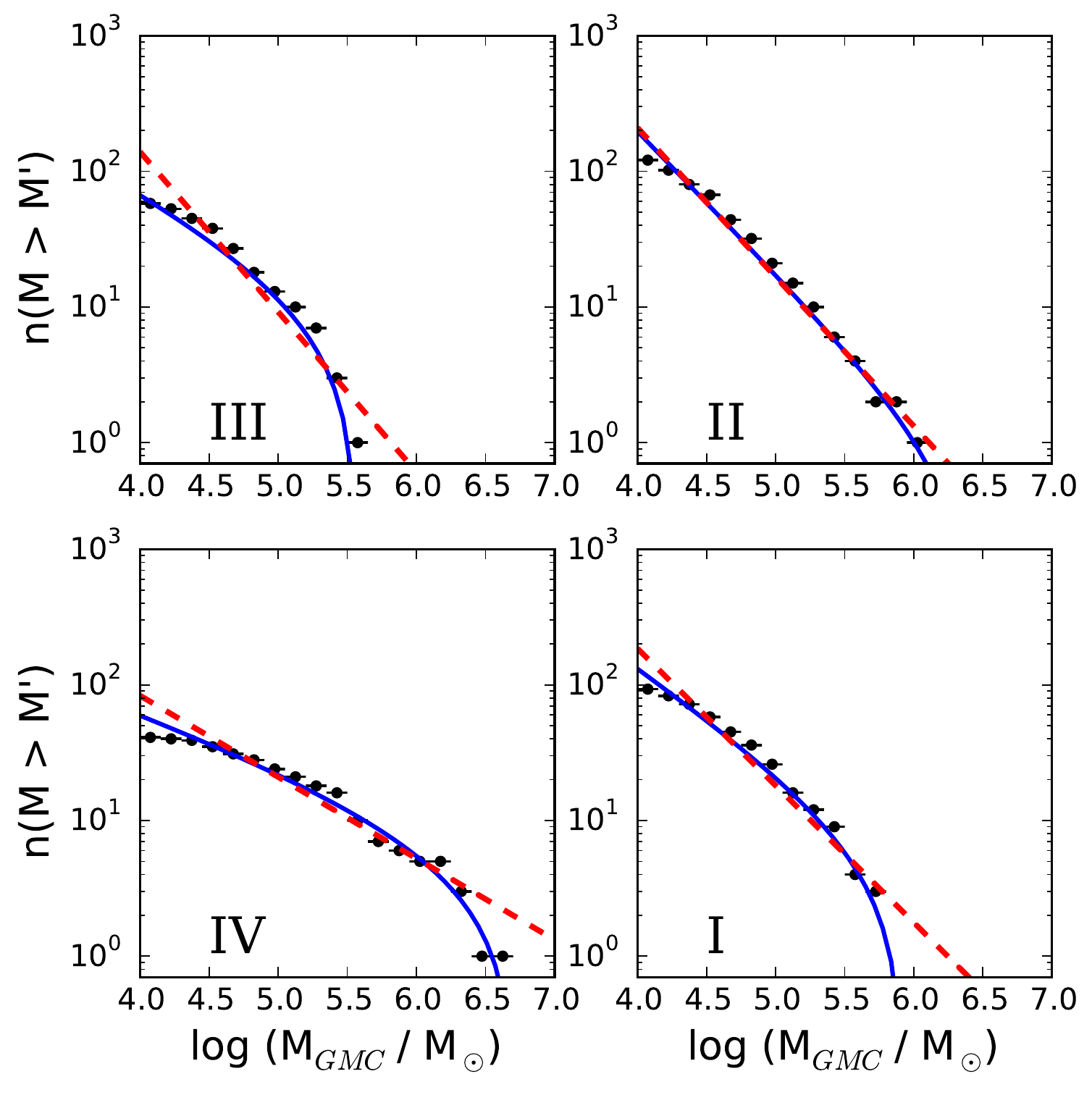}{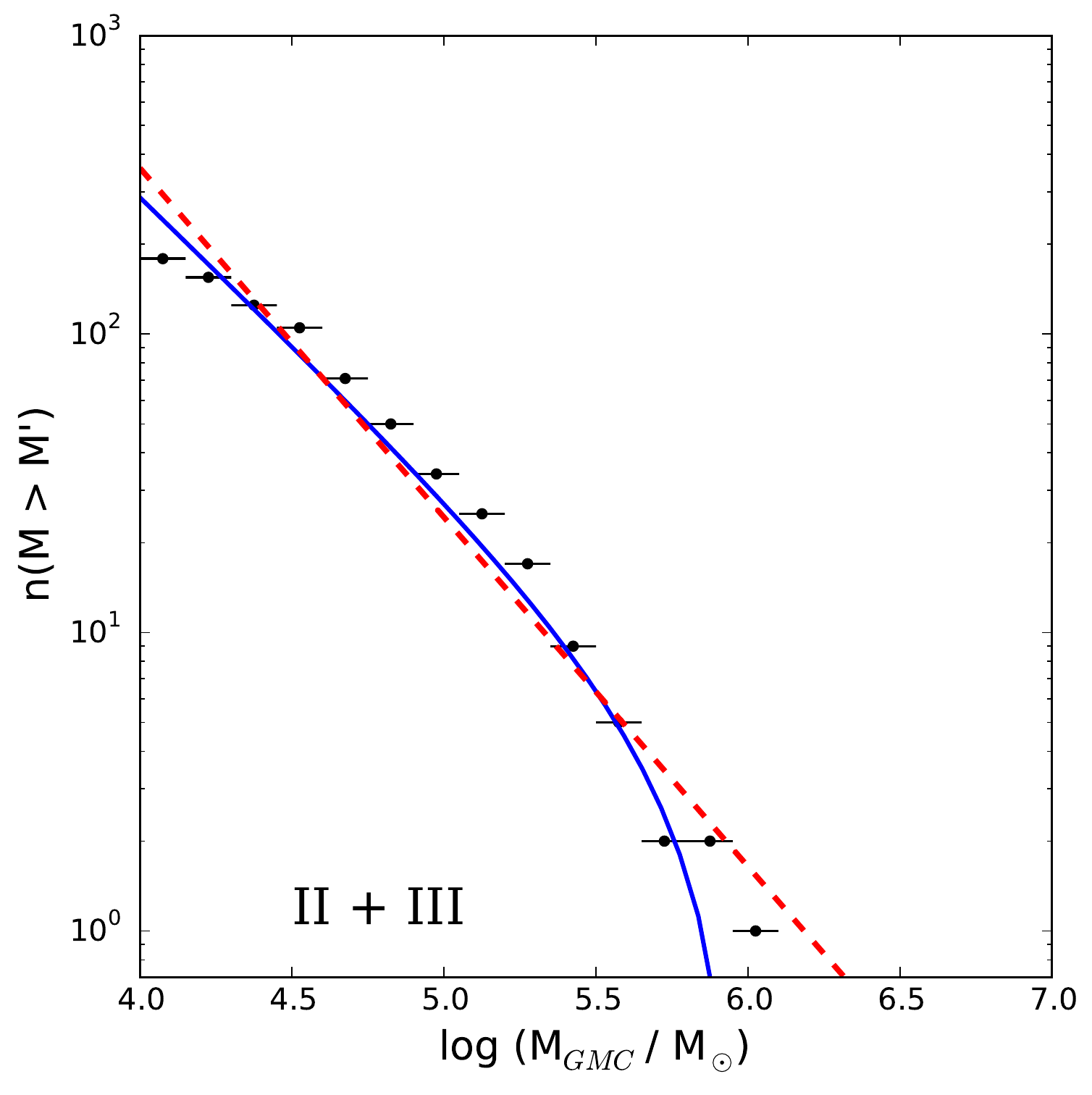}
  \caption{Cumulative mass functions for clouds beyond the solar circle.
  In each panel, both a truncated (solid blue) and non-truncated (dashed red) power law mass spectrum is fit.
  For the outer Galaxy, the non-truncated, dashed red power law functional form is preferred.
  The parameters of the fit for each region are displayed in Table \ref{tab:massfunction}.
  \textit{Left}: Quadrant-by-quadrant fit. \textit{Right}: Combined fit for the second and third quadrants, which contain the most reliable outer Galaxy clouds.
  \label{fig:outer_cmf}}
\end{figure}

\begin{figure}
  \plottwo{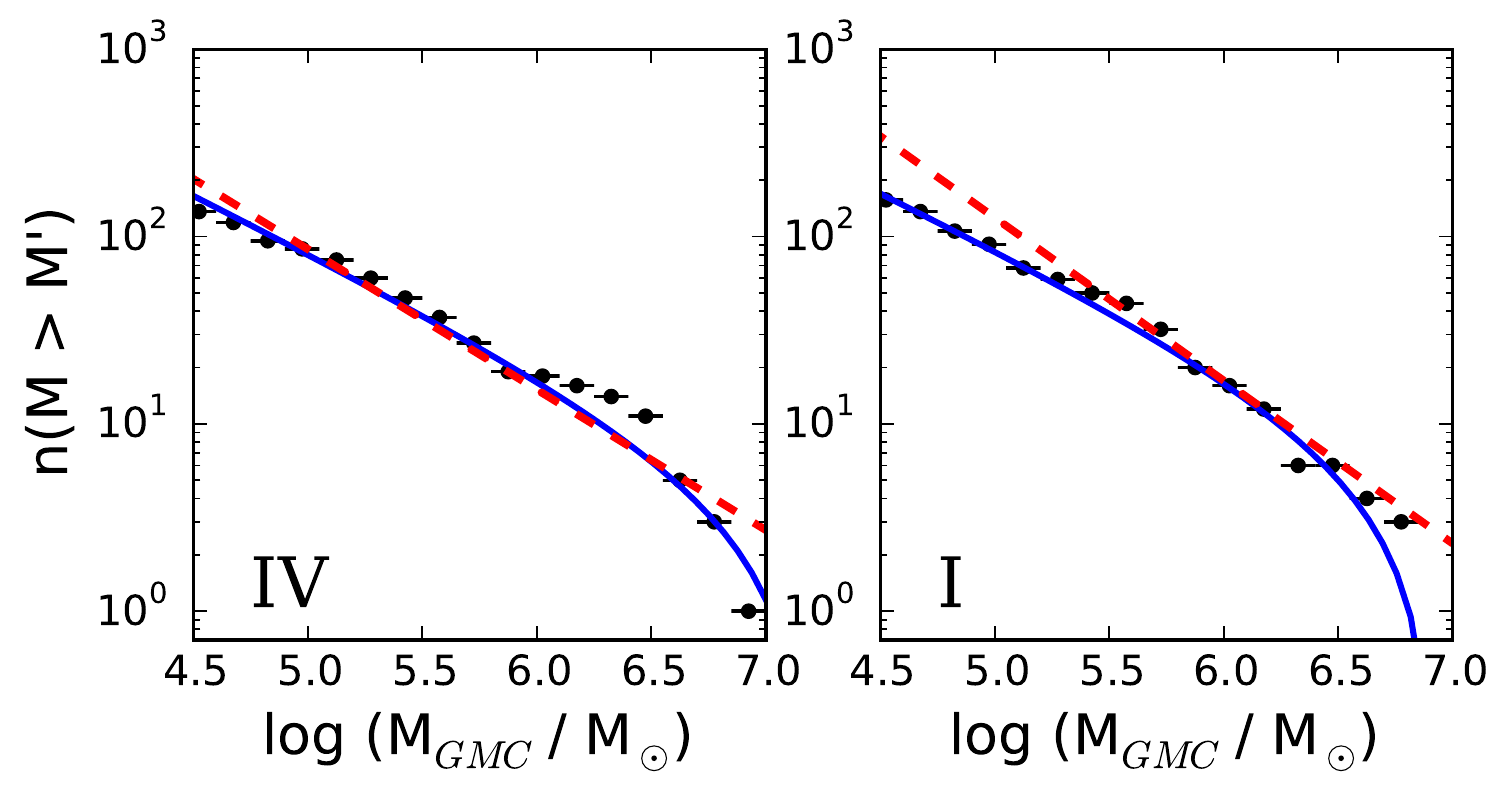}{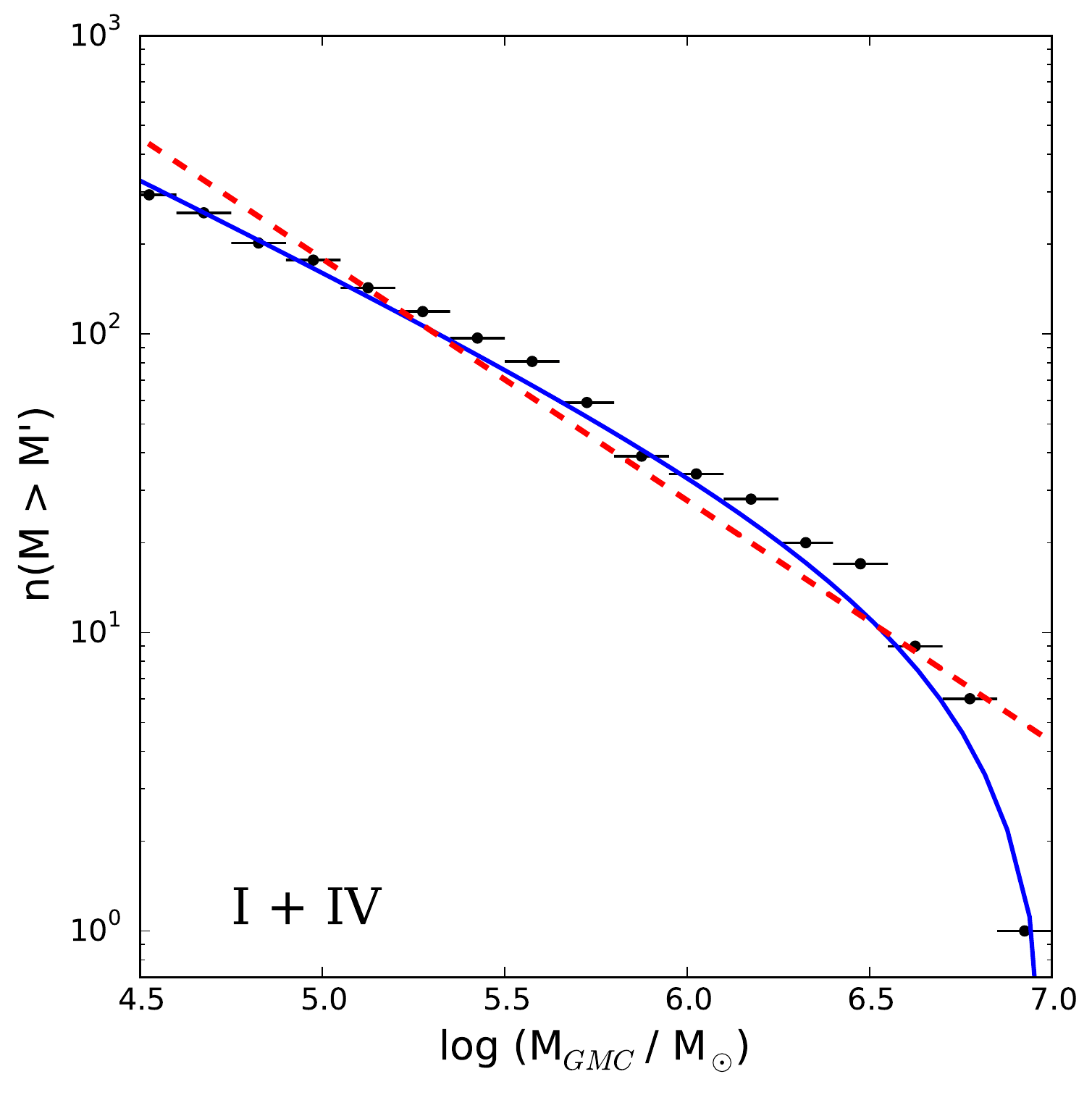}
  \caption{Cumulative mass functions for clouds within the solar circle.
  In each panel, both a truncated (solid blue) and non-truncated (dashed red) power law mass spectrum is fit.
  The parameters of the fit for each region are displayed in Table \ref{tab:massfunction}.
  \textit{Left}: Quadrant-by-quadrant fit. \textit{Right}: Combined fit.
  \label{fig:inner_cmf}}
\end{figure}

Our mass spectra fits are shown in Figures~\ref{fig:outer_cmf} and \ref{fig:inner_cmf}, and presented in Table \ref{tab:massfunction}.
In the outer Galaxy, we find that the truncated power law fits are generally disfavored due to low values of $N_0$ (except for clouds in the third quadrant, which has only 33 clouds suitable for the mass spectrum analysis), and so we adopt the non-truncated power law values for our overall outer Galaxy analysis.
The combined outer Galaxy mass spectrum fit, which we restrict to the second and third quadrants for data fidelity (similarly to Section \ref{sec:larson}), 
is $M_0 = (1.5 \pm 0.5) \times 10^6 \Msun$, and $\gamma=-2.2 \pm 0.1$.

In the inner Galaxy, we find that the truncated power law mass spectrum is preferred: the combined first and fourth quadrant fit yields $N_0 = 11 \pm 6$, which is unambiguously $\gg 1$. 
We derive a truncation mass of $(1.0 \pm 0.2) \times 10^7 \Msun$ and a power index $\gamma = -1.6 \pm 0.1$ here.
In the first quadrant alone, we find $M_0 = (8.6 \pm 2.6) \times 10^6 \Msun$ and $\gamma= -1.6 \pm 0.1$, 
while the fourth quadrant fit constrains $M_0$ somewhat poorly, with $M_0 = (1.5 \pm 0.6) \times 10^7 \Msun$ and $\gamma =-1.6 \pm 0.1$.

We find that the slope $\gamma$ of the mass function is much steeper in the outer Galaxy ($\gamma \approx -2.2$) than the inner Galaxy ($\gamma \approx -1.6$). 
This environmental dependence of the GMC mass spectrum may indicate that the higher densities of the inner Galaxy systematically encourage high-mass clouds to form and thrive, whereas these processes are absent in the outer Galaxy.
While the outer Galaxy mass spectrum does not exhibit a sharp truncation like the inner Galaxy, almost no clouds above $10^6 \Msun$ exist outside the solar circle, whereas clouds above this mass account for most of the mass inside the solar circle.

Previous measurements of Galactic cloud mass spectra yield results broadly consistent with the results presented here.
From a synthesis of first quadrant inner Galaxy cloud catalogs, \citet{williams97} found $M_0=6 \times 10^6 \Msun$, which is within the error bars of the first quadrant inner Galaxy results presented here, and $\gamma=-1.6$, which is consistent with our results throughout the inner Galaxy.
\citet{garcia14} fit clouds above $10^6 \Msun$ to a simple power law mass function and find $\gamma=-1.5\pm0.4$;
the largest cloud in the \citet{garcia14} catalog has a molecular mass of $8.7 \times 10^6 \Msun$, higher than the $6\times 10^6$ upper limit of the aforementioned first quadrant studies but consistent with the truncation mass of $M_0=(1.5 \pm 0.6) \times 10^7 \Msun$ presented here for the fourth quadrant.
Perhaps most importantly, \citet{rosolowsky05} studied mass spectra in the Milky Way (both inner and outer) and the Local Group using the aforementioned maximum likelihood method and a synthesis of cloud catalogs from the literature, including \citet{solomon87} and \citet{scoville87} (in the inner Galaxy), and \citet{heyer01} and \citet{brunt03} (in the outer Galaxy).
For the inner Galaxy, \citet{rosolowsky05} measured a truncated power law mass spectrum with $\gamma=-1.5 \pm 0.1$ and $M_0 \approx 3 \times 10^6 \Msun$; this $\gamma$ is entirely consistent with our inner Galaxy result of $-1.6 \pm 0.1$, and the $M_0$ is within an order of magnitude to our result.
In the outer Galaxy, \citet{rosolowsky05} measured a power law mass spectrum with no truncation and $\gamma=-2.1 \pm 0.2$, which also matches our measurement of $-2.2 \pm 0.1$.
It is worth noting that the results from \citet{rosolowsky05} are in almost exact agreement with the results presented here, even though the data used by the catalogs compiled in \citet{rosolowsky05} are entirely different from the DHT survey used here.

Cloud mass functions in the extragalactic environment of M51 were studied by \citet{colombo14}, who fit truncated power law mass functions to various regions throughout the disk of M51, focusing on clouds $10^6 \Msun$ and above.
They found that the power index $\gamma$ was steeper (more negative) in the external regions of the galaxy, ranging from 
$\gamma=-1.33$ in the nuclear bar to $\gamma=-2.55$ downstream of the outer spiral arm regions.
The truncation masses $M_0$ varied from $5.2\times10^6 \Msun$ (in the nuclear bar) to $1.6 \times 10^8 \Msun$ in the external parts of the spiral arms (although no clouds with masses at this extreme value were present; $N_0 \approx 0$ in this region).
They also found that the outer regions, outside of spiral arms, were better described by a pure power-law mass spectrum instead of a truncated power law spectrum, similarly to the outer Galaxy results we present above. 
They interpret these results to indicate that high density regions (spiral arms and the inner regions of M51's disk) have processes that actively accumulate clouds into high-mass GMCs (with an upper limit on mass), while the diffuse outer and inter-arm regions lacked these accumulation processes and encouraged the survival of smaller, low-mass clouds.
Both local gravitational instabilities and large-scale dynamical effects are indicated to play a role in the formation of clouds in M51.
The results seen in M51 are similar to the present study, as we also find that the slope $\gamma$ varies between high-density and low-density regions: the outer Galaxy preferentially hosts lower-mass clouds than the inner Galaxy, with ramifications for how the Milky Way forms and grows clouds.
Similarly, the inner regions of both M51 and the Milky Way have mass spectra with high truncation masses, whereas the outer, less dense regions do not have a sharp truncation and are better described by pure power laws.

In M33, \citet{rosolowsky05} report an even steeper mass spectrum based on the catalog of \citet{engargiola03}, with $\gamma = -2.9 \pm 0.4$, although this measurement may be biased by only sampling the high-mass end of the mass spectrum where the slope gets steeper.
\citet{rosolowsky05} report a mass spectrum in the LMC similar to that in the inner Milky Way, with $\gamma = -1.7 \pm 0.2$ and a truncation mass of $\sim 3 \times 10^6 \Msun$.
Mass spectra in extragalactic regions have been reported for IC 10 \citep[$\gamma=-1.71 \pm 0.06$;][]{leroy06}, the LMC \citep[$\gamma=-1.74 \pm 0.08$;][]{fukui08}, and M31 \citep[$\gamma=-1.55 \pm 0.20$;][]{rosolowsky07a}.
These extragalactic results are summarized in \citet{fukui10}, where the similarity in the power index $\gamma$ has been interpreted to mean that GMC properties are relatively consistent between galaxies.

%% file: test_table_thing.tex
\begin{deluxetable}{ccccccccc}
\tablecaption{Cloud Catalog \label{tab:catalog}}
\tablehead{\colhead{$l$} & \colhead{$b$} & \colhead{$v_{\textrm{LSR}}$} & \colhead{$\sigma_r$} & \colhead{$\sigma_v$} & \colhead{Distance} & \colhead{KDA} & \colhead{$R$} & \colhead{Mass}\\ \colhead{$\mathrm{{}^{\circ}}$} & \colhead{$\mathrm{{}^{\circ}}$} & \colhead{$\mathrm{km\,s^{-1}}$} & \colhead{$\mathrm{{}^{\circ}}$} & \colhead{$\mathrm{km\,s^{-1}}$} & \colhead{$\mathrm{kpc}$} & \colhead{ } & \colhead{$\mathrm{pc}$} & \colhead{$\mathrm{M_{\odot}}$}}
\startdata
13.29 & -0.10 & 97.17 & 0.08 & 3.63 & 10.37 & F & 28.90 & $4.54 \times 10^4$ \\
13.37 & -0.07 & 113.19 & 0.07 & 2.06 & 10.00 & F & 24.84 & $3.15 \times 10^4$ \\
13.83 & -0.11 & 85.76 & 0.17 & 9.08 & 10.69 & F & 59.50 & $6.50 \times 10^5$ \\
13.85 & -0.04 & 115.69 & 0.09 & 2.40 & 9.95 & F & 28.50 & $3.63 \times 10^4$ \\
13.94 & 0.22 & 47.01 & 0.07 & 1.36 & 3.87 & N & 8.62 & $3.36 \times 10^4$ \\
14.07 & 0.76 & 52.35 & 0.06 & 4.38 & 11.84 & F & 22.87 & $1.19 \times 10^5$ \\
14.14 & -0.59 & 20.08 & 0.18 & 1.69 & 2.06 & N & 11.96 & $9.48 \times 10^4$ \\
14.21 & 0.40 & 122.42 & 0.22 & 4.23 & 6.28 & N & 46.74 & $1.95 \times 10^5$ \\
14.22 & -0.20 & 39.72 & 0.17 & 2.15 & 3.44 & N & 19.75 & $2.77 \times 10^5$ \\
14.38 & 0.41 & 23.53 & 0.07 & 4.56 & 13.43 & F & 32.10 & $3.88 \times 10^5$ \\
14.54 & -0.06 & 142.79 & 0.33 & 8.45 & 9.34 & F & 102.47 & $9.90 \times 10^5$ \\
14.61 & -0.34 & 59.13 & 0.13 & 1.75 & 4.34 & N & 18.41 & $6.27 \times 10^4$ \\
14.68 & -0.00 & 66.42 & 0.06 & 2.30 & 11.33 & F & 23.05 & $1.89 \times 10^5$ \\
14.70 & -2.24 & -7.96 & 0.17 & 0.95 & 16.79 & U & 97.40 & $7.86 \times 10^4$ \\
14.87 & -0.71 & 75.86 & 0.35 & 4.51 & 4.95 & N & 58.20 & $2.14 \times 10^5$ \\
14.90 & -0.05 & 26.97 & 0.07 & 1.47 & 13.21 & F & 29.88 & $3.47 \times 10^5$ \\
15.00 & -0.25 & 106.99 & 0.34 & 8.37 & 10.16 & F & 114.77 & $9.75 \times 10^5$ \\
15.35 & -1.09 & 55.91 & 0.13 & 1.93 & 4.12 & N & 17.18 & $5.72 \times 10^4$ \\
15.54 & -0.06 & -8.67 & 0.06 & 1.53 & 16.79 & U & 33.99 & $7.01 \times 10^4$ \\
15.60 & -0.39 & 57.83 & 0.15 & 1.61 & 4.18 & N & 21.34 & $6.55 \times 10^4$ \\
15.62 & -0.01 & 119.93 & 0.13 & 3.75 & 9.83 & F & 41.77 & $7.74 \times 10^4$ \\
15.76 & -2.23 & -8.40 & 0.08 & 0.81 & 16.73 & U & 43.53 & $2.89 \times 10^4$ \\
15.82 & -0.46 & 81.65 & 0.13 & 4.82 & 10.86 & F & 46.47 & $2.53 \times 10^5$ \\
16.03 & -3.67 & -8.80 & 0.09 & 1.14 & 16.76 & U & 49.65 & $2.49 \times 10^4$ \\
16.07 & -0.62 & 66.31 & 0.04 & 2.79 & 11.37 & F & 13.67 & $6.91 \times 10^4$ \\
16.13 & 0.36 & 142.39 & 0.08 & 2.23 & 9.26 & F & 24.78 & $3.14 \times 10^4$ \\
16.24 & -1.02 & 56.80 & 0.17 & 2.56 & 4.07 & N & 23.30 & $1.93 \times 10^5$ \\
\enddata
\caption{ Table of the 1064 clouds identified in this work. 
    Column ``KDA'' refers to the resolution of the kinematic distance ambiguity; 
    `U' refers to unambiguous distances, while `N' and `F' denote near and far.
    The full table appears in the online version of the journal.}
\end{deluxetable}

%% file: mass_function_fits_table.tex



\begin{deluxetable}{cc|cccc|cc}

  \tabletypesize{\scriptsize}
  
  \tablecaption{Parameters of Mass Spectra throughout the Milky Way \label{tab:massfunction}}
  \tablewidth{0pt}

  \tablehead{

    \colhead{Region} &
    \colhead{ } & 
    \multicolumn{3}{c}{Truncated Power Law} &
    \colhead{ } &
    \multicolumn{2}{c}{Power Law}  \\
    \cline{3-5}  
    \cline{7-8}  \\

    \colhead{ } &
    \colhead{Number} &
    \colhead{$N_0$} & 
    \colhead{$M_0$} & 
    \colhead{$\gamma$} &
    \colhead{} & 
    \colhead{$M_0$ } &
    \colhead{$\gamma$ } \\
  }
  
  \startdata 

   outer & & & & & & &  \\
   \hline



   I & 57 &
       $5.61 \pm 4.45$ & $(8.46 \pm1.90) \times 10^5$ & $-1.72 \pm 0.20$ & &
       $(1.76 \pm 0.66) \times 10^6$ & $-2.01 \pm 0.11$ \\


   II & 61 &
       $0.56 \pm 1.15$ & $(2.67 \pm1.56) \times 10^6$ & $-2.05 \pm 0.15$ & &
       $(1.29 \pm 0.58) \times 10^6$ & $-2.10 \pm 0.18$ \\



   III & 32 &
       $10.53 \pm 6.04$ & $(3.76 \pm0.68) \times 10^5$ & $-1.55 \pm 0.27$ & &
       $(6.56 \pm 1.91) \times 10^5$ & $-2.18 \pm 0.10$ \\


   IV & 33 &
       $6.52 \pm 5.94$ & $(5.19 \pm 2.57) \times 10^6$ & $-1.37 \pm 0.13$ & &
       $(1.59 \pm 1.03) \times 10^7$ & $-1.60 \pm 0.06$ \\



   II+III & 93 &
       $3.48 \pm 4.52$ & $(9.13 \pm 7.53) \times 10^5$ & $-1.98 \pm 0.19$ & &
       $\mathbf{(1.53 \pm 0.52) \times 10^6}$ & $\mathbf{-2.17 \pm 0.12}$ \\

	\hline

   inner & & & & & &  \\
   \hline


   I & 74 &
       $6.64 \pm 5.02$ & $(8.19 \pm 2.57) \times 10^6$ & $-1.59 \pm 0.13$ & &
       $(2.60 \pm 1.21) \times 10^7$ & $-1.82 \pm 0.07$ \\


   IV & 79 &
       $3.89 \pm 3.32$ & $(1.53 \pm 0.62) \times 10^7$ & $-1.61 \pm 0.12$ & &
       $(3.78 \pm 1.93) \times 10^7$ & $-1.75 \pm 0.05$ \\

   I+IV & 153 &
       $\mathbf{11.19 \pm 5.70}$ & $\mathbf{(1.02 \pm 0.23) \times 10^7}$ & $\mathbf{-1.59 \pm 0.11}$ & &
       $(6.04 \pm 1.95) \times 10^7$ & $-1.81 \pm 0.04$ \\

	\hline

   all & & & & & &  \\
   \hline


   all & 218 &
       $8.54 \pm 3.99$ & $(1.05 \pm 0.21) \times 10^7$ & $-1.72 \pm 0.09$ & &
       $(5.12 \pm 1.55) \times 10^7$ & $-1.89 \pm 0.05$ \\


  \enddata
  \tablecomments{ Mass function parameter fits, computed via the maximum-likelihood method of \citet{rosolowsky05}. 
  To each region, we fit both a traditional power law and truncated power law mass function; where $N_0 \gg 1$, the truncated form is appropriate, while $N_0 \lesssim 1$ indicates a traditional power-law is more appropriate.
  The quoted errors on parameter fits are computed as the median absolute deviations from 100 iterations of bootstrap trials. }


\end{deluxetable}


%% file: caveats.tex

In this section we present a discussion of the caveats of our methods.
As part of this discussion, we present a quantitative analysis of how the results of our study are sensitive to additional random noise in the emission data.

\subsection{Noise analysis}

Some readers may be concerned that, because the relationships identified by dendrograms can be influenced by the presence of noise, the conclusions of this work may be noise-dependent.
To demonstrate that the presence of noise does not substantially bias our conclusions, we have performed a series of 25 ``noise-added'' analyses of a subset of our data, and re-run the analysis software for each trial.
In these trials, we added random noise to the first quadrant dataset before processing the data, identifying clouds, and repeating our analysis from Sections \ref{sec:quadbyquad} and \ref{sec:all_galaxy_analysis}.
By compiling the results of these 25 trials, we can show that our conclusions are robust to the addition of substantial noise, even when this noise exceeds the already-present noise by a factor 2.

The procedure of this noise analysis is described here.
For each trial of this analysis, we started with the interpolated first quadrant DHT08 datacube and added normally-distributed noise with rms intensity at a given level, which varied from trial to trial between 0.045 K and 0.36 K\footnote{The noise per channel inherent in the raw DHT08 cube is 0.18 K, as previously noted in Table \ref{tab:surveys}.}. 
In this survey, the pixel spacing (0\fdg125) closely matches the beam FWHM (0\fdg14) of the 1.2 m CfA telescope, so for the purposes of this demonstration we consider the noise to be un-correlated from pixel to pixel.
We then performed moment-masking on this noise-added data, masking at a level determined by the noise present in a smoothed version of the datacube (as recommended in \citealt{dame11_moment}).
With the moment-masked data, we then performed the full dendrogram-based decomposition and analysis, as described in Section \ref{sec:methods} and illustrated in Figure \ref{fig:flowchart}.

\begin{figure}
	\plotone{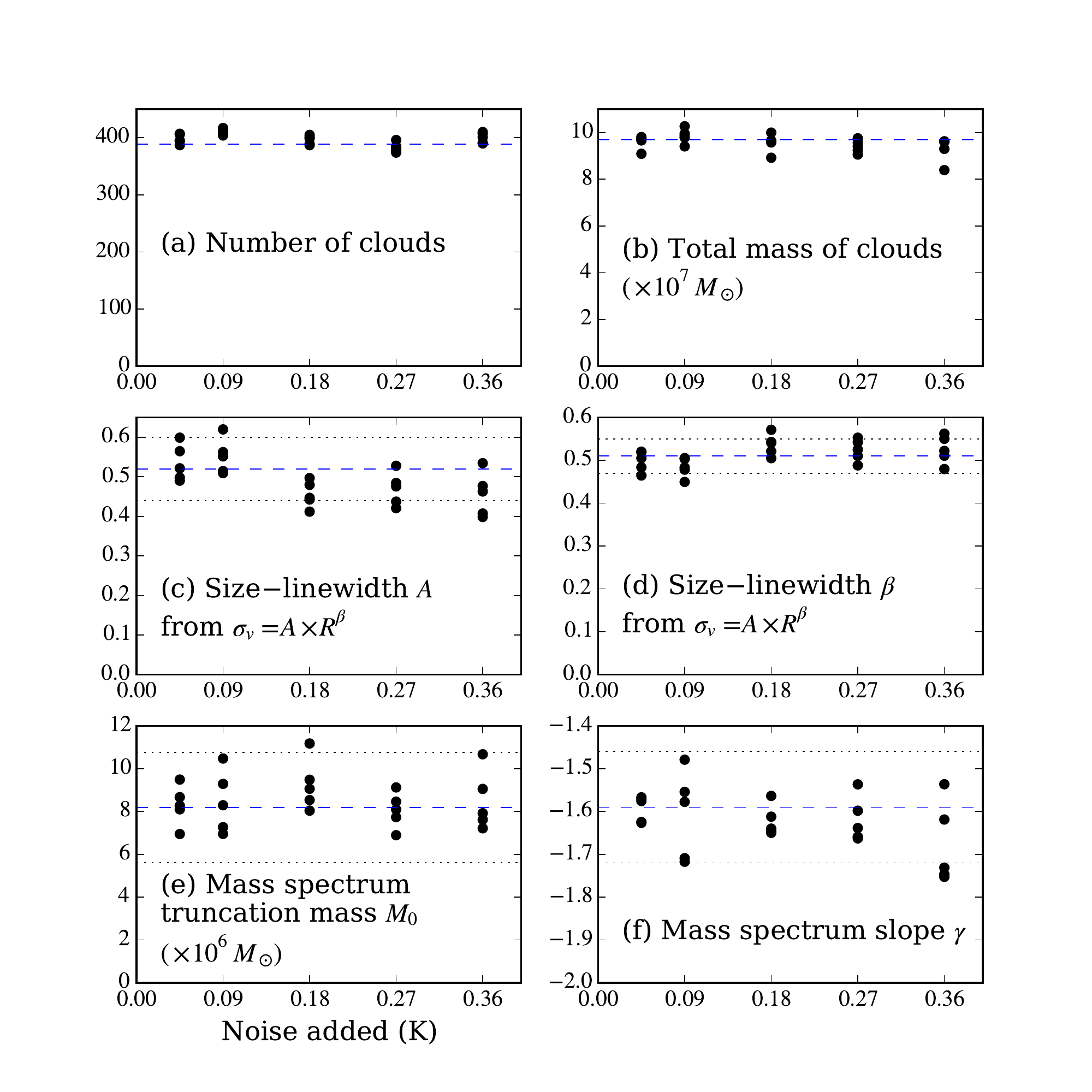}
	\caption{Results of re-running our analysis on a modified version of the first quadrant data, in which random noise was added. 
	In each panel, the dashed blue line indicates the value derived from the first quadrant data \textit{without} noise added; in panels (c) through (f), a pair of dotted lines above and below the blue dashed line indicates the estimated error bars on that value, as presented in Section \ref{sec:all_galaxy_analysis}. (a)~--~the number of clouds identified in the first quadrant. (b)~--~the total mass of those clouds. (c)~--~the multiplicative coefficient $A$ of the size-linewidth relation. (d)~--~the power-law slope $\beta$ of the size-linewidth relation. (e)~--~the truncation mass $M_0$ of the truncated power-law cloud mass spectrum. (f)~--~the power index $\gamma$ of the mass spectrum. In panels (c) through (f), only inner galaxy parameters are shown, for simplicity. \label{fig:noise_experiment}}
\end{figure}

From running this analysis on 25 trials using various levels of added noise, we have extracted the following parameters for comparison:
the number of clouds identified in the first quadrant, the total mass of those clouds, the size-linewidth fits for clouds within the solar circle, and the mass spectrum fits for clouds within the solar circle.
These results are presented in Figure \ref{fig:noise_experiment}.
In each case, even when the added noise is twice (0.36 K) the original noise level, the added noise induces only slight deviations from the values derived from the original data. 
In the parameters for which error estimates are available, the scatter in these noise-added analyses matches those error bars.

One caveat of this analysis is that we lack a ``noise-free'' dataset to compare against, and so this analysis only probes the presence of random \textit{additional} noise.
A direct test of the effects of adding noise to noise-free emission data in dendrograms was presented by \citet{beaumont13}; in a result that reinforces the above analysis, they concluded that the presence of noise made little difference in identifying dendrogram structures or measuring their properties. 
They noted that the high opacity and crowding of \ce{^{12}CO} emission in dense regions had a more severe effect on cloud identification.

\subsection{Binarity}

In this analysis we use a dendrogram technique that enforces binary mergers, as opposed to allowing an arbitrary number of leaves to merge at a given branch. 
This binary approach is the standard and most common way to use dendrograms (starting with \citealt{rosolowsky08}, and including \citealt{goodman09}, \citealt{pineda09}, \citealt{shetty10}, \citealt{beaumont13}, and \citealt{burkhart13}).
In the presence of noise, binary dendrograms face the following issue: if three (or more) leaves merge together at a nearly identical intensity level, then the dendrogram's choice of which structures to merge first is randomly determined by the noise in the data.

One example of a way to deal with this was seen in \citet{storm14}, who implemented a non-binary dendrogram algorithm to study structure of \ce{N2H^+} emission in the Barnard 1 cloud.
Non-binary dendrograms were particularly relevant for the analysis in \citet{storm14} due to the relatively small area coverage, low number of emission structures (which placed a premium on carefully classifying the few structural relationships available), and their focus on analyzing tree statistics.
\citet{storm14} note that tree statistics are dependent on whether dendrograms are constructed to be binary or non-binary. 
However, they also caution that even non-binary dendrograms can be adversely affected by noise: 
for example, when one structure's ``ideal'' merge level is very near 1-$\sigma$ above or below the merge level of two other structures, then noise will determine whether the one structure merges together or separately from the two other structures.

In this work, instead of adopting non-binary dendrograms, we have dealt with this issue by setting the minimum number of pixels in any given structure ($N_{min}$) at 20, to minimize the impact that noise in individual pixels can have in influencing the inferred structure.
As mentioned in Section \ref{sec:methods}, this choice has a trade-off of limiting our sensitivity to small clouds, especially when they lie at large distances.
Because our work is more concerned with identifying clouds than on analyzing their hierarchical substructure on a statistical level, there is less need to use non-binary dendrograms.

\subsection{Astronomical dendrograms are not a clustering method}

Though an astronomical dendrogram is a hierarchical abstraction of structure within an emission datacube, it is conceptually distinct from hierarchical clustering methods applied to abstract datapoints (e.g., as discussed in ``Cluster Analysis,'' \citealt{book:856838}). Specifically, an astronomical dendrogram does not cluster datapoints; rather, it segments the pixels within an image or datacube.
A number of caveats which are relevant to hierarchical clustering techniques -- for example, the existence of many incompatible ways to quantify ``similarity'', of which none can be defined as ``optimal'' -- are therefore not applicable to our analysis.

\subsection{Interpreting dendrograms of PPV data}

\citet{rosolowsky08} present three paradigms for measuring flux in dendrograms: the ``clipping'', ``bijection'', and ``interpolation'' paradigms.
As noted in Section \ref{sec:methods}, we have chosen the ``bijection'' paradigm for extracting flux from dendrogram structures. 
This paradigm assigns the total flux contained within that structure (i.e., it sums the intensities of all pixels associated with a structure), and does not subtract out the background flux (as done in ``clipping'') nor does it attempt to infer the flux between the boundaries of the structure and a zero-intensity level (``extrapolation'').
\citet{rosolowsky08} note cases in which either ``clipping'' or ``extrapolation'' might be more appropriate; for example, clouds superposed on a bright emission background might be better measured using the ``clipping'' paradigm (although \citealt{rosolowsky08} notes that this would be a very conservative flux estimate); properties of a bright cloud superposed on a faint background (or a faint cloud in a noisy dataset extracted with a high $T_{min}$) might be more accurately measured in the ``extrapolation'' paradigm, where the faint outer regions of such a cloud are excluded from the associated dendrogram structure.
We have chosen the bijection paradigm as the simplest and likely most accurate paradigm on average, given the very large number of structures considered in this study, but note that a more refined approach might be worthwhile for detailed study of small numbers of clouds.

\citet{burkhart13} note that in turbulent environments, structures in PPV data do not strictly map to PPP structures in a one-to-one way.
This issue is worst for subsonic turbulent motions.
They find, however, that supersonic turbulent environments give a better mapping between PPP and PPV structures, and that the bright leaves of a PPV dendrogram map to the high-density regions of a PPV cube.
They conclude that ``structures in PPV space at the level of the leaves can generally be interpreted as 3D density structures.''
Because GMCs exhibit supersonic linewidths (e.g., \citealt{ballesteros-paredes11}), this gives confidence that the brightest emission  structures in data such as the DHT survey represent real density structure, although smaller subsonic regions would be distorted in PPV space.

\citet{beaumont13} also study projection effects between PPP and PPV space.
They note that in dense regions, \ce{^{12}CO} shows the worst effects of crowding and overlap, while \ce{^{13}CO} is more faithful, especially in small, dense regions within a cloud.
They find that the effects of superposition from PPP to PPV space induces a $\sim 40\%$ scatter in masses, sizes, and velocity dispersions in the \ce{^{13}CO} emission.
Despite these caveats, they note that gravity can act to modestly reduce confusion; on the size and mass scales analyzed in the present study, clouds are generally affected by gravity \citep{rosolowsky08}.

%% file: conclusions.tex

In this study, we have created a catalog of massive molecular clouds throughout the Galactic plane,
 using a consistent dendrogram-based decomposition of the \ce{^{12}CO} CfA-Chile survey \citep*{dame01},
 which is by far the most uniform survey of molecular gas in the Galaxy.
This catalog, with 1064 massive clouds totaling $2.5\times 10^8 \Msun$, contains $25^{+10.7}_{-5.8} \%$ of the molecular gas mass of the Galaxy, and traces several spiral arms, most notably the Sagittarius, Perseus, Outer, Carina, and Scutum-Centaurus arms.
We display a ``face-on'' Galactic view of the clouds in this catalog.
From this distribution of clouds, we have created ``simulated'' images of what an extragalactic observer would see of the Milky Way's CO emission.

We measure Larson's first law, the size-linewidth relationship ($\sigma_v = A R^\beta$), in all Galactic regions. We find that its power index $\beta$ is near $0.5$ everywhere, which may (as discussed in Section \ref{sec:larson}) be an observational bias, but that the scaling coefficient $A$ is significantly higher in the inner Galaxy $(A=0.50\pm0.05)$ than in the outer Galaxy $(A=0.38\pm0.05)$, indicating that Galactic environment plays some role in setting the linewidth of clouds.

The cloud mass spectrum varies substantially by Galactic region.
We find that in the inner Galaxy, it is best described by a truncating power law with power index $\gamma=-1.6\pm0.1$ and truncation mass $M_0 = (1.0 \pm 0.2) \times 10^7 \Msun$.
In the outer Galaxy, the mass spectrum is better described by a non-truncating power law
with $\gamma=-2.2\pm0.1$ and an upper mass $M_0 = (1.5 \pm 0.5) \times 10^6 \Msun$.
This indicates that the inner Galactic environment is much more favorable for the production and survival of massive molecular complexes, while the outer Galaxy systematically fosters less massive clouds.

This study can be used to give global context to nearby studies of GMCs.
Further, this study places molecular clouds in the Galaxy, for the first time in such a global catalog, in direct comparison with extragalactic studies of Larson's laws and the mass spectrum.
Differences between the inner Galaxy and the outer Galaxy presented here help corroborate similar differences seen, for example, in M51:
inner Galaxy clouds have higher linewidths than comparably sized outer Galaxy clouds, and the inner Galaxy mass spectrum is less steep until an upper truncation mass.